\documentclass[useAMS,usenatbib,referee]{mn2e}
\usepackage{lipsum}
\usepackage{latexsym}
\usepackage{amssymb}
\usepackage{graphicx}
\usepackage[utf8x]{inputenc}
\usepackage{longtable}
\usepackage{multicol}
\usepackage{supertabular}
\usepackage{float} 
\newcommand{\ion}[2]{{\textrm{#1}}\,{\textrm{\sc #2}}}

\usepackage{wrapfig}
\usepackage{mathtools}
\usepackage{graphicx}
\usepackage{wrapfig}
\usepackage{acronym}
\usepackage[acronym]{glossaries}
\usepackage[utf8x]{inputenc}
\usepackage{caption}
\DeclareCaptionFont{mysize}{\fontsize{8}{9.6}\selectfont}
\usepackage[titletoc,title]{appendix}
    \apptocmd{\appendices}{\let\LaTeXStandardTheSection\thesection
    \apptocmd{\figure}{\let\thesection\LaTeXStandardTheSection}{}{}
\apptocmd{\thesection}{:}{}{}}{}{}

\usepackage{xparse}
\makeatletter
\DeclareDocumentEnvironment{appendixfig}{}{
   \par\medskip\noindent
   \begin{minipage}{\linewidth}
   \def\@captype{figure}
   \centering
}
{
\end{minipage}
\par\bigskip
}
\makeatother

\title{Interaction effects on galaxy pairs with Gemini/GMOS-\,III: Stellar population synthesis}

\author[Krabbe et al.]
{A.~C.~Krabbe$^{1}$\thanks{E-mail:angelak@univap.br}, D.~A.~Rosa$^1$, M.~G.~Pastoriza$^2$,
G.~F. H\"agele $^{3,4}$, M.~V. Cardaci$^{3,4}$, 
\newauthor{O.~L.~Dors Jr.$^1$,  C.~Winge$^2$.}\\
$^1$ Universidade do Vale do Para\'iba, Av. Shishima Hifumi, 2911, Cep
12244-000, S\~ao Jos\'e dos Campos, SP, Brazil\\
$^2$ Instituto de F\'{\i}sica, Universidade Federal do Rio Grande do Sul, Av. Bento Gon\c calves, 9500, Cep 91359-050, Porto Alegre, RS, Brazil\\
$^3$ Instituto de Astrof\'isica de La Plata (CONICET La Plata--UNLP), Argentina. \\
$^4$ Facultad de Ciencias Astron\'omicas y Geof\'{\i}sicas, Universidad Nacional de La Plata, Paseo del Bosque s/n, 1900 La Plata, Argentina\\
}
\begin{document}

\date{Accepted -. Received -.}

\pagerange{\pageref{firstpage}--\pageref{lastpage}} \pubyear{2014}

\maketitle

\label{firstpage}

\begin{abstract}
We present an observational study of the impacts of the interactions on the stellar population in a sample of galaxy pairs. 
Long-slit spectra  in the wavelength range 3440-7300\,\AA\  obtained with the Gemini Multi-Object Spectrograph (GMOS) at Gemini South 
for fifteen galaxies in nine close pairs were used.
The spatial distributions of the stellar population contributions were obtained using the stellar population synthesis code {\scriptsize\,STARLIGHT}. 
Taking into account the different contributions to the emitted light, we found that 
most of the galaxies in our sample 
are dominated by the young/intermediate stellar populations. 
This result differs from the one derived for isolated galaxies where the old stellar population  dominates the disc surface brightness. 
We interpreted such different behavior as being due to the effect of gas inflows along the disk of interacting galaxies
 on the  star formation in a time scale of the order of  about  2\,Gyr.
We also found that, in general,  the secondary galaxy of the pairs
has a higher contribution of the young stellar population than the primary one.
We compared the estimated values of the stellar and nebular extinctions derived from the synthesis method and the H$\alpha$/H$\beta$
 emission-line ratio finding that the nebular extinctions are systematically higher  than stellar ones by about a factor of 2.
We did not find any correlation between nebular and stellar metallicities. 
We neither found a correlation between stellar metallicities and ages while a positive correlation between nebular metallicities and 
stellar ages was obtained, with the older regions being the most metal-rich.

\end{abstract}


\section{Introduction}
Interactions and mergers of galaxies  have a deep influence on the  star formation  pattern and on the  chemical enrichment of the interstellar medium  of these objects. 

It has been well known that  interacting galaxies show enhanced star formation when compared them with isolated galaxies
(e.g. \citealt{larson78,sekiguchi92,donzelli97,barton03,woods07,scudder12,patton13}).
The enhancement of the star formation rate (SFR) has been observed as being a function of the separation of the close pair galaxies. 
In fact, \citet{patton13}, by using spectroscopic data of about 211\,000 star forming galaxies taken from the  
Sloan Digital Sky Survey  Data Release 7 \citep{abazajian09}, found a clear enhancement of the SFR for galaxies with projected separations up to 150 kpc, being the strongest enhancements for separations 
lower than $\sim 20$ kpc (see also \citealt{barton00,lambas03,nikolic04,scudder12,casteels13,satyapal14}). 
The effect of the interaction on the SFR  seems  to be more  pronounced for galaxy pairs with low-mass  \citep{woods07,krabbe08} and gas-rich \citep{chien07}.

The SFR enhancements are accompanied of other events such as perturbations in the radial velocity field and dilution of the metallicity gradient. 
Interaction-induced  flows of gas with low metallicity from the outer parts of the disk of a galaxy can decrease the metallicity in
the inner regions and modify the radial abundance gradients. 
In fact, \citet{krabbe08}, \citet{kewley10}, \citet{rupke10}, and \citet{rosa14}, analyzing spectroscopic  data 
of \ion{H}{ii} regions located in galaxy pairs,  found shallow metallicity gradients.  In particular, \citet{rosa14} 
found a break in the oxygen gradient (generally used as metallicity tracer)
at a certain  galactocentric distance in four interacting galaxies of their sample: AM\,1219A, AM\,1256B, AM\,2030A and 
AM\,2030B.  Interestingly, these authors reported that the extreme SFR values estimated from the H$\alpha$ emission-line fluxes, 
(minimum SFR for AM1219A and AM2030B, and maximum for AM\,1256B and AM\,2030) are located very close to the oxygen gradient break zones. 
Rosa and collaborators also suggested that for the AM1219A and AM2030B systems, the minimum values of the SFRs and the break zones could be associated with corotation radii.   

Determinations of SFRs based on H$\alpha$ emission-line fluxes give information of the ionizing star forming clusters. To access to the history of the stellar population, we should perform studies based on stellar population synthesis. Studies of the spatial distribution of the stellar population components in  interacting galaxies yield important insights on the  effects of the interaction on the kinematics of the galaxies involved, allowing to know the periods on which occurred the bursts of star formation as well as the stage of the interaction \citep{krabbe08}. 

On the previous papers of this series, we presented observational studies of the spatial variation 
of the electron density (\citealt{krabbe14}, hereafter Paper\,I) and of the metallicity (\citealt{rosa14}, hereafter Paper\,II) of \ion{H}{ii} regions located in a sample of interacting systems of galaxies. 
In the present work, we used these data to obtain the stellar population across the disk of these galaxies and to study the impacts of the interaction on the stellar population of the galaxies in our sample.  
The main goals of our study are:\\
i) To estimate the main epoch of star formation along the disk of the interacting galaxies.\\
ii) To analyze  the relation between the stellar population  distributions
with the metallicity  of the gas phase.\\
iii) To investigate the  relation  between the age of the dominant stellar population  with 
several nebular parameters (e.g. metallicity, extinction).

This paper is organized as follows. In Section~\ref{obser_data} we summarize the observations and data reduction. 
In Section.~\ref{deter} the method used to perform the stellar population synthesis
is described. Section~\ref{res3} presents the detailed results of the 
stellar population synthesis. The discussion of our results is given in Section~\ref{disc} while the conclusions are presented in Section~\ref{conc}.

\begin{table*}
\footnotesize
\caption{Observed sample}
\label{tabs}
\label{sample}
\scriptsize
\begin{tabular}{lllccllll}
\hline
\noalign{\smallskip}
  System name  & Individual names	    & Designation    &$\alpha$(2000)                  & $\delta$(2000)                  & PA (\degr)& $\Delta\,\lambda$ (\AA)&\textit{E}(B-V) & NS(kpc)\\                 
\hline
\noalign{\smallskip}
AM\,1054-325         & ESO\,376-IG\,027		   &  AM\,1054A           & $10^{\rm{h}}56^{\rm{m}}58\fs2$ & $-33^{\rm{h}}09^{\rm{m}}52^{\rm{s}}$ & 77   & 4280-7130              &0.084           & -   \\ 
                     & ESO-LV\,37660271		   &  AM\,1054B           & 10 ~56 ~59.0		   & $-$33 ~09 ~39		     & 77        & 4280-7130              &0.083           & 17   \\ 
% %%%%%%%%%%%%%%%%%%%%%%%%%%%%%%%%%%%%%%%%%%%%%%%%%%%%%%%%%%%%%%%%%%%%%%%%%%%%%%%%%%%%%%%%%%%%%%%%%%%%%%%%%%%%%%%%%%%%%%%%%%%%%%%%%%%%%%%%%% %%%%%%%%%%%%%%%%%%%%%%%%%%
AM\,1219-430         & ESO\,267-IG\,041            &AM\,1219A             & 12 ~21 ~57.3	           & $-$43 ~20 ~05                   & 162,341	 & 3350-7130              &0.109           & -   \\
                     & FAIRALL\,0157		   &AM\,1219B             & 12 ~22 ~04.0	           & $-$43 ~20 ~21	             & 25        & 3350-7130              &0.110           & 33.7    \\
%%%%%%%%%%%%%%%%%%%%%%%%%%%%%%%%%%%%%%%%%%%%%%%%%%%%%%%%%%%%%%%%%%%%%%%%%%%%%%%%%%%%%%%%%%%%%%%%%%%%%%%%%%%%%%%%%%%%%%%%%%%%%%%%%%%%%%%%%%%%%%%%%%%%%%%%%%%%%%%%%%%%
AM\,1256-433         & ESO\,269-IG\,023\,NED01	   &AM\,1256B             & 12 ~58 ~57.6		   & $-$43 ~50 ~11                   & 292,325   & 4280-7130              &0.091           & 91.6  \\
%%%%%%%%%%%%%%%%%%%%%%%%%%%%%%%%%%%%%%%%%%%%%%%%%%%%%%%%%%%%%%%%%%%%%%%%%%%%%%%%%%%%%%%%%%%%%%%%%%%%%%%%%%%%%%%%%%%%%%%%%%%%%%%%%%%%%%%%%%%%%%%%%%%%%%%%%%%%%%%%%%%%
AM\,1401-324         & ESO\,384-G\,041  	   &AM\,1401A             &14 ~04  ~14.7	           & $-$33  ~01 ~32	             & 294,41    & 3350-6280              &0.078           & 23.4    \\
 %%%%%%%%%%%%%%%%%%%%%%%%%%%%%%%%%%%%%%%%%%%%%%%%%%%%%%%%%%%%%%%%%%%%%%%%%%%%%%%%%%%%%%%%%%%%%%%%%%%%%%%%%%%%%%%%%%%%%%%%%%%%%%%%%%%%%%%%%%%%%%%%%%%%%%%%%%%%%%%%%%%
AM\,2030-303         & ESO\,463-IG\,003\,NED01	   &AM\,2030A             & 20 ~33  ~56.3		   & $-$30  ~22 ~41		     & 75        & 4280-7130              &0.070           &  -     \\	    
                     & ESO\,463-IG\,003\,NED02	   & AM\,2030B            &20 ~33  ~59.7	           & $-$30  ~22 ~29                  & 75,22     & 4280-7130              &0.060           & -    \\	   
                     & ESO\,463-IG\,003\,NED03	   & AM\,2030B            &20 ~33  ~59.7	           & $-$30  ~22 ~23		     & 22        & 4280-7130              &0.060           &  40.5   \\	   
%%%%%%%%%%%%%%%%%%%%%%%%%%%%%%%%%%%%%%%%%%%%%%%%%%%%%%%%%%%%%%%%%%%%%%%%%%%%%%%%%%%%%%%%%%%%%%%%%%%%%%%%%%%%%%%%%%%%%%%%%%%%%%%%%%%%%%%%%%%%%%%%%%%%%%%%%%%%%%%%%%%%
AM\,2058-381         & ESO\,341-G\,030		   & AM\,2058A            &21 ~01 ~39.1			   & $-$38 ~04 ~59	             &42,125,350 & 3350-7130              &0.050           & -   \\
                     & ESO\,341-G\,030\,NOTES01	   & AM\,2058B            &21 ~01 ~39.9			   & $-$38 ~05 ~53	             &94,350     & 3350-7130              &0.050           & 44  \\
%%%%%%%%%%%%%%%%%%%%%%%%%%%%%%%%%%%%%%%%%%%%%%%%%%%%%%%%%%%%%%%%%%%%%%%%%%%%%%%%%%%%%%%%%%%%%%%%%%%%%%%%%%%%%%%%%%%%%%%%%%%%%%%%%%%%%%%%%%%%%%%%%%%%%%%%%%%%%%%%%%%%
AM\,2229-735         & AM\,2229-735\,NED01	   &AM\,2229A             &22 ~33 ~43.7			   & $-$73 ~40 ~47                   & 134,161   & 4280-7130              &0.037           &  24.5   \\
%%%%%%%%%%%%%%%%%%%%%%%%%%%%%%%%%%%%%%%%%%%%%%%%%%%%%%%%%%%%%%%%%%%%%%%%%%%%%%%%%%%%%%%%%%%%%%%%%%%%%%%%%%%%%%%%%%%%%%%%%%%%%%%%%%%%%%%%%%%%%%%%%%%%%%%%%%%%%%%%%%%
AM\,2306-721         & ESO\,077-G\,003		   &AM\,2306A             &23 ~09 ~39.3			   & $-$71 ~01 ~34	             &238,190    & 3350-7130              &0.030           & -  \\
                     & ESO\,077-IG\,004		   &AM\,2306B             &23 ~09 ~44.5			   & $-$72 ~00 ~04		     &118,190    & 4280-7130              &0.030           & 52.6  \\
%%%%%%%%%%%%%%%%%%%%%%%%%%%%%%%%%%%%%%%%%%%%%%%%%%%%%%%%%%%%%%%%%%%%%%%%%%%%%%%%%%%%%%%%%%%%%%%%%%%%%%%%%%%%%%%%%%%%%%%%%%%%%%%%%%%%%%%%%%%%%%%%%%%%%%%%%%%%%%%%%%%%
AM\,2322-821         & ESO\,012-G\,001, NGC\,7637  &AM\,2322A           &23 ~26 ~27.6			   & $-$81 ~54 ~42	             & 59,28,318 & 3450-7130              &0.181           & -  \\
                     & ESO\,012-G\,001\,NOTES01	   &AM\,2322B             &23 ~25 ~55.4			   & $-$81 ~52 ~41	             & 318       & 3350-7130              &0.179           & 33.7  \\
\hline
\noalign{\smallskip}
\end{tabular}
\begin{minipage}[c]{1\textwidth}
{Notes: Column (1): System identification from the  Arp-Madore catalogue \citep{arp87}. Column (2): Individual galaxy names from Eso/Uppsala catalogue \citep{eso82}, New General Catalogue \citep{ngccatalogue} and from Spectroscopic survey of southern compact and bright-nucleus galaxies of \cite{fairall79}. Column (3): Adopted designation. Columns (4) and (5): equatorial coordinates for the center of each observed galaxy. Column (6)-(9) Slit position angles, spectral wavelength coverage, B-V color excess and nuclear separation between components.}
\end{minipage}
\end{table*}

\section{Observational data}
\label{obser_data}

The present study is based on long slit spectroscopy of a sample of nine systems of galaxies in interaction, obtained with the 
Gemini Multi-Object Spectrograph (GMOS) attached to the 8\,m Gemini South telescope.
Spectra in the range of 3450  to 7130 \AA\ were obtained with two settings with the B600 grating,   slit width of 1$\arcsec$ and a spectral resolution of $\sim 5.5$ \AA. 
The blue setting provided a wavelength coverage from 3450 to 6280 \AA\ and the red setting of 4280 to 7130 \AA,  
both with  about the same spectral resolution.

In Table~\ref{tabs}   we listed the basic information of the observed galaxies: the name of the galaxy system, the names of the individual galaxies, our abbreviated designations, the celestial equatorial coordinates (J2000), the slit position angles, 
the wavelength range of the spectra, 
the  Galactic extinction E(B-V) as listed in NED\footnote{http://ned.ipac.caltech.edu/} of each galaxy, and the nuclear separation (NS) between the individual galaxies of the pairs.
%In Figures \ref{am1054_fenda} - \ref{am2229_fenda} of the Appendix A, the slit positions observed for this sample, superimposed on the GMOS-S r$\arcmin$ acquisition image are shown. 
Detailed information about the galaxy systems and the slit positions observed for the objects in this sample 
were provided in Paper\,I and II, with the exception of AM\,1401-324 system, that was not included in the  
previous works due to the wavelength coverage does not include the [N{\sc II}]$\lambda$6584 \AA\ emission-line.
 For the AM\,1401-324  system, the spectra were taken at two slit position on the sky (PA\,=\,294\degr\,  and PA\,=\,41\degr\,), with the goal of observing the nucleus and 
the brightest regions of the  main galaxy of the  system (see Fig.\,\ref{am1401_fenda}).
% In Figure \ref{fendas_1} the slit positions observed for this object, superimposed on the GMOS-S r$\arcmin$ acquisition image is shown.
The secondary galaxy of the AM\,1401-324 system was not observed.

The data reduction followed the standard procedures and it was  detailed  discussed in Paper\,I and Paper\,II. 
Basically, the  spectra comprise the flux contained in an aperture of
1$\arcsec$ $\times$1.152 $\arcsec$ which, considering a spatially flat cosmology with  $H_{0}$\,=\,71 $ \rm km\:s^{-1} Mpc^{-1}$ \citep{wright06} 
and the distances to the systems of our sample (see Paper\,II),
corresponds to apertures from  $\sim$200 to $\sim$1100\,pc on the galaxies plane. 
The spectra taken on red and blue setting were combined applying a median filter with the IRAF task {\scriptsize\,LSCOMBINE}.

\section{Stellar population synthesis}
\label{deter}

In order to obtain the stellar population contribution of our sample of interacting galaxies, we 
used the {\scriptsize\,STARLIGHT}  stellar population synthesis code \citep{cid04,cid05,mateus06,cid07,asari07}, which is discussed in details in  \citet{cid04,cid05}.
Briefly, the code fits the observed spectrum (\textit{O}$_{\lambda}$) of a galaxy 
using a combination of simple stellar populations\,(SSPs) 
obtained from evolutionary synthesis models by \citet{bruzual03}. 
These models are composed of spectra with a resolution of 3 \AA\ across the wavelength range 3200-9500 \AA\ with a wide range of metallicities. 
The {\scriptsize\,STARLIGHT} assumes the Padova 1994 tracks, as recommended 
by \citet{bruzual03}, and an initial mass function (IMF) by \citet{chabrier03} for stars with masses  between 0.1 and 100\,\textit{M}$_{\odot}$. 

The synthetic spectrum \textit{M}$_{\lambda}$  is solved by {\scriptsize\,STARLIGHT} according to the following equation:

\begin{equation}
\label{my}
\centering
M_\lambda= M_{\lambda_{0}} \left[\sum_{j=1}^{N_\star} \vec{x}_j b_j,_{\lambda} r_{\lambda} \right]  \otimes G(\nu_{\star},\sigma_{\star}),
\end{equation}
where b$_{j}$,$_{\lambda}$ is the reddened spectrum of the $j^{th}$ SSP normalized at $\lambda_{0}$=5870\AA; 
$r_\lambda$\,$\equiv$\,10$^{-0.4(A_{\lambda}\,-\,A_{\lambda_{0}})}$ 
is the extinction term; $\otimes$ represents the convolution operator; G$(\nu_{\star},\sigma{\star})$
is the Gaussian distribution used to model the line-of-sight stellar motions, centered at velocity $\nu_{\star}$ and with a dispersion $\sigma_{\star}$;
\textit{M}$_{\lambda_{0}}$ is the synthetic flux at the normalization wavelength and $\vec{x}$ is the population vector. 
The population vector represents the fractional contribution of the SSPs in terms of age and metallicity ($t_{j}$, $Z_{j}$) over the synthetic 
flux at $\lambda_{0}$.  These spectral components can also be expressed as a function of the mass of the population, 
represented by the vector $\vec{m}$.

The fit between the observed and modeled spectra
 is calculated with the support of an algorithm that searches for the minimum value of 
\begin{equation}
\label{ma}
\centering
\chi^{2}\,(x,\nu_{\star},\sigma{\star},A_{V}, M_{\lambda_{0}})=\sum_{\lambda=1}^{N_\lambda}[w_{\lambda}(\textit{O}_{\lambda}-\textit{M}_\lambda)]^{2} ,
\end{equation}
where \textit{w}$_\lambda^{-1}$ is the error in \textit{O}$_{\lambda}$. 
Emission lines and spurious pixels are excluded of the fits establishing \textit{w}$_\lambda$\,=\,0. 
The intrinsic reddening is modeled by {\scriptsize\,STARLIGHT} as due to foreground dust model, 
using the extinction  law by \citet{cardelli89}, with R$_{V} =3.1$ and parametrized by the 
V-band extinction, A$_{V}$\,=\,R$_{V}$\,E(B-V). 
The SSPs considered in this work takes into account 15 ages, t\,=\,0.001, 0.003,0.005, 0.01, 0.025, 0.04, 0.1, 0.3, 0.6, 0.9, 1.4, 2.5, 5, 11,
and 13\,Gyr, and three metallicities, $Z$\,=\,(0.2, 1, and 2.5\,$Z_{\odot}$), summing up  N$_{\star}$\,=\,45 SSP components.
The stellar spectra of the SSPs  were convolved with an elliptical Gaussian function to achieve
the same spectral resolution of the observed one; transformed into the rest frame;  
normalized to $\lambda$5870 \AA; and  corrected by foreground Galactic extinction as 
in \citet{schlegel98}. 

The {\scriptsize\,STARLIGHT} code provides individual population vectors. However, according to \citet{cid05}  the individual components
of $\vec{x}$ are very uncertain. To troubleshoot this problem, \citet{cid05} proposed that the  population contribution  should be combined in age bins. 
Following the prescription of these authors, the population vectors are binned according to the flux contribution in:
young, $x_{y}$ ($t\leq4\times10^{7}$ years), intermediate-age, $x_{i}$ (4$\times10^{7}<t\leq2.5\times10^{9}$ years) 
and old, $x_{o}$ ($t>2.5\times10^{9}$ years) components, which have uncertainties lower than 0.05, 0.1 and 0.1,
respectively, for signal-to-noise S/N$>$10. The same bins are to be used for the fractional contribution to the stellar mass 
$\vec{m}$ ($m_{y}$, $m_{i}$, $m_{o}$).
The quality of the  fitting result is quantified by the parameters $\chi^{2}$ and $adev$. The latter gives 
the relative mean deviation 
$|(\textit{O}_{\lambda}-\textit{M}_\lambda)|/\textit{O}_{\lambda}$ over all fitted pixels, between the 
observed and model spectra.

\begin{figure*}
\centering
\includegraphics*[angle=-90,width=0.7\textwidth]{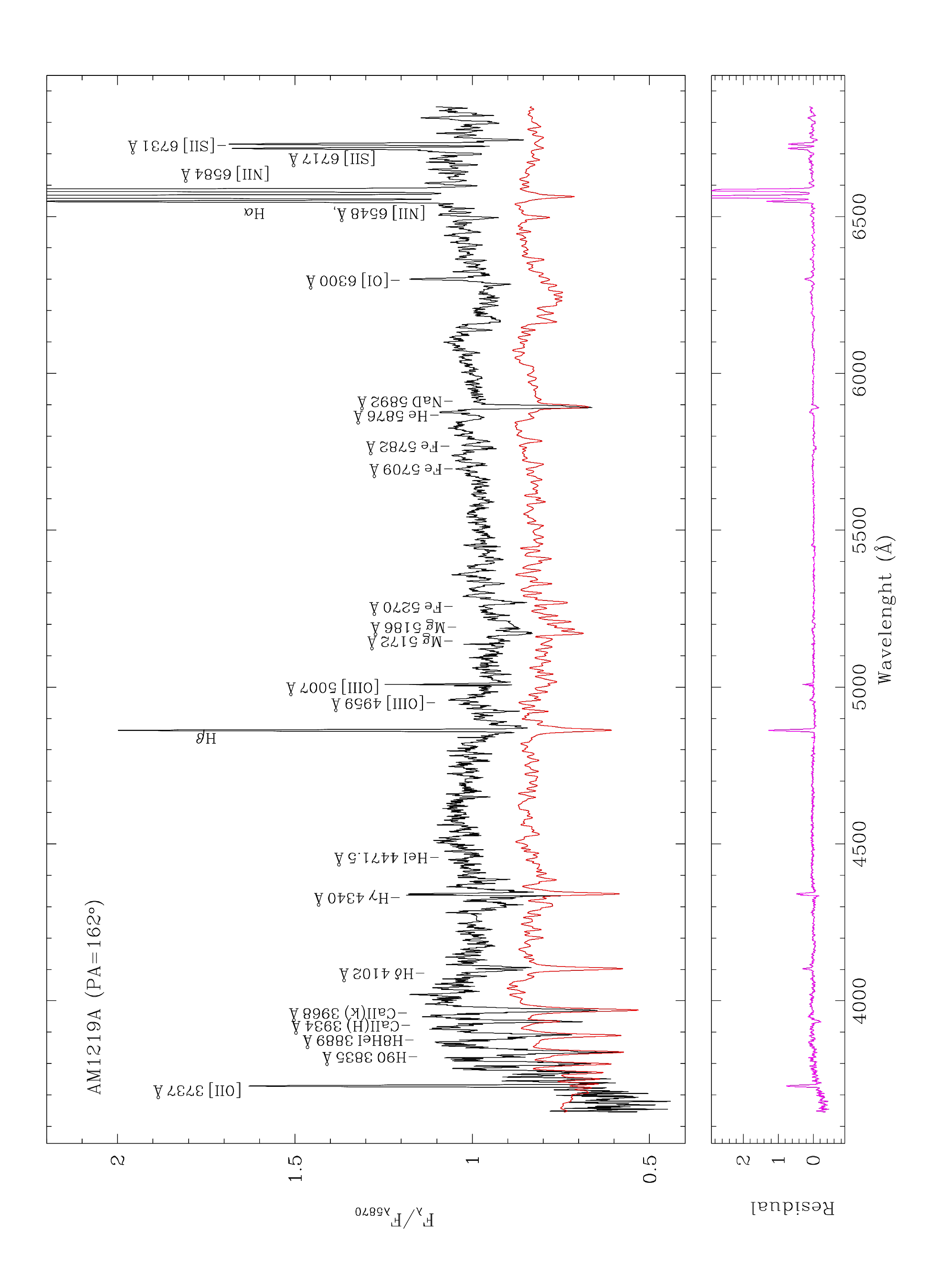}
\includegraphics*[angle=-90,width=0.7\textwidth]{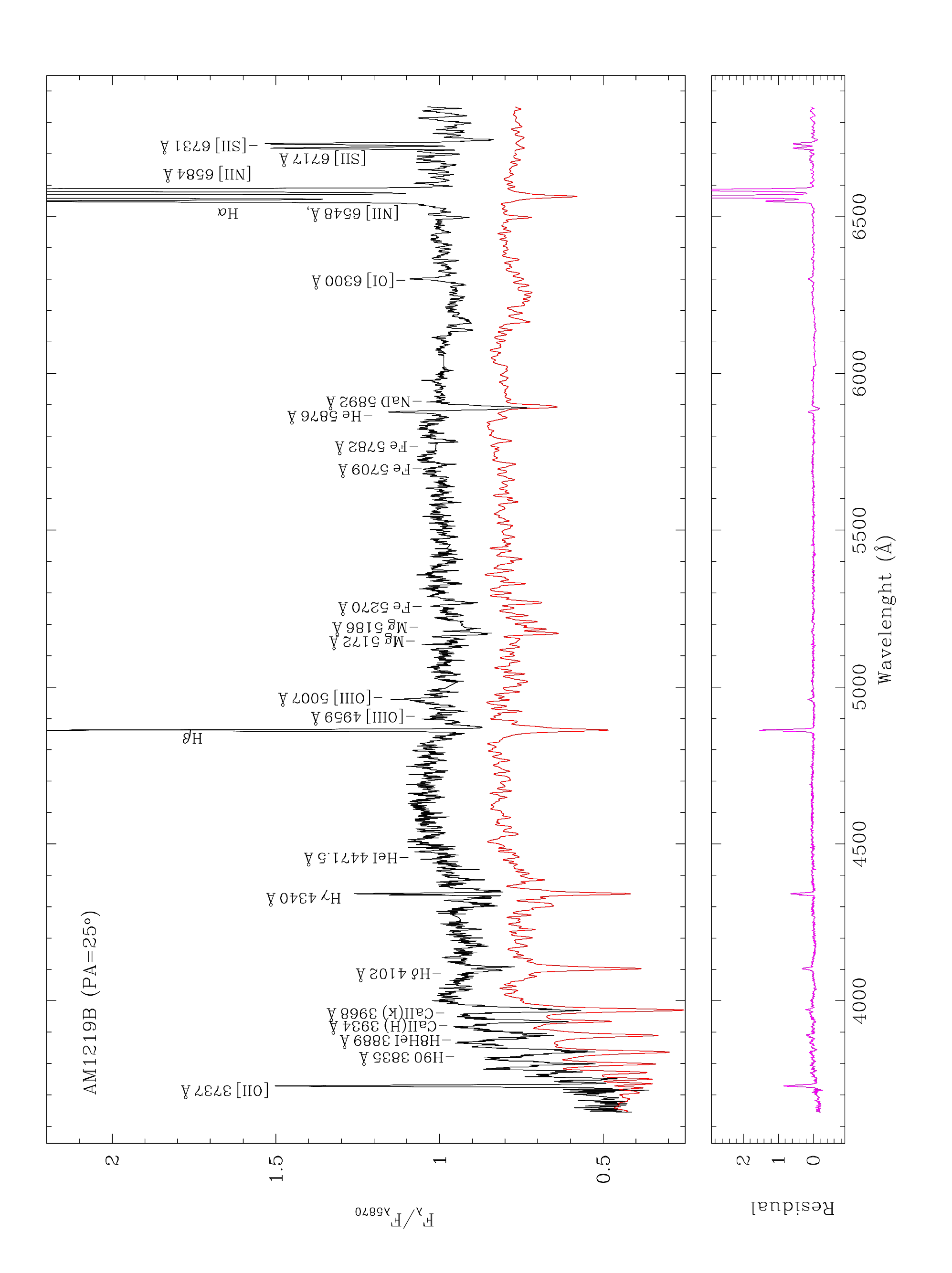}
\caption{Stellar population synthesis for the nuclear region of AM\,1219A and AM\,1219B.
For each object we plotted in the top panel the observed (in black) and the synthesized
spectra (in red), and in the bottom panel the pure emission spectrum corrected for reddening. The main absorption and emission features have been identified.}
\label{sintese_002}
\end{figure*}

\begin{table*}
\caption{Stellar-population synthesis results for AM\,1054A galaxy. Results for the whole sample of galaxies is available in the electronic version.}
\label{tcorr1}
%\vspace{0.3cm}
\begin{tabular}{|lccccccccc}
\hline
Position\,(kpc)&$x_{y}$\,(\%)&$x_{i}$\,(\%)&$x_{o}$\,(\%)&$m_{y}$\,(\%)&$m_{i}$\,(\%)&$m_{o}$\,(\%)&$\chi^{2}$&$adev$&A$_{V}$\,(mag)\\    
 \hline  
                            &     \multicolumn{7}{c}{AM\,1054A} \\
\noalign{\smallskip}			    
-0.564  &   80.3  &   13.9  &	0.5  &   52.9  &   39.0  &   8.1   &   0.8  &  4.54  &  0.37 \\
-0.282  &   68.7  &   17.1  &	10.8 &   14.0  &   8.4   &   77.6  &   0.9  &  2.85  &  0.33 \\
 0.000  &   70.5  &   15.1  &	11.4 &   13.1  &   12.7  &   74.2  &   0.9  &  2.28  &  0.40 \\
 0.282  &   59.5  &   19.9  &	16.5 &   5.4   &   4.3   &   90.3  &   0.8  &  2.58  &  0.19 \\
 0.564  &   68.5  &   17.2  &	13.0 &   7.3   &   7.7   &   85.0  &   1.3  &  3.00  &  0.12 \\
 0.846  &   75.4  &   21.1  &	0.8  &   53.0  &   40.5  &   6.5   &   1.2  &  2.99  &  0.46 \\
 1.128  &   57.4  &   38.8  &	0.0  &   17.7  &   82.3  &   0.0   &   0.6  &  3.84  &  0.16 \\
 1.410  &   35.5  &   58.2  &	0.0  &   3.1   &   96.9  &   0.0   &   1.0  &  4.94  &  0.00 \\
 1.692  &   42.7  &   43.1  &	0.0  &   22.8  &   77.2  &   0.0   &   1.1  &  4.13  &  0.34 \\
 1.974  &   50.6  &   34.2  &	8.1  &   7.2   &   11.8  &   81.0  &   0.7  &  4.23  &  0.22 \\
\hline
\end{tabular}
\begin{minipage}[l]{13.0cm}
 %[*] Logarithm of the \azul{H$\beta$ observed flux} in  erg s$^{-1}$ cm$^{-2}$.
\end{minipage}
\end{table*}

\section{Results}
\label{res3}

\begin{figure*}
\includegraphics*[angle=0,width=0.7\textwidth]{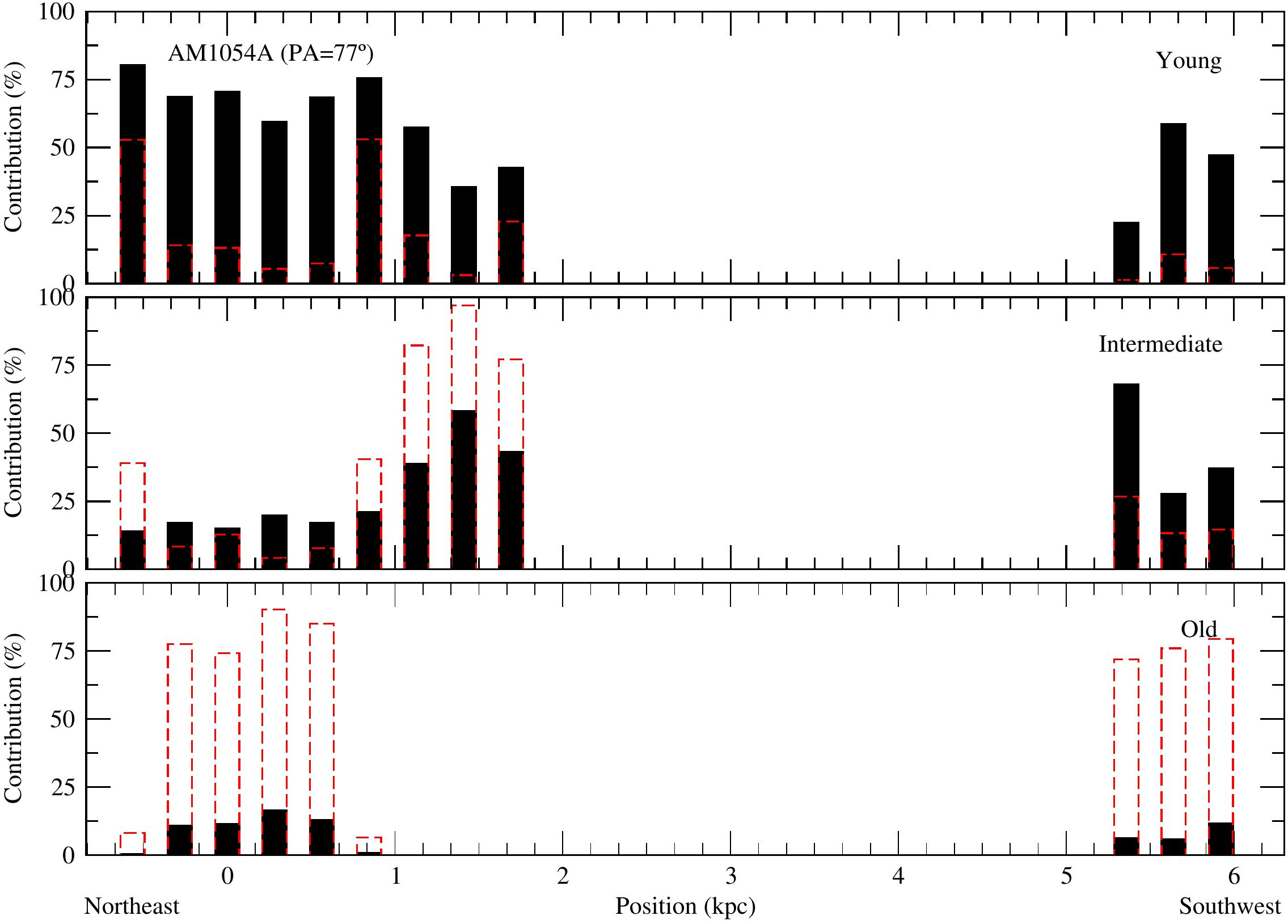} 
\includegraphics*[angle=0,width=0.7\textwidth]{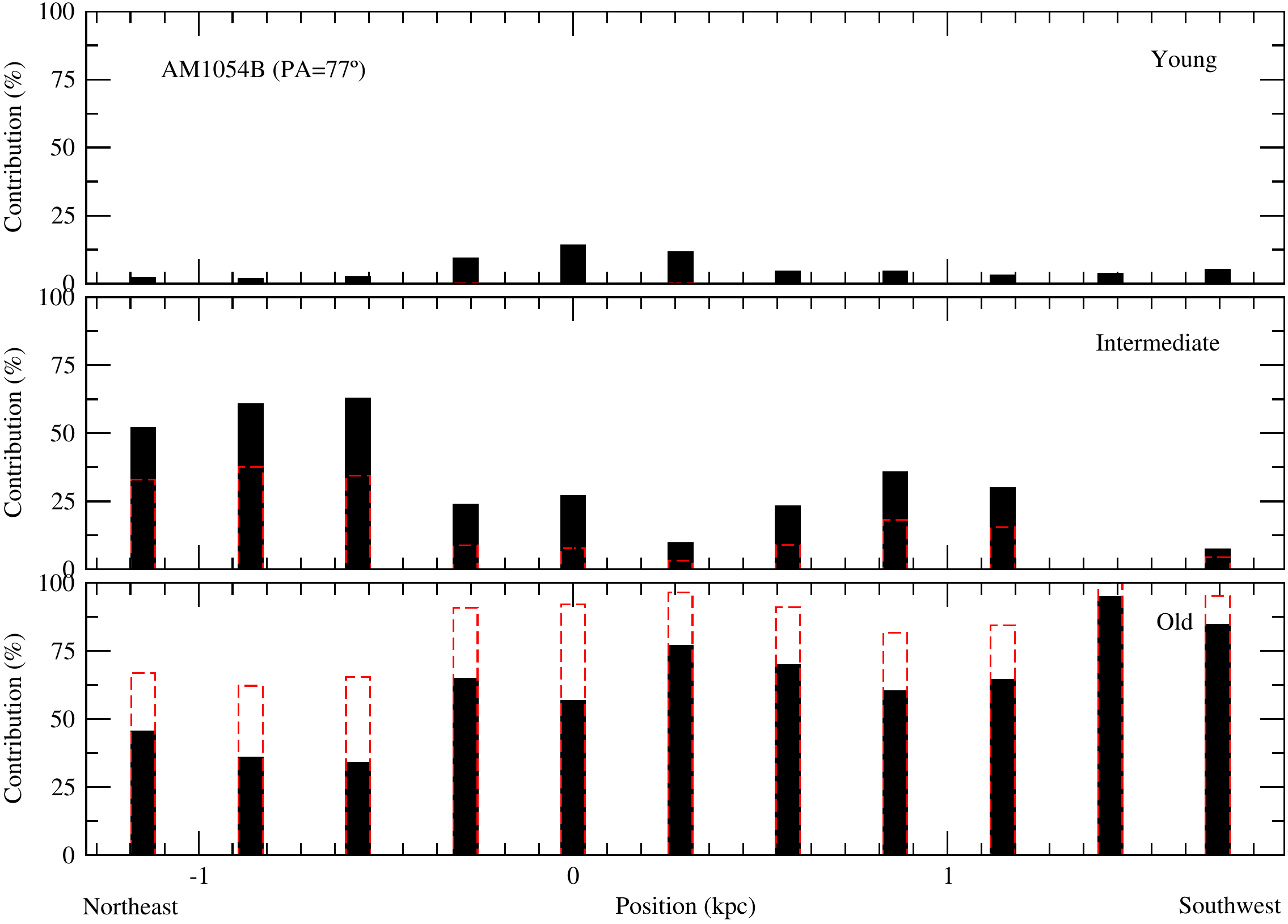}
\caption{Synthesis results represented in terms of the flux  (black) and mass fractions (red dashed) as 
a function of the distance to the center of AM\,1054A (top) and AM\,1054B (bottom) along the PA\,=\,77\degr. 
For each object, the top panel corresponds to the young stellar population; the middle panel to 
the intermediate stellar population; and the bottom panel to the old one.}  
\label{sin_1054AB}
\end{figure*}

\begin{figure*}
\includegraphics*[angle=-90,width=0.7\textwidth]{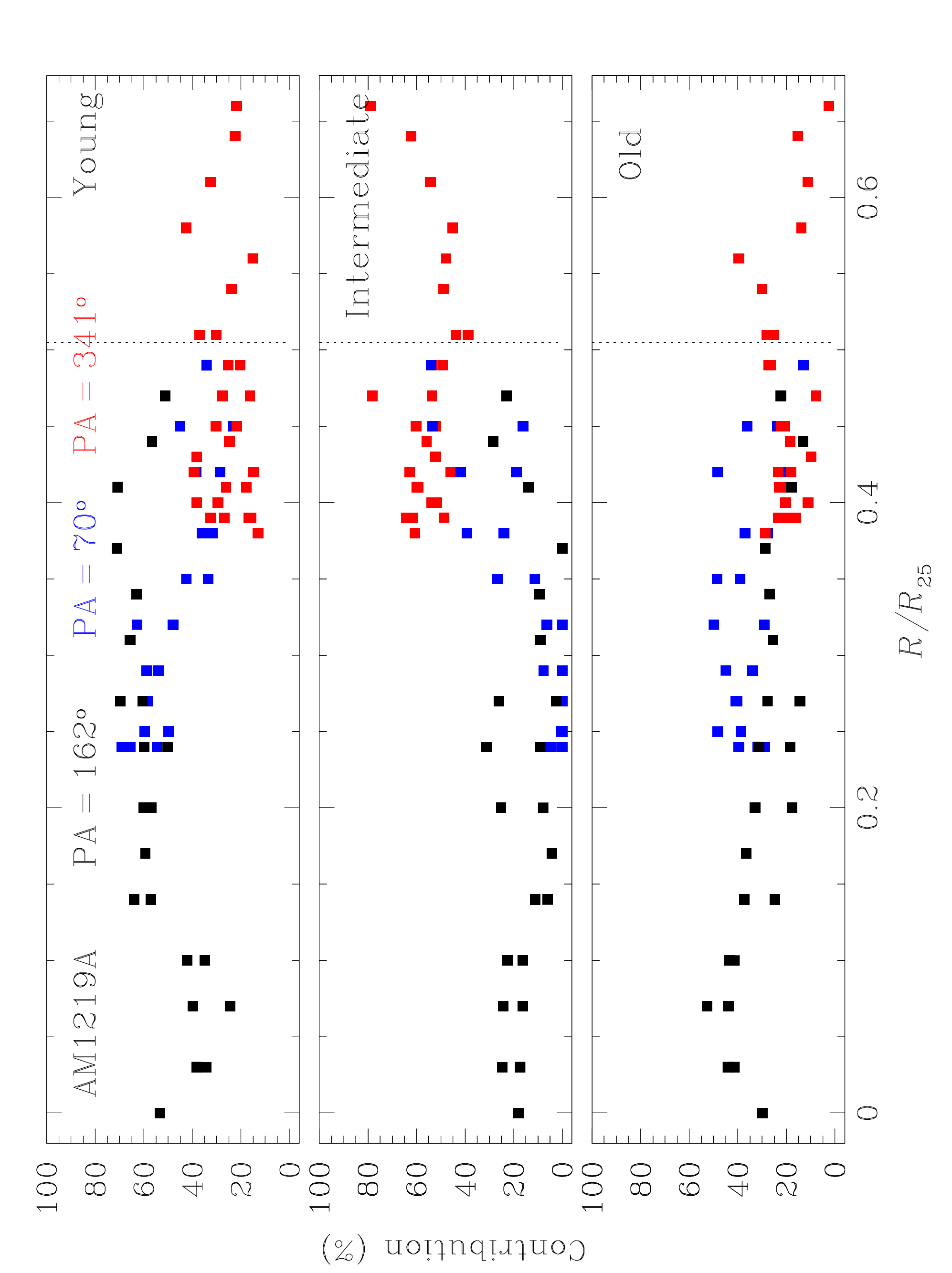}
\includegraphics*[angle=-90,width=0.7\textwidth]{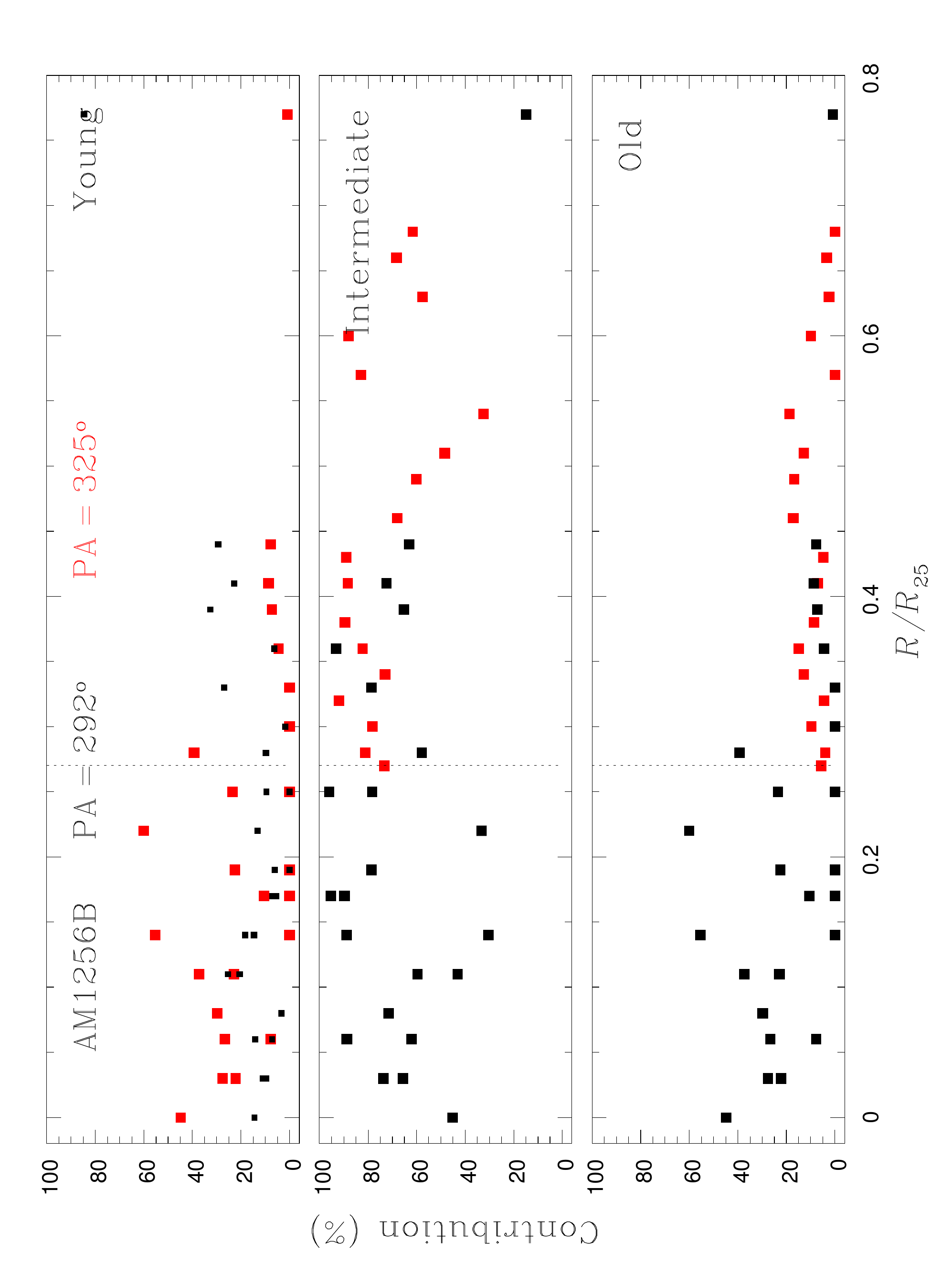} 
\caption{Synthesis results in flux fraction as a function of the galactocentric distance for AM1219A and  AM1256B.
Colors correspond to the different position angles.}
\label{corotation}
\end{figure*}

In  Table~\ref{tcorr1} the  stellar-population results are presented for the regions 
located along the disk of each galaxy in our sample. These regions are the same considered in Papers I and II.
Fig.~\ref{sintese_002} shows an example of the results of the synthesized spectrum and the pure emission for the nuclear 
region of AM\,1219A and AM\,1219B, while  the results for the nuclear regions of the other objects in our sample
are shown in Appendix \ref{ap_synth1} (Figs.~\ref{sintese_003} and \ref{sintese_004}).
The results of the synthesis are summarized in Table~\ref{tcorr1}, for the individual spatial bins in each galaxy, 
stated as the percentage contribution of each base element weighted by flux and mass.
In Fig.~\ref{sin_1054AB} the contribution of the stellar population components  along 
the slit position  of AM\,1054A and AM\,1054B are shown.  The nominal center of each slit was
chosen to be the continuum peak at $\lambda$5870\AA\ flux.
 The remaining synthesis results are showed in Appendix \ref{ap_synth2}, from Figs.~\ref{sin_1219AB} to \ref{sin_2322A}, where
the regions with intersection on the spectrograph slit positions are highlighted in the plots with different filling patterns.
In Fig.\ \ref{corotation} the synthesis results for  AM\,1219A and AM\,1256B  are plotted as a function of the galactocentric radius $R$ normalized by $R_{25}$ 
(galactocentric distance with surface brightness of 25 mag arcsec$^{-2}$, see Table\,2 of Paper\,II).
In what follows, the results for each system are discussed separately.

\begin{figure*}
\includegraphics*[angle=-90,width=0.7\textwidth]{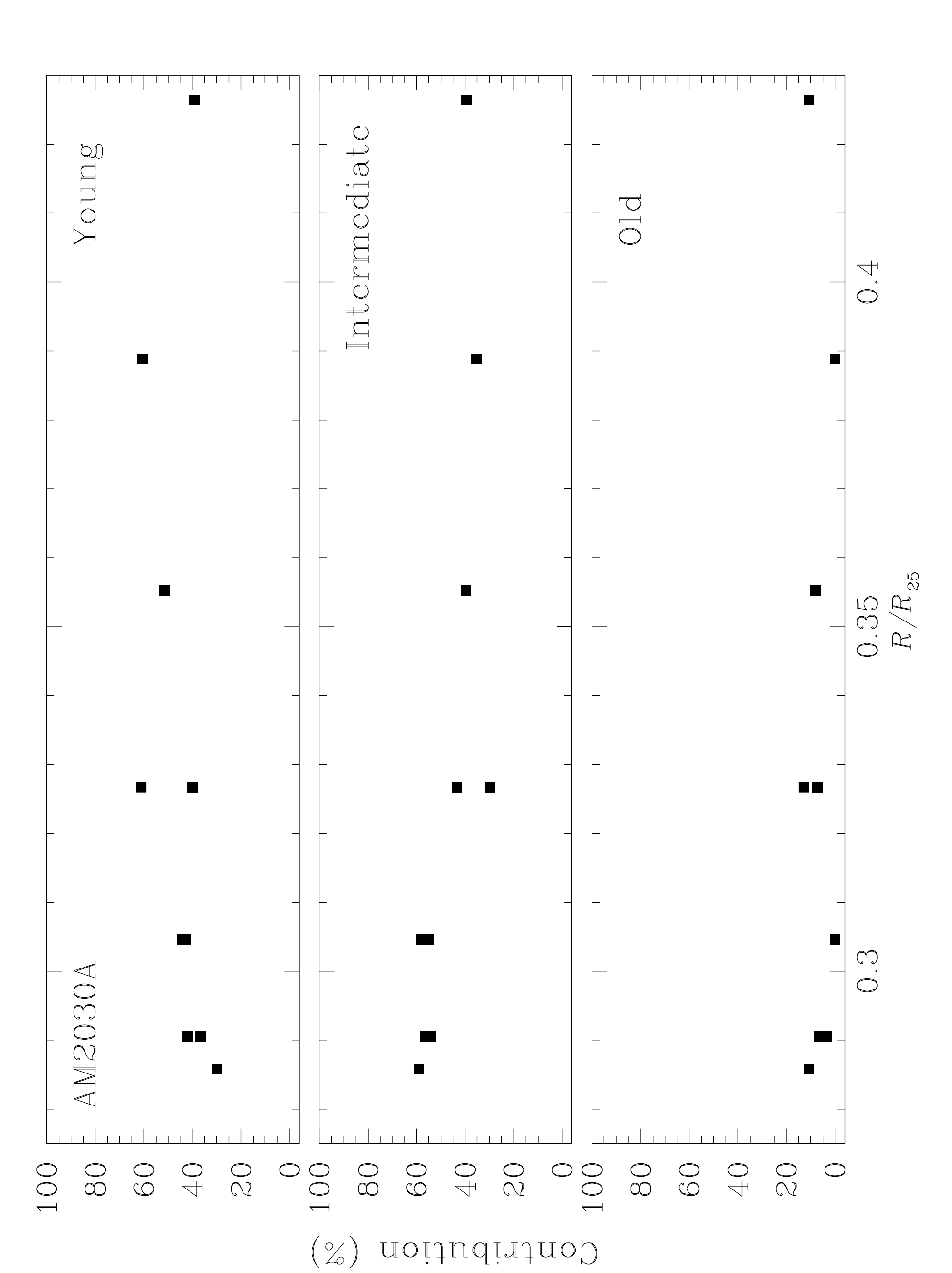} 
\includegraphics*[angle=-90,width=0.7\textwidth]{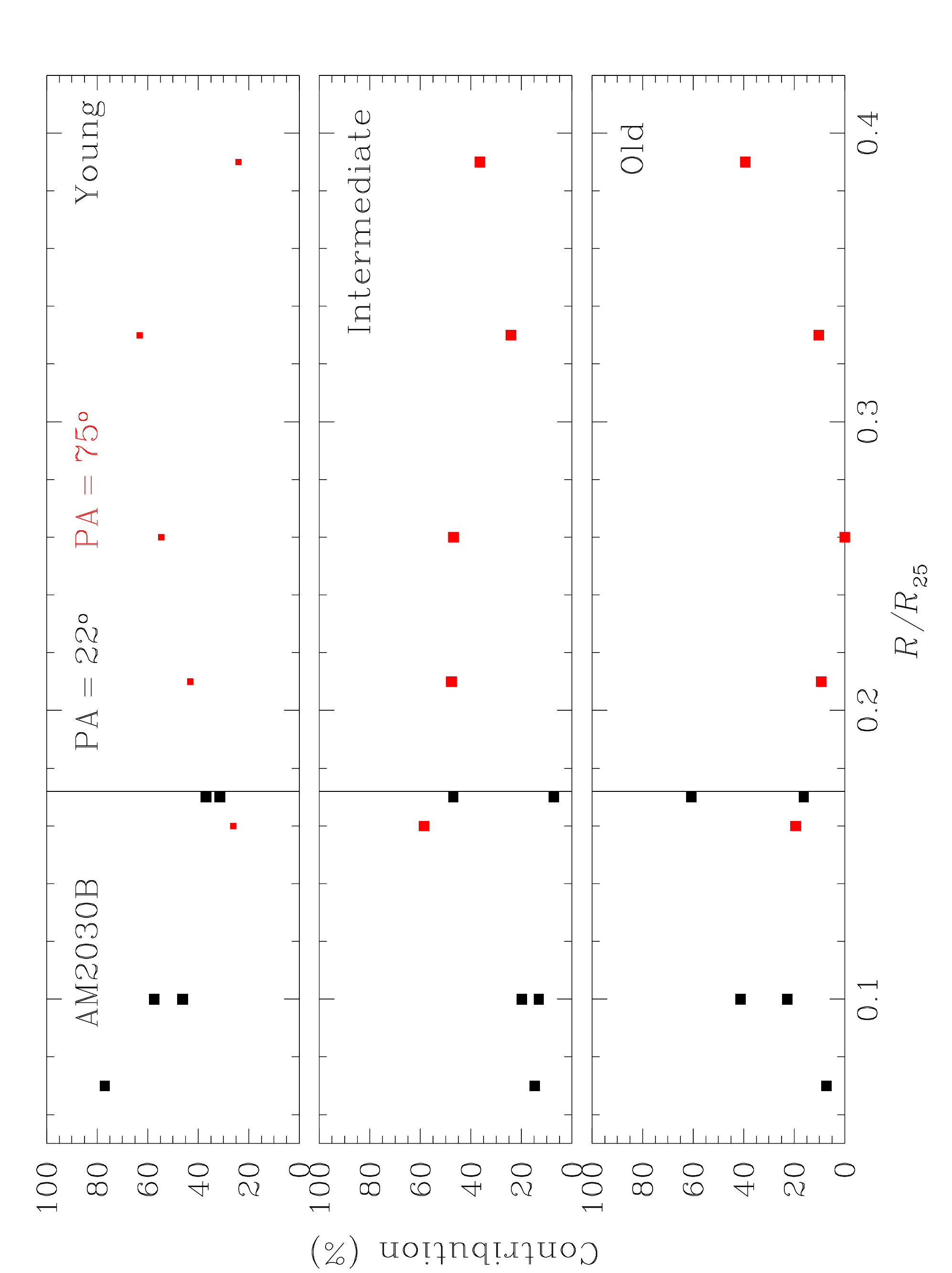}
\caption{Same as Fig. \ref{corotation}, but for  AM2030A and AM2030B.}
\label{corotation2}
\end{figure*}

\subsection{AM\,1054-325}

This system is composed by a main spiral galaxy (hereafter AM\,1054A) and a companion galaxy (hereafter AM\,1054B). 
AM\,1054A is an optically bright spiral in the early stages of merging  \citep[$\sim$ 85\,Myr; ][]{mullan11}.
As can be seen in Fig.~\ref{sin_1054AB} (top panel) the population synthesis results indicate that the light  along the disk of AM\,1054A,  
is mainly dominated by a young stellar population component  ($t\leq4\times10^{7}$\,yr), with a not negligible contribution of an intermediate population.
This spatial profile corresponds to a slit position  crossing the supposed 'second' nucleus (or starburst region?) of AM1054\,A (see Paper\,I). 

In the bottom panel of Fig.~\ref{sin_1054AB},  we can see that AM\,1054B is mainly dominated by the old stellar population component,  
both in the contribution of light and mass, with a mean value of  62\% for light and  84\% for mass.
The Intermediate stellar population contributes significantly  for this galaxy, with a mean value of about 30\%  for a light-weighted 
context and  16\% for a mass-weighted one.

\subsection{AM\,1219-430}

This system is composed by two galaxies, a main spiral galaxy (hereafter, AM\,1219A) and a secondary one (hereafter, AM\,1219B). 
The primary galaxy presents a normal arm to the south-west and a strong tidal arm,  with several \ion{H}{ii} regions along it. AM\,1219B has a 
bright nuclei with smooth open spiral arms. 
A bridge of material connecting both galaxies and a lens structure in addition to the bulge and disc of AM\,1219B were reported by 
\citet{jose14}. A rough estimate of stellar population based on equivalent
widths of absorption lines and on continuum for the integrated spectrum of these galaxies shows  that both components have a strong flux contribution from
stellar populations younger than $10^{8}$ years   \citep{pastoriza99}. For the main galaxy, the estimations obtained in Paper\,I for electron densities show 
an increase of these  ones  towards the outskirts of this galaxy ($R/R_{25} \approx 0.5$). Interestingly, in this region an oxygen abundance 
break, that could be associated with a corotation radius, was obtained in Paper\,II. 
%These measurements \azul{only correspond} to regions along the PA\,=\,341\degr.

The  distribution of the stellar population in both galaxies are shown in Fig.~\ref{sin_1219AB}. 
Three different slit positions were observed in AM\,1219A: PA\,=\,162\degr\ and PA\,=\,341\degr\ along the North-South direction crossing the nuclei 
and the disturbed arm, respectively; and 
PA\,=\,70\degr\ in the  Northeast-Southwest direction crossing two very bright H{\sc ii} regions. 
For  AM\,1219B, we only have observations with a  PA\,=\,25\degr crossing its nucleus: PA\,=\,25\degr.

For the slits with PA\,=\,162\degr\ and PA\,=\,70\degr\ the light emitted by  AM\,1219A is dominated by the contribution of the young stellar component
($\sim$54\%), with a significant contribution ($\sim$35\%) of the old component. 
The  intermediate stellar population is absent around the closest zone to the nucleus mapped by the slit with a PA\,=\,70\degr. This very 
bright zone (see Fig.\,1 of Paper\,II) corresponds to the continuum peak of an  \ion{H}{ii} region complex. The stellar population spatial 
profile associated with 
this slit position seems to indicate an increase in the young stellar component towards this \ion{H}{ii} region complex. 
Considering the slit with a  PA\,=\,341\degr, we can see that the intermediate stellar population dominates the light, with  56\% in average,
followed by the young component, with 26\% in average.  
One possible interpretation to these results is that the intermediate stellar component could be  associated with the perigalacticum passage.  
Preliminary estimations indicate that it occurred about 220\,Myr ago (Hernandez-Jimenez et. al., private communication). 
Hence, along the disturbed arm, most of the stellar population could have been triggered  during the perigalacticum passage, remaining a 
lower reservoir of gas to form the younger generation of stars
than in the other mapped parts of the galaxy, and consequently decreasing the current star formation rate.  
As can be seen in Fig.~\ref{corotation},  none  variations of stellar population were found along the PA\,=\,341\degr.
It is interesting to note that the zones sampled by two different slits, which  are marked in Fig.~\ref{sin_1219AB}, show estimations 
of their stellar population ages in reasonable agreement to each other. 

The stellar population estimated for AM\,1219B is roughly homogeneous along the observed slit. In light-weighted fraction, the stellar 
population is distributed, in average, in the three population 
components as: 23\% of young, 39\% of intermediate and 40 \% of old.

Taking into account the contribution of the different stellar populations to the mass, we found that in both 
components of this system the old stellar population is the clear dominant component.

\subsection{AM\,1256-433}

This system is composed by three galaxies, from which  we only observed the secondary galaxy (hereafter AM\,1256B). 
A break in the oxygen abundance gradient was  obtained in Paper\,II at $R/R_{25}\approx0.27$, which  suggested  
 that  this break could be associated with a corotation radius. Interestingly, 
at about the same position it was found the maximum star formation rate (see Paper\,II). 

As can be seen in  the Fig.~\ref{sin_1256B},
the  stellar population is predominantly composed by the intermediate component, in average 70\%, for both slit positions.
We  observed variations of the stellar population across the galaxy,  that in general are not dependent on the position, with exception of
the region between 4 and 7 kpc along the PA\,=\,325\degr. In this region, the young component increases while  the intermediate one decreases.
 It could be probably due to the slit is not completely aligned with the curved arm. In that area, the slit cross the arm and  therefore a young population is observed.

For the PA\,=\,292\degr\ and a distance of about 18\,kpc from the center of AM\,1256B in the Northwest direction (see Fig.\,1 of Paper\,I),  
there is a region where the young population represents about 80\% of the  $\lambda$5870\AA\ flux.
In this zone, there is an   \ion{H}{ii} region  whose age estimated by \citet{ferreiro04} is about 6 Myr, in agreement with our estimation.
For this slit position mainly the light-dominant population is also the intermediate one. In four cases, the old population contribution to the light is comparable or even greater than that of the intermediate population. Analyzing the mass-contribution we found that in the inner zone the old population dominates.
The intersection of the spectrograph slit positions (marked in  Fig.~\ref{sin_1256B}) 
shows stellar population estimations in good agreement mainly for the light.

\subsection{AM\,1401-324}

The main galaxy of this system, AM\,1401A (ESO 384-G 041), has a very bright nucleus and an arm giving the appearance of a ring. Its companion is the smaller galaxy  AM\,1401B (ESO 384-G 041 NOTES01/PGC682060).
Only the primary galaxy was observed. In Fig.~\ref{am1401_fenda} the slit positions are shown. 

The results of the stellar population synthesis for AM\,1401A are presented in   Fig.~\ref{sin_1401A}.
AM\,1401A is completely light-dominated by intermediate stellar populations, with a contribution at $\lambda$5870 \AA\ flux of 
about  50-75\% in the central region and higher than  90\% in the outer regions. 
For the slit that cross the nucleus (PA\,=\,294\degr), there is a spatial variation of the stellar population and the light-contribution of the intermediate one  
decreases towards the center of the galaxy.
The fraction of old stellar population component is found to be increasing from the outer regions to the center.

\subsection{AM\,2030-303}

This is a system formed by a spiral galaxy as the main component (hereafter AM\,2030A) and probably a triplet of irregular galaxies (see Fig.\ 1 of Paper\,II) as the secondary component (hereafter AM2030B).
Two different slit positions were used to acquired the data for this system:  PA\,=\,22\degr\, and  PA\,=\,75\degr. 
The first slit maps a complex of H{\sc II} regions located in an outer arm of AM\,2030A and cross the one of the irregular galaxies of AM\,2030B.
The second slit maps part of the secondary triplet complex. 

%Unfortunately, any  slit is  crossing the nuclei of AM\,2030A.

For AM\,2030A,  we found  a light-dominant  stellar population with  young/intermediate age, summing up more than 90\% of the optical flux at $\lambda$5870 \AA\ 
(see Fig.\ \ref{sin_2030A}).
This result is expected because the slit crosses a complex of  \ion{H}{ii}
regions, whose estimated  age by \citet{ferreiro04} is about  6.5\,Myr.
As can be seen in Fig.\ \ref{sin_2030B}, the AM\,2030B spatial profile is derived  mainly from two distinct regions  along the PA\,=\,22\degr. 
One towards the northeast side and another towards the southwest, whose measurements are from ESO 463-IG 003 NED03 and ESO 463-IG 003 NED02, respectively. 
ESO 463-IG 003 NED03 is light-dominated by a young stellar population with a significant contribution of the old stellar population in a couple of bins. On the other hand,  
ESO 463-IG 003 NED02 is light-dominated by an intermediate stellar population  with a significant contribution of old and young populations. 
For PA\,=\,75\degr, the stellar population of AM\,2030B is distributed mainly in young and intermediate populations.
The spectra obtained from regions on the intersection of the slits presented low signal-to-noise (S/N$<$10) and therefore they were not used to 
performed the synthesis.

\subsection{AM\,2058-381}

This system is composed by two arms spiral galaxy as the main component and an irregular one as the secondary, hereafter AM\,2058A and AM\,2058B respectively. 
The stellar population distributions in flux and mass of both components are shown in Fig.~\ref{sin_2058A}.

AM\,2058A is dominated by an intermediate population component except for the inner zone where the dominant is the older one. 
For the slit positions that cross the center of this galaxy (PA\,=\,42\degr\, and PA\,=\,350\degr),  
systematic variations are  observed for the intermediated stellar component, i.e.  its  contribution to the flux is  increasing outwards. 
On the other hand, a decreasing in the older component can be seen in both contributions, flux and mass, from 
the center to the outer regions. A small contribution (generally negligible) from the young population is observed. 
Along the spatial profile of PA\,=\,125\degr, we found two two zones on which the young stellar component is prominent.  One  at
$\sim$3 kpc towards Southeast and another at $\sim$7.5 kpc towards the Northwest. As can be seen in Fig.\ 1 of Paper\,I  these regions are 
located in one spiral arm, and therefore a young population is expected to be observed.
 
It can be seen in the  bottom right panel of Fig.\ \ref{sin_2058A} that AM\,2058B is mainly dominated by the contribution of the
intermediate stellar population component in light as well as in mass, 
with a mean value of 69\%  and 74\%, respectively.  No significant variations along the spatial profile are present in this galaxy.

\subsection{AM\,2229-735}

AM\,2229-735 is a system composed by a main spiral galaxy (hereafter AM\,2229A) interacting with a smaller disc galaxy (hereafter AM\,2229B).
Spectra obtained along AM\,2229B are very noisy ($\rm S/N<<10$) and they were not considered.

The contribution of the different stellar-population components along the disk of AM\,2229A for the two distinct slits are shown in Fig.~\ref{sin_2229A}. 
For both slit positions, there is a zone whose behavior is noticeable. For the PA\,=\,161\degr\ the region between   
-4 and -8 kpc (hereafter called region A) the light and mass are mainly dominated by the young and intermediate populations.
Same behavior is shown by the zone between 0 and -3 kpc (hereafter called region B) for the PA\,=\,134\degr.
These areas are marked in Fig.\ \ref{am2229_fenda}
and  correspond to two complexes  of \ion{H}{ii} regions. The origin of these complexes is probably related to the interaction 
between the galaxies of this pair.

\subsection{AM\,2306-721}

This pair is formed by a spiral galaxy (hereafter AM\,2306A) and an irregular galaxy companion (hereafter AM2306B).
As can be seen in Figs.~\ref{sin_2306A}, the optical flux of both galaxies are mainly dominated by the intermediate population component.
 
AM\,2306A shows systematic variations for old and intermediate stellar components along the two slit positions observed.
While the intermediate component is decreasing towards the center of the galaxy, the old component is increasing.
Likewise it can be seen that the young stellar population is increasing towards the 
outer regions of AM\,2306A. This behavior is more conspicuous for the regions mapped with the slit with  PA\,=\,238\degr, for which the more 
significant contributions from the young stellar component are located between -11 and -6 kpc, and from 10 to 12 kpc,  that corresponds to the regions on the spiral arms.
\citet{krabbe08} derived the stellar population for this interacting system using the stellar population synthesis method developed by \citet{bica98}.
This method employs the equivalent widths of several spectral absorption features and the measured continuum fluxes at different wavelengths, comparing  them to those of a model computed from a base of simple
stellar population (SSP) elements with known ages and metallicities.
Our estimated stellar populations are in good agreement with the results previously obtained by \citet{krabbe08}.
In the case of the slit with PA\,=\,190\degr, this increment towards the outer zones is also 
observed, although the variation of the light contribution from the young population is not so significant.

Interestingly, for AM2306B there is a completely different behavior. Around the center of this galaxy we can see that the old stellar population 
dominates the mass content although the intermediate and young populations dominate the light.  Towards the Northwest of the galaxy there is a zone, 
between $\sim$3.5 and 8 kpc (PA\,=\,118\degr), without old stellar population component.
At about 3\,kpc there is a zone for which the old stellar population dominates the mass content and around it (mapped by both slits) there are zones 
completely dominated (in mass and light) by the intermediate and young stellar populations. Probably, this enhancement of the stellar formation was 
triggered by the strong interaction with AM2306A.

\subsection{AM\,2322-821}

The system AM\,2322-821 (see Fig.\ 1 of Paper\,II) is composed by a spiral galaxy classified as SA(r)c showing disturbed arms (hereafter AM\,2322A) in interaction with an irregular galaxy (hereafter AM\,2322B). 
Four slit position angles were used to map this system: PA\,=\,59\degr\, crossing the nucleus of AM2322A; PA\,=\,28\degr\, crossing the main body of the
primary but not across the nucleus (an offset of about 8 arcsec
north-west, NW, from the nucleus); PA\,=\,60 \degr\ mapping the NW spiral
arm of the main component, located between the main and secondary component; PA\,=\,318\degr\ along the main axis of the
secondary component AM\,2322B and also along the
north-east (NE) spiral arm of AM\,2322A.
As was pointed out in Paper\,II, the spectra obtained using PA\,=\,60\degr\ present low signal-to-noise (S/N$<$10). 
Therefore they were not used to performed the synthesis.
Indeed, the spectra from the regions located in the intersections of the slits present low signal-to-noise and therefore they were not considered in the synthesis.

The  distribution of the stellar population in both galaxies of the AM\,2322-821 system
is shown in Fig.\ \ref{sin_2322A}.
The contribution to the light by the stellar populations is heterogeneous in AM\,2322A for the three slit positions considered (PA\,=\,28, 59, 318\degr).
There are no hints of age gradients along this galaxy: the young, intermediate and old populations contribute significantly to the optical flux at $\lambda$ 5870 \AA. 
It is worth noting that the mass is clearly dominated by the old stellar population component in the majority of the zones.

The light emitted by AM2322B is dominated by the young stellar population component in all the galaxy except in its center, where contribution of the three 
populations are comparable. This young component shows a systematic variation along the slit,
increasing outwards. In the outskirts the old component barely contributes to the light although it clearly dominates the mass along all the galaxy.

The stellar population in this system was studied by \citet{krabbe11} using the same data and a similar methodology that in this work. 
Even though they considered four population vectors components
%$x_{vy}$ ($t\leq4\times10^{8}$ years); young, %$x_{y}$(4$\times10^{8}<t\leq5\times10^{8}$ years); intermediate age,
%$x_{i}$ (5$\times10^{8}<t\leq2\times10^{9}$ years); and old,  %$x_{o}$($t>2\times10^{9}$ years) components. 
the results in both works are compatible.
The contribution of the young population in AM\,2322B was related with the perigalactic passage occurred at  $\sim$90\,Myr after perigalacticum 
\citep{krabbe11}. 

\section{Discussion}
\label{disc} 

\subsection{Stellar population distribution}

One of the main effects of interactions on the evolution of galaxies is  modify the stellar formation history
of the objects involved. Our results, presented in the previous section, indicate  that the light at $\lambda$ 5870 \AA\ emitted by most of the galaxies in our sample are dominated 
by young/intermediate stellar populations, i.e. by stellar populations with ages lower than $2\times10^{9}$ years. Similar result was derived for other interacting
galaxies by  \citet{dametto14}, whom used near-infrared data to investigate  the spatial distribution of the stellar populations
in  three interacting starburst galaxies.  Indeed, young  star clusters   (ages lower than 300 Myr) have been 
found in tidal tails of interacting galaxies (e.g. \citealt{mulia15, degris03, bastian03}).
In contrast, in the disks of isolated galaxies the light is dominated by an older stellar population. In fact,  \citet{morelli15} investigated the
properties of the stellar populations in the discs of ten spiral galaxies, mostly not belonging to interacting systems. These authors found  
that the old stellar populations (ages higher than 4\,Gyr) usually dominates the disc surface brightness. 
The difference in the light-dominant population between interacting and isolated galaxies could be due to the induced gas inflow 
that increases the bursts of star formation along the disk during and after the interaction.

Other important issue is to investigate the differences between the stellar populations in the main and secondary galaxies belonging to interacting systems. 
Numerical
simulations by \cite{cox08}  showed 
that during the interaction process in minor mergers,  the fractional SFR enhancement is higher in the  secondary galaxy 
due to it is more susceptible to the tidal forces induced by the interaction. 
This theoretical result was confirmed
by observational studies of the minor mergers AM\,2306-721 by \citet{krabbe08} and NGC\,7771+NGC\,7770 by \citet{alonso12}.
Inspection in  Figs.~\ref{sin_2058A}, \ref{sin_2306A} and \ref{sin_2322A},
shows clearly that the secondary galaxies of the systems AM\,2058-381, AM\,2306-721 and AM\,2322-821 have higher contributions of young stellar populations than the primary galaxies.
 For AM\,1054-325 and AM\,1219-430 systems an opposite result is derived (see Fig.\ \ref{sin_1054AB} and \ref{sin_1219AB}, respectively).  However, the secondary galaxy of the system
AM\,1219-430  and the primary one of AM\,1054-325 have few positions with available synthesis results. 
Moreover, only  AM\,1054B has an old population
($t>2\times10^{9}$\,years) as being the main source of the 
flux at $\lambda$5870 \AA. In Fig.\ 1 of Paper\,I we can see that this object has
an elliptical form, being the old population predominant on it.
 
It is largely accepted that the  growth of spiral galaxies follow  the inside-out scenario, in the sense
that inner regions of the disk   are   most chemically evolve and present
a higher percentage of old population than the outer regions \citep[see e.g.][]{minchev14, brook12,  few12, pilkington12, molla05}.
In fact, recently,  \citet{rosa15}, by using the data of the CALIFA survey \citep{sanchez12}, obtained the  radial structure of stellar population properties of 
300 galaxies, mostly not interacting,  and found negative age gradients 
in agreement with an inside-out growth of galaxies. Other support for the assumption that spiral galaxies
are formed following the inside-out scenario is that 
negative metallicity gradients are generally observed in isolated spiral galaxies (e.g. \citealt{pilyugin15, sanchez14, dors05, pilyugin04}).
For  interacting galaxies,  the metallicity gradients are modified by the gas flow and these objects present gradients
significantly flatter than the ones observed in isolated galaxies (see e.g. Paper\,II and references therein). 
However,  it is unclear if  negative age gradients are  
maintained in interacting galaxies. In other words, it is ill-defined if the imprints of the inside-out scenario, in the case of
a negative gradient in the stellar population age along disks, are observed in interacting galaxies.
 
In Figs.~\ref{cen1}-\ref{cen4} the logarithm of the  average of the stellar population age weighted by the light contribution of each population 
($\rm \log[Age_{light}]$)
versus the galactocentric distance $R/R_{25}$ for our sample of objects is shown. 
The  $\rm Age_{light}$ gives information about the globally averaged star formation history in each galaxy \citep{rosa15}.
Taking into account the expected perturbations in a galaxy suffering stellar formation triggered by an external agent, we can note that 
for the most spiral components (AM\,1219A, AM\,1256B, AM\,1401A, AM\,2058A, AM\,2306A   and AM\,2322A) of the interacting pairs belonging to our 
sample, the $\rm Age_{light}$ decreases with the increase of the galactocentric distance,
 indicating  that the interactions seem do not destroy the global imprints of the galaxy formation considering the inside-out  scenario. 
AM\,1054A and AM\,2229A,  both spiral galaxies, not show this behavior, however for AM\,1054A  the slit crosses only their inner region 
and AM\,2229A is a spiral very disturbed.

In  Paper\,II, we found oxygen abundance breaks along the disk of AM\,1219A, AM\,1256B, AM\,2030A and AM\,2030B galaxies
at galactocentric distances $R/R_{25}$ between 0.2 and 0.5.  In particular, for  AM\,1219A and AM\,2030B, 
 we  found a minimum in the  instantaneous star formation rate (SFR)  close to the break region, and for AM\,1256B and AM\,2030A  maximum values were derived. In particular, AM\,1219A  presents
 also a minimum value of the metallicity in the break region, which could be associated with a corotation radius, in agreement with the scenario proposed by \citet{mishurov02}. 
As can be seen in  Fig.~\ref{corotation}, we  do not find variations of the stellar population  associated with the corotation radius.

\subsection{Stellar versus nebular extinction}
\label{extinction}

The stellar population synthesis process gives us the extinction form the continuum  A$_{V}[\rm Synthesis]$.
We derive the nebular extinction A$_{V}[\rm Nebular]$ from the H$\alpha$/H$\beta$ emission-lines ratio adopting the \citet{calzetti94} law:
\begin{equation}
A_{V}[\rm Nebular]=7.96\log\left[\frac{(H\alpha/H\beta)_{\rm obs}}{(H\alpha/H\beta)_{\rm int}}\right]   
\end{equation}
where (H$\alpha$/H$\beta)_{\rm obs}$ and (H$\alpha$/H$\beta)_{\rm int}$ are the observed and the intrinsic ratios, respectively. We adopted the theoretical value of 2.86 for 
the intrinsic emission-line ratio \citep{osterbrock06}.

The comparison between the synthesis and the nebular extinctions is plotted in Fig.~\ref{extinc} and was done for the objects in our sample.
As can be seen in this figure, 
the nebular extinctions are higher than the stellar ones, up to about twice. 
These results are in agreement with those earlier published by \citet{calzetti94, asari07,martins13b,dametto14} and was interpreted by 
\citet{calzetti94} as the result of the fact that the hot (young) ionizing stars are associated with dustier regions than the cold stellar population.

\subsection{Gas versus stellar parameters}
\label{comparation}
We compared the nebular and stellar metallicities for our sample of galaxies, in order to search for a relation between these two quantities. 
We assume the oxygen abundance as a nebular metallicity estimator. For our sample, we estimated the oxygen abundances in Paper\,II using empirical calibrations based on strong 
emission-lines, and the stellar metallicitty was obtained from the synthesis results.

Fig.~\ref{gradiente_gas_estelar} shows the nebular abundance versus stellar metallicity. As can be seen, there is no correlation between 
nebular and stellar metallicities. Similar results were found by \citet{cacho14} in a study about the gaseous-phase metallicities and stellar populations in 
the centers of barred galaxies. 
In a study of the stellar populations using Sloan Digital Sky Survey galaxies,  \citet{cid05} stated to have found a correlation between a mass-weighted stellar 
metallicities and nebular oxygen abundances, although the dispersion is very large, with a Spearman rank correlation coefficient (R) of  0.48.  

We also compared nebular and stellar metallicities as a function of the stellar age. These comparisons are plotted in Fig.~\ref{gradiente_age_gas_stellar}.
As can be seen in the bottom panel of this figure, there is no correlation between the stellar metallicities and stellar ages.  
In contrast, a relationship between the nebular metallicities and stellar ages is found (see upper panel of Fig.~\ref{gradiente_age_gas_stellar}), 
in the sense that older regions are more metal-rich.

\begin{figure}
\begin{center}
\includegraphics*[angle=270,width=0.8\columnwidth]{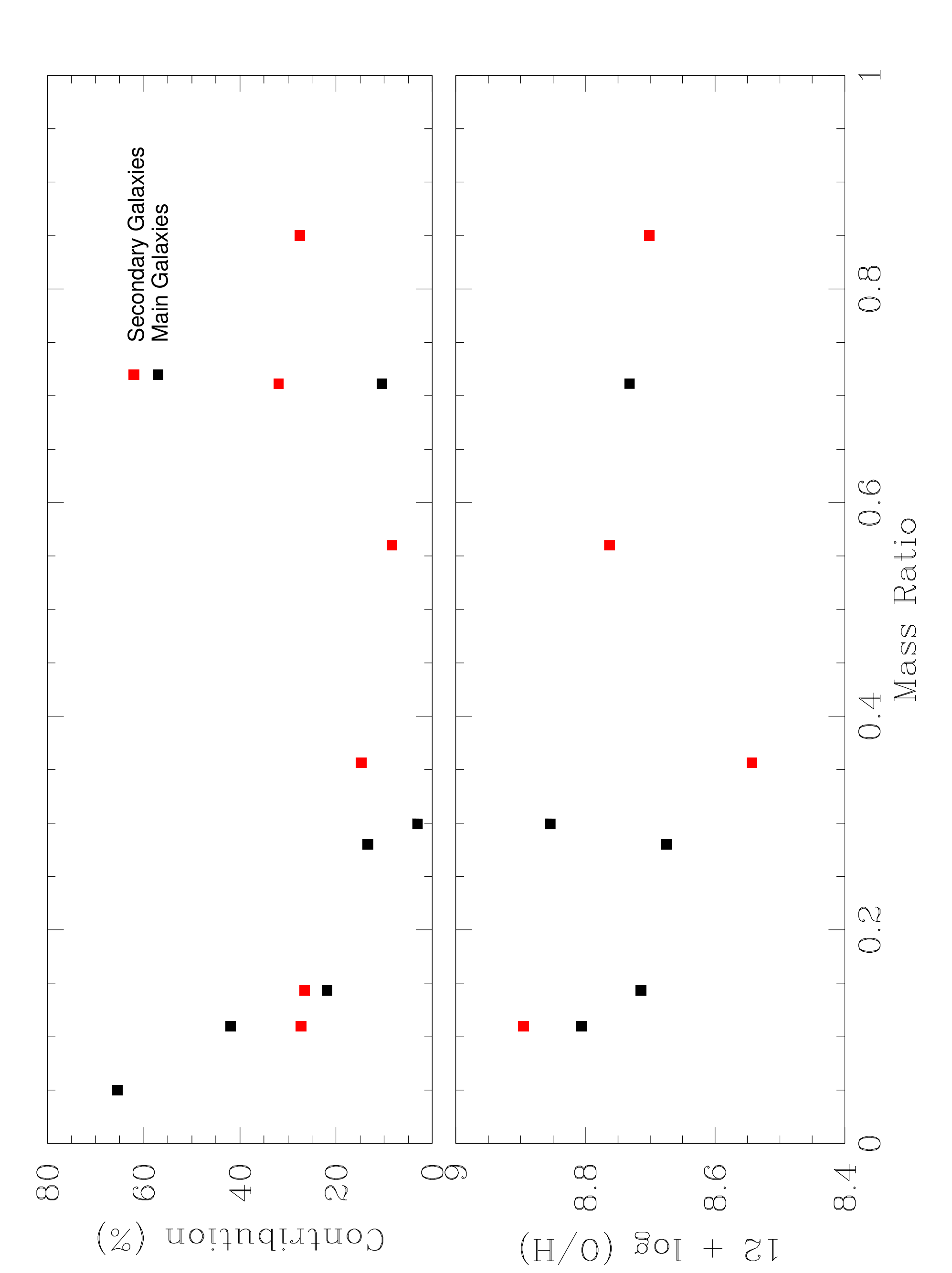}
\caption{Contribution of the young stellar population component to the light and estimated oxygen abundance as a function of the mass ratio between the components
($M_{secondary}/M_{primary}$), top and bottom panels, respectively. Points represent the central region of the galaxies inside 1\,kpc in diameter.}
\label{mass_ratio}
\end{center}
\end{figure}

\begin{figure}
\begin{center}
\includegraphics*[angle=270,width=0.8\columnwidth]{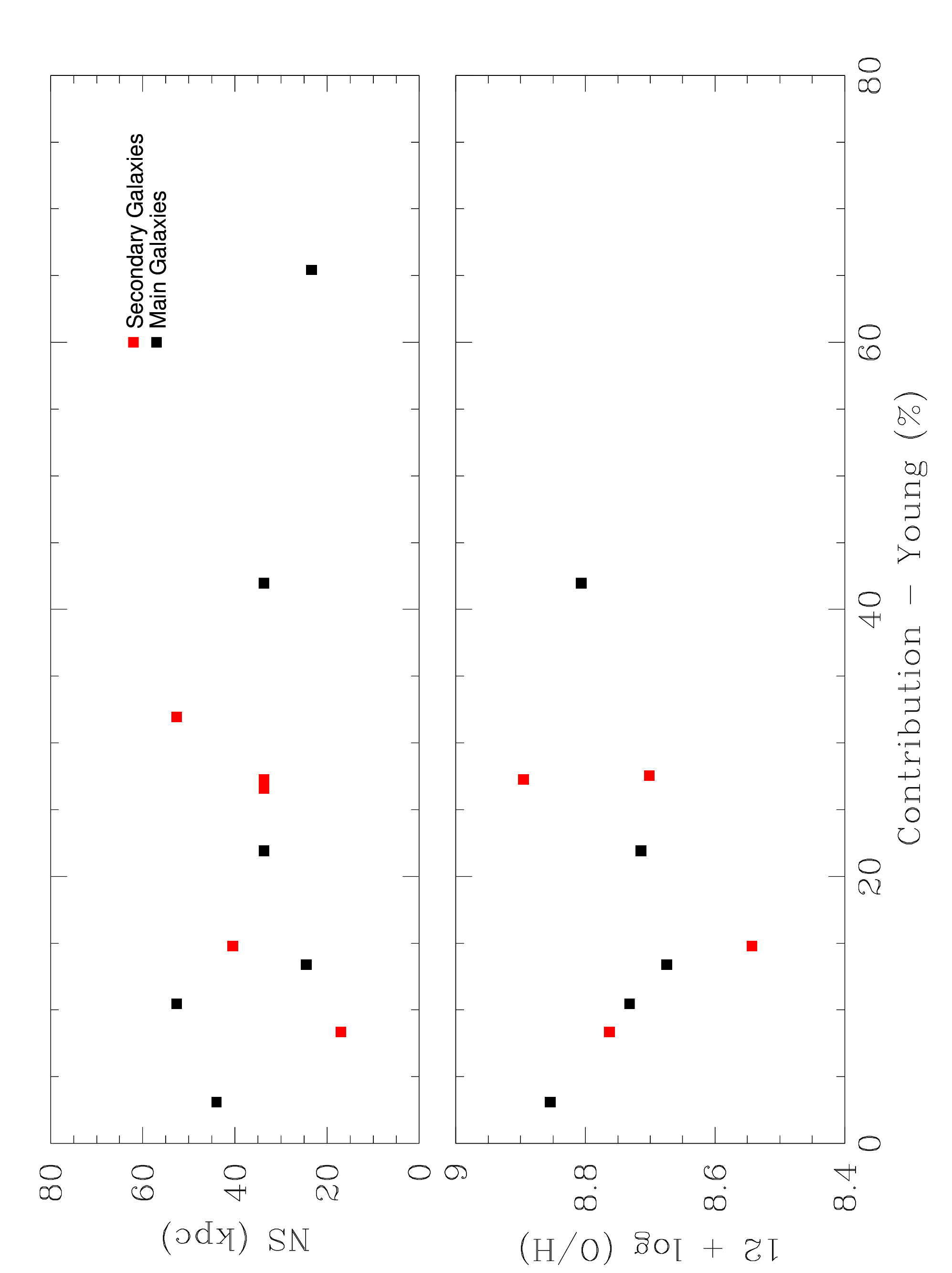}
\caption{ Nuclear separation between pair components and estimated oxygen abundances as a function of the fraction of the young stellar population component to the light, top 
and bottom panels, respectively.  Points represent the central region of the galaxies inside 1\,kpc in diameter.}
\label{mass_ratio2}
\end{center}
\end{figure}

We analysed some properties of the nebular gas and the stellar populations belonging to the central region (1 kpc in diameter) of the galaxies.
A comparison of the contribution (to the light) of the young stellar population component and the estimated oxygen abundances with the mass ratio between the components ($M_{secondary}/M_{primary}$) was performed (see upper and lower panels of Fig.~\ref{mass_ratio}, respectively). 
The mass ratio values were taken from \citet{ferreiro04}. 
We also compared the nuclear separation between the components  and the estimated oxygen abundances with the contribution (to the light) of the young stellar population component (see upper and lower panels of Fig.~\ref{mass_ratio2}, respectively). 
We did not find any correlation between the studied parameters. However, a more statistically significant sample is needed to reach any conclusive result.

\section{Conclusions}

\label{conc} 

We present an observational study about the stellar population in 
interacting galaxies using a synthesis method. Long-slit spectra  in the azul{spectral} range 3440-7300\,\AA\ were obtained with the GMOS at Gemini South 
for fifteen galaxies in nine close pairs.
The stellar population contribution was obtained  
using the {\scriptsize\,STARLIGHT} stellar population synthesis code. The main results are summarized in what follows: 
\begin{enumerate}
\item 
The contribution of the stellar components in relation to the optical flux at $\lambda$5870\,\AA\ for most of the galaxies in 
our sample: AM\,1054A, AM\,1219A, AM\,1256B, AM\,1401A, AM\,2030A, AM\,2030B, AM\,2058A, AM\,2058B, AM\,2306A, AM\,2306B, and AM\,2322B, 
is dominated by young/intermediate stellar populations.
\item
None variations in the stellar population components were found for AM\,1219A, AM\,1256B, AM\,2030A and AM\,2030B at the oxygen gradient break zones 
which could be associated with corotation radii (see Paper\,II).
\item 
We compared the stellar extinction (A$_{V}[\rm Synthesis]$) given by the population synthesis method with 
the nebular extinction (A$_{V}[\rm Nebular]$) estimated from the H$\alpha$/H$\beta$ emission-line ratio.
We found that for most of the objects the nebular extinction is higher than the stellar extinction, up to about twice.
\item 
Non correlation was found between nebular and stellar metallicities.
\item
We compared the nebular and stellar metallicities as a function of the stellar ages. 
Non correlation between stellar metallicities and stellar ages was found and a positive correlation 
between nebular metallicities and stellar ages was obtained showing, as expected, that 
the older regions are more metal-rich.
%Similar results have also been
%found in a studies about the gaseous-phase metallicities and stellar populations in the centers of barred galaxies by \citet{cacho14}.
%These authors investigate whether there is a relation between these components and whether bars induces episodes of star formation in the 
%nucleus of galaxies. The results show that did not observed any relation between gas and stars properties. No correlation between the 
%stellar age and metallicity was observed, since that age-metallicity degeneracy were broken.

\item
For the central regions of the studied galaxies, we did not find any correlation between the mass ratio and the contribution of the young stellar population component to the light or the estimated oxygen abundances. We did not either find any correlation between the contribution of the young stellar population to the light and the nuclear separation of the pairs or the estimated oxygen abundances. However more observations are needed to confirm these results.

\end{enumerate}

\section*{Acknowledgements}

Based on observations obtained at the Gemini Observatory, which is operated by 
the Association of Universities for Research in Astronomy, Inc., under a 
cooperative agreement with the NSF on behalf of the Gemini partnership: the 
National Science Foundation (United States), the Science and Technology
Facilities Council (United Kingdom), the National Research Council (Canada), CONICYT
(Chile), the Australian Research Council (Australia), 
Minist\'erio da Ciencia e Tecnologia (Brazil), and SECYT (Argentina).

D. A. Rosa and A. C. Krabbe thanks the support of FAPESP, process 2011/08202-6 and  2010/01490-3, respectively.
We also thank Ms. Alene Alder-Rangel for editing the English in this manuscript.

\newpage
\begin{appendices}

\section{Observed slit positions of AM\,1401-324}

\begin{appendixfig}
\includegraphics[width=0.7\columnwidth]{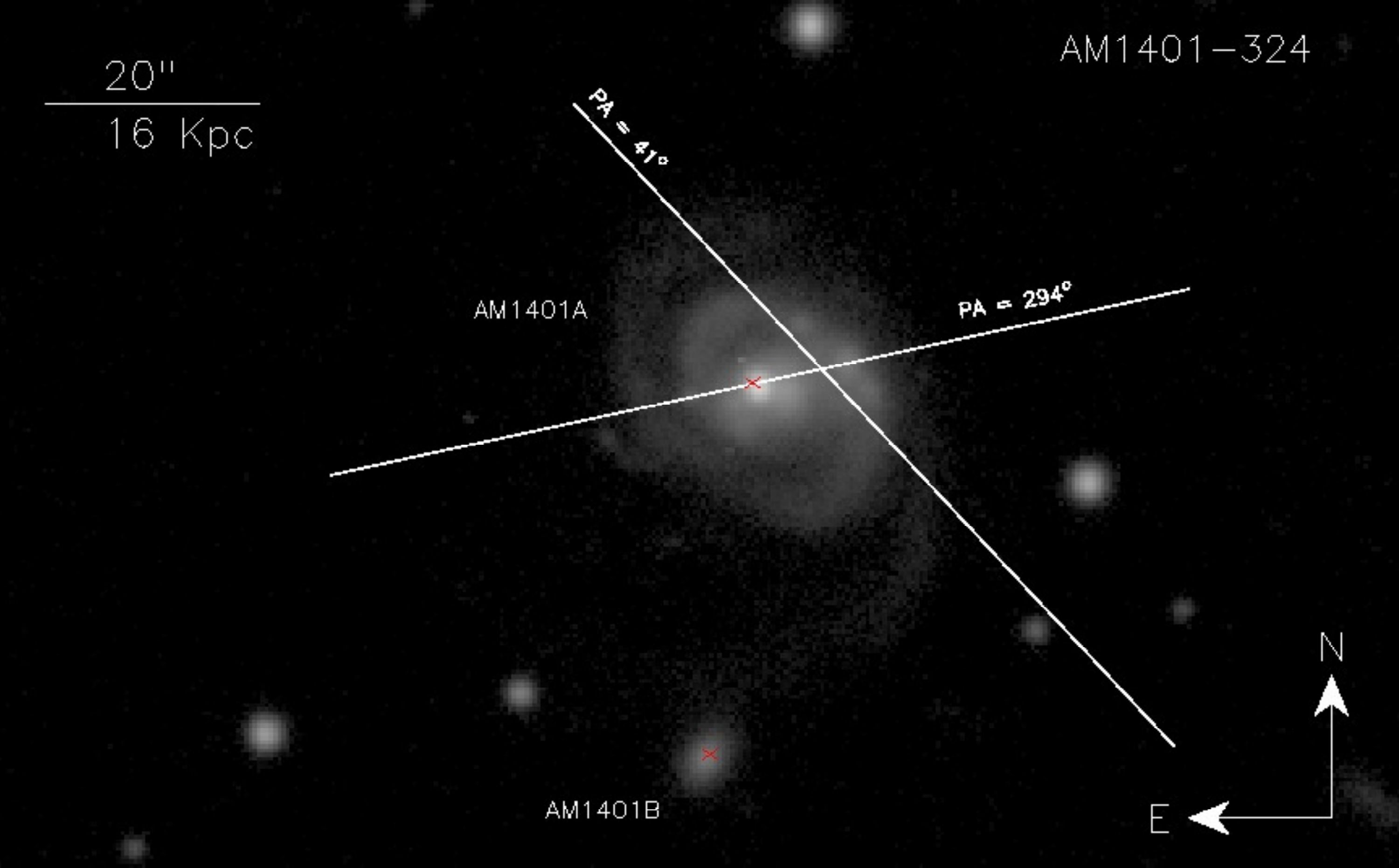}
\caption{Slit positions for AM\,1401-324 showed overimposed on the GMOS-S r$\arcmin$ acquisition image.}
\label{am1401_fenda}
\end{appendixfig}

\section{Regions A and B of  AM\,2229-735}

\begin{appendixfig}
\includegraphics*[angle=0,width=0.7\columnwidth]{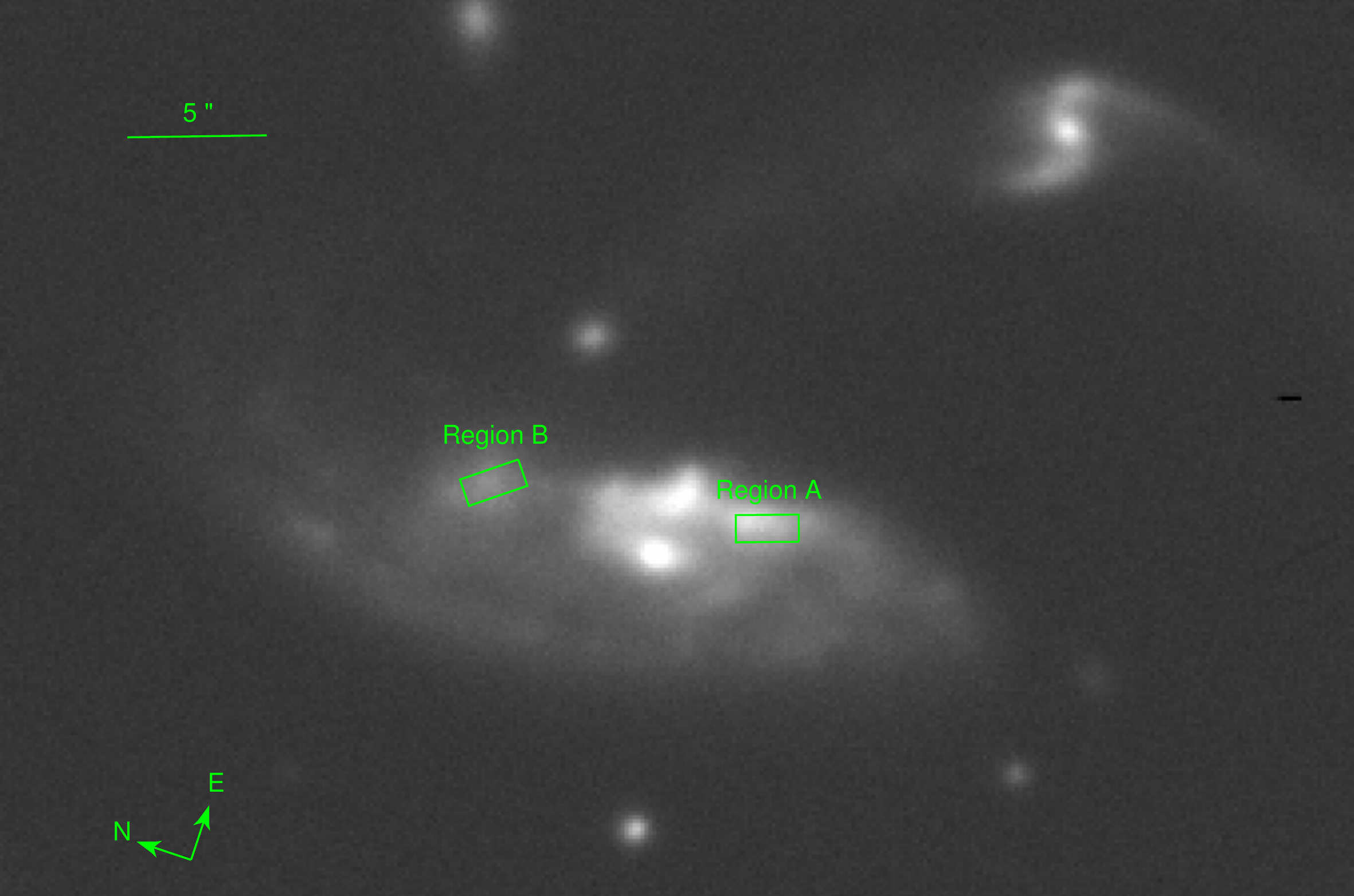}
\caption{Image Regions A and B of  AM\,2229-735 showed overimposed on the GMOS-S r$\arcmin$ acquisition image.}
\label{am2229_fenda}
\end{appendixfig}

\newpage

\section{Logarithm of the  average of the  age stellar population}

\begin{appendixfig} 
\includegraphics*[angle=-90,width=0.45\columnwidth]{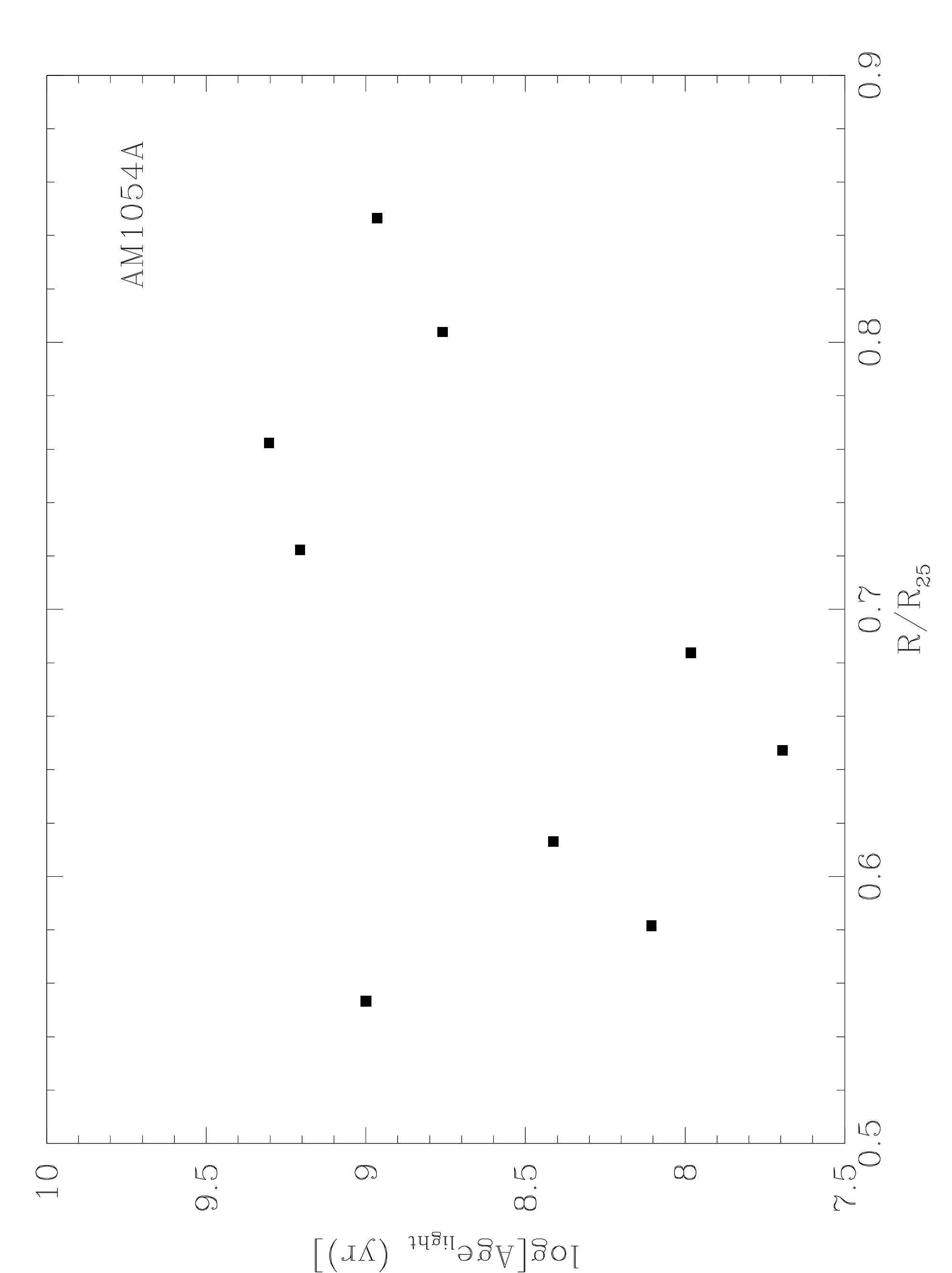}\hspace{0.4cm}
\includegraphics*[angle=-90,width=0.45\columnwidth]{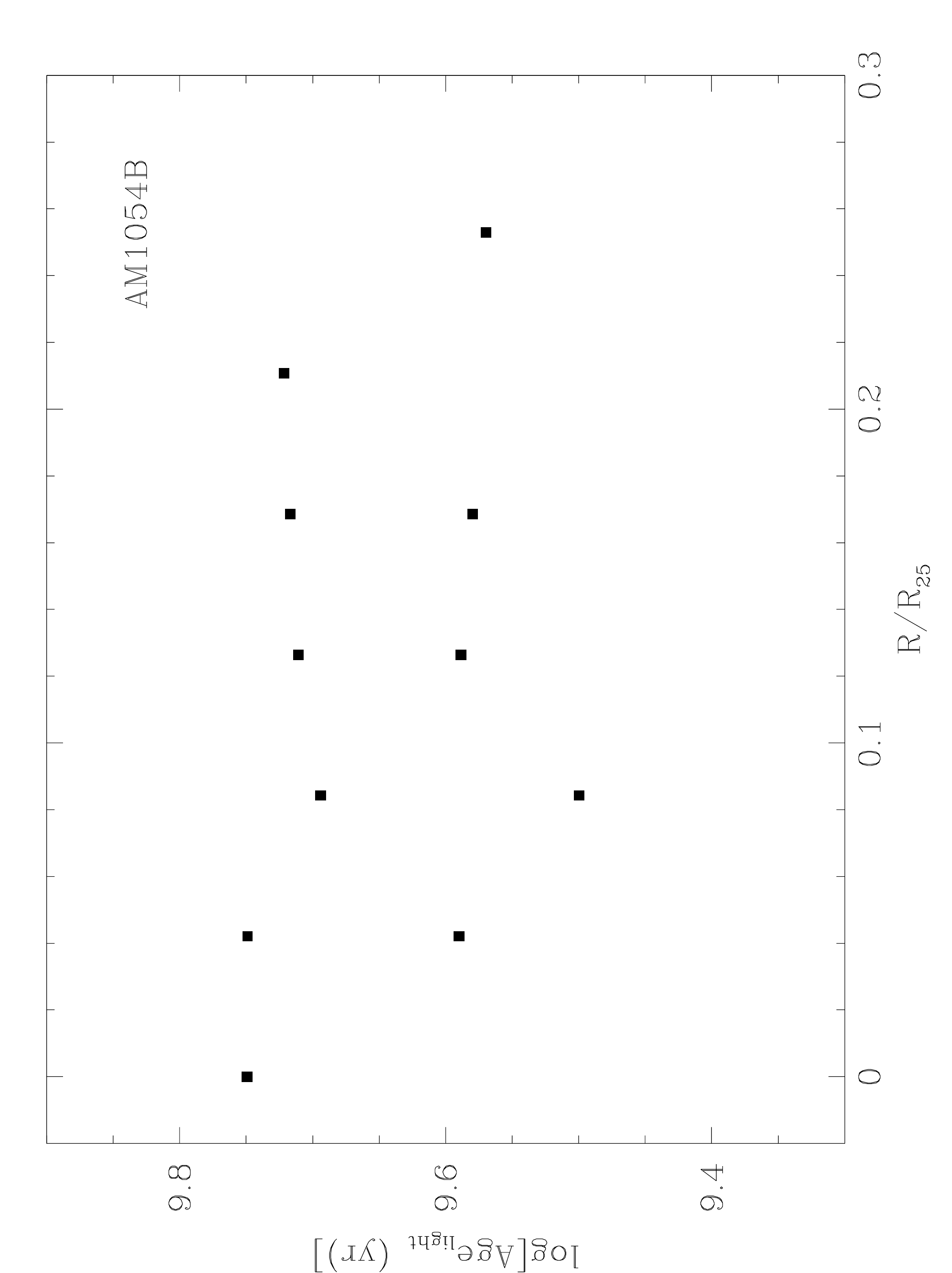}\\\vspace{0.4cm}
\includegraphics*[angle=-90,width=0.45\columnwidth]{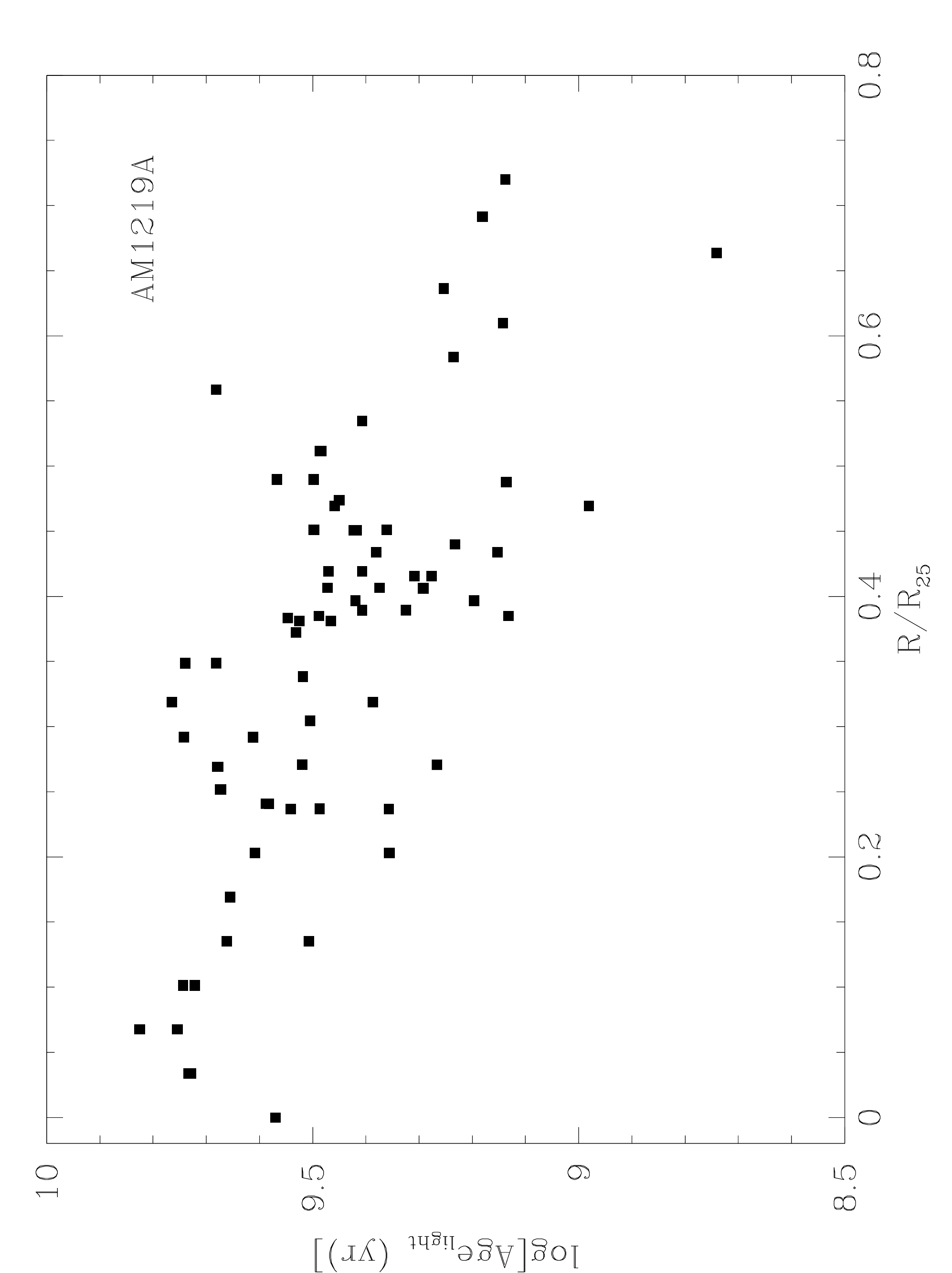}\hspace{0.4cm}
\includegraphics*[angle=-90,width=0.45\columnwidth]{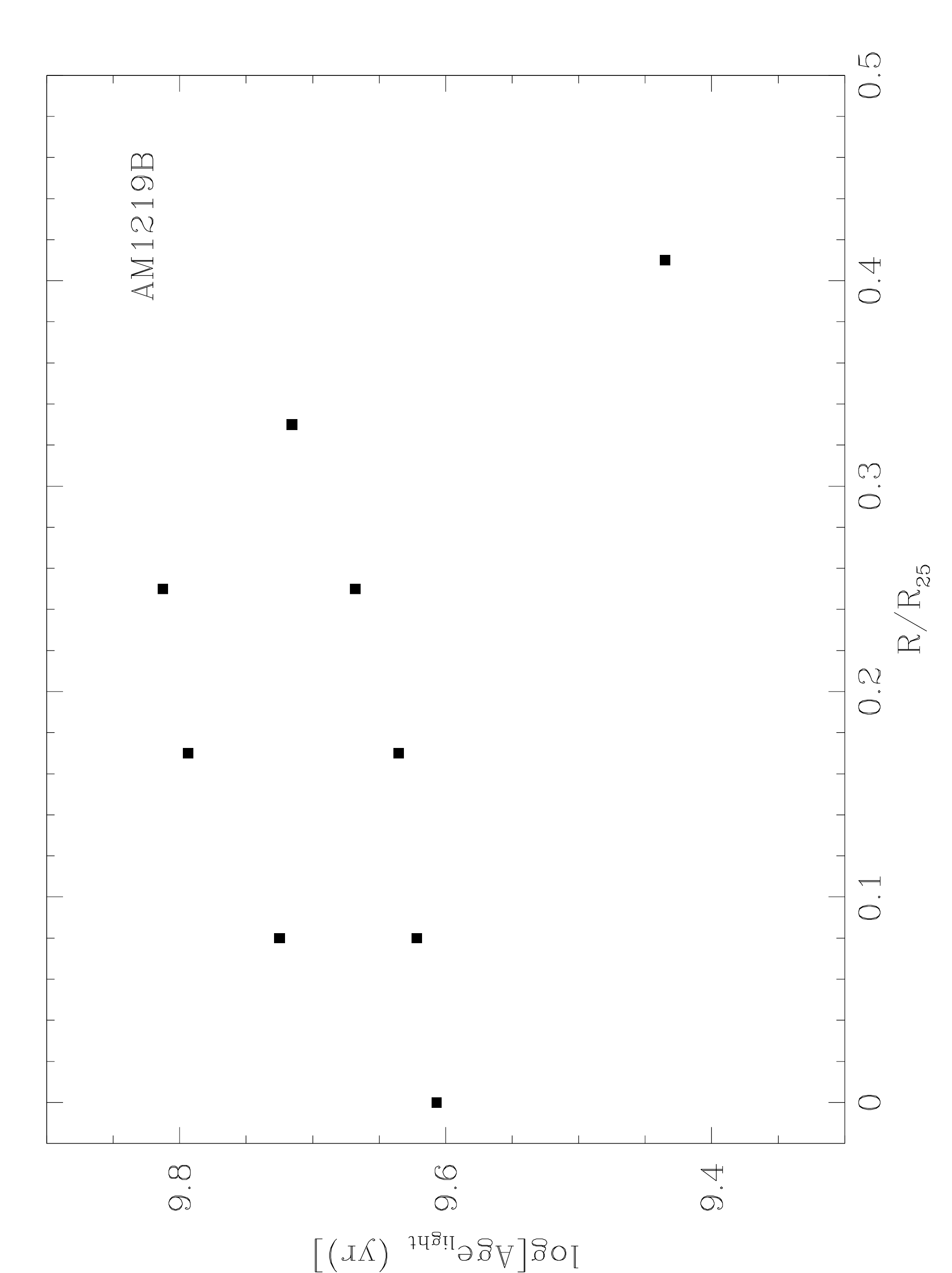}
\includegraphics*[angle=-90,width=0.45\columnwidth]{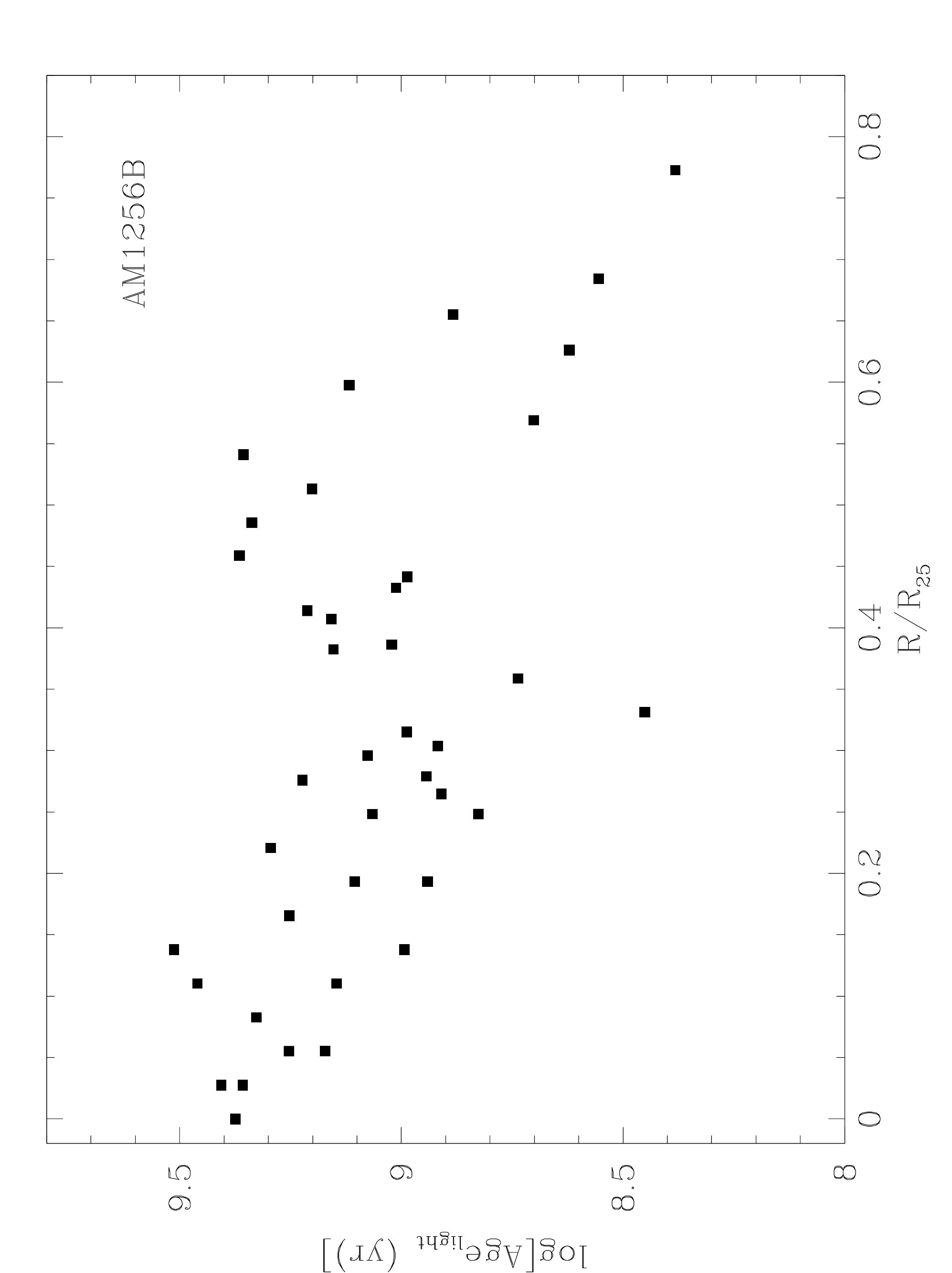}\hspace{0.4cm}
\includegraphics*[angle=-90,width=0.45\columnwidth]{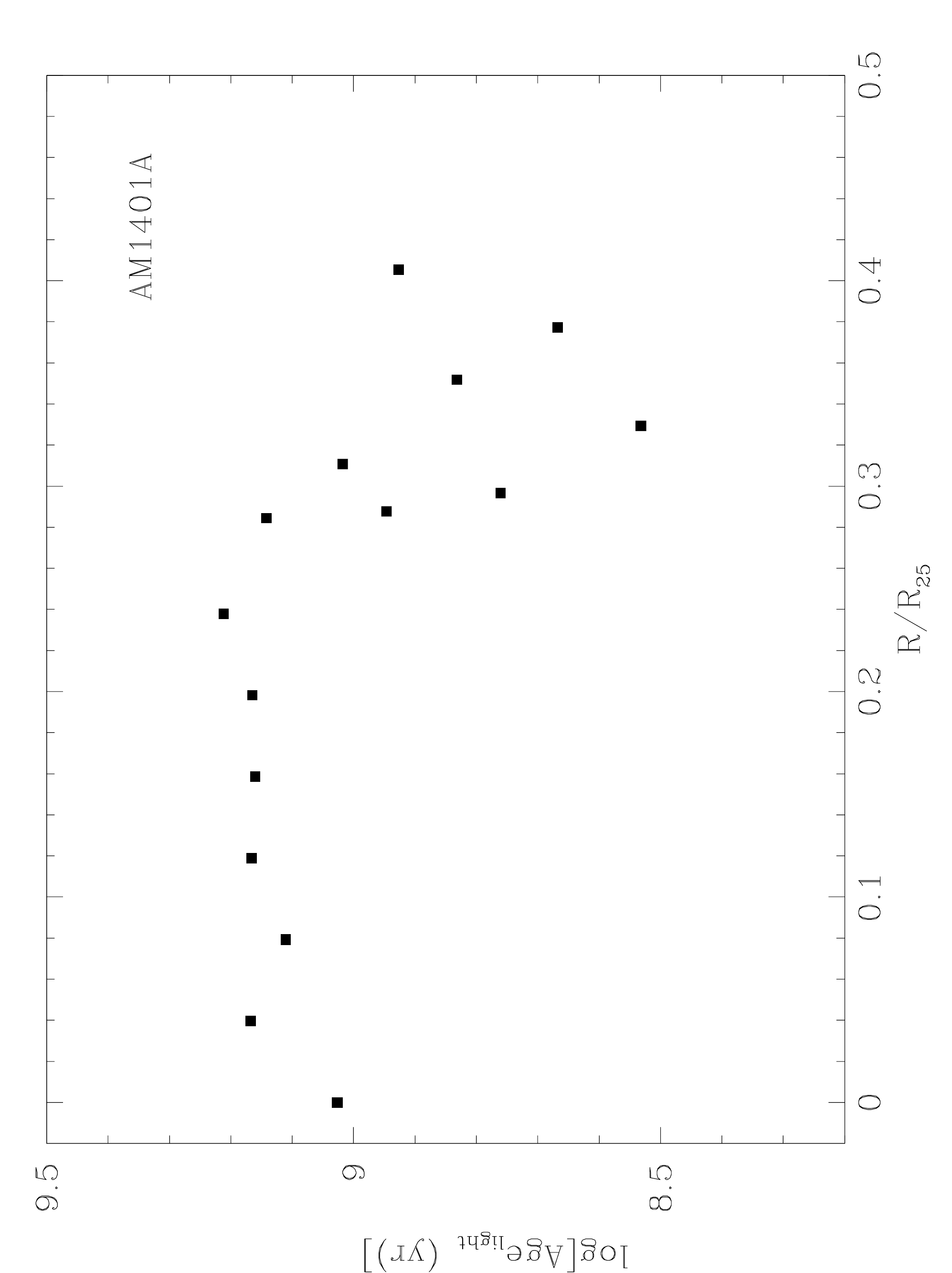}\\\vspace{0.4cm}
\caption{Logarithm of the  average of the  age 
 stellar population ($\rm \log[Age_{light}]$)  weighted by the  flux contribution of each population 
 versus the galactocentric distance $R/R_{25}$ for the objects of our sample as indicated. Points represents
 estimations for the regions along the disk of each galaxy.} 
\label{cen1}
\end{appendixfig}

\begin{appendixfig} 
\includegraphics*[angle=-90,width=0.45\columnwidth]{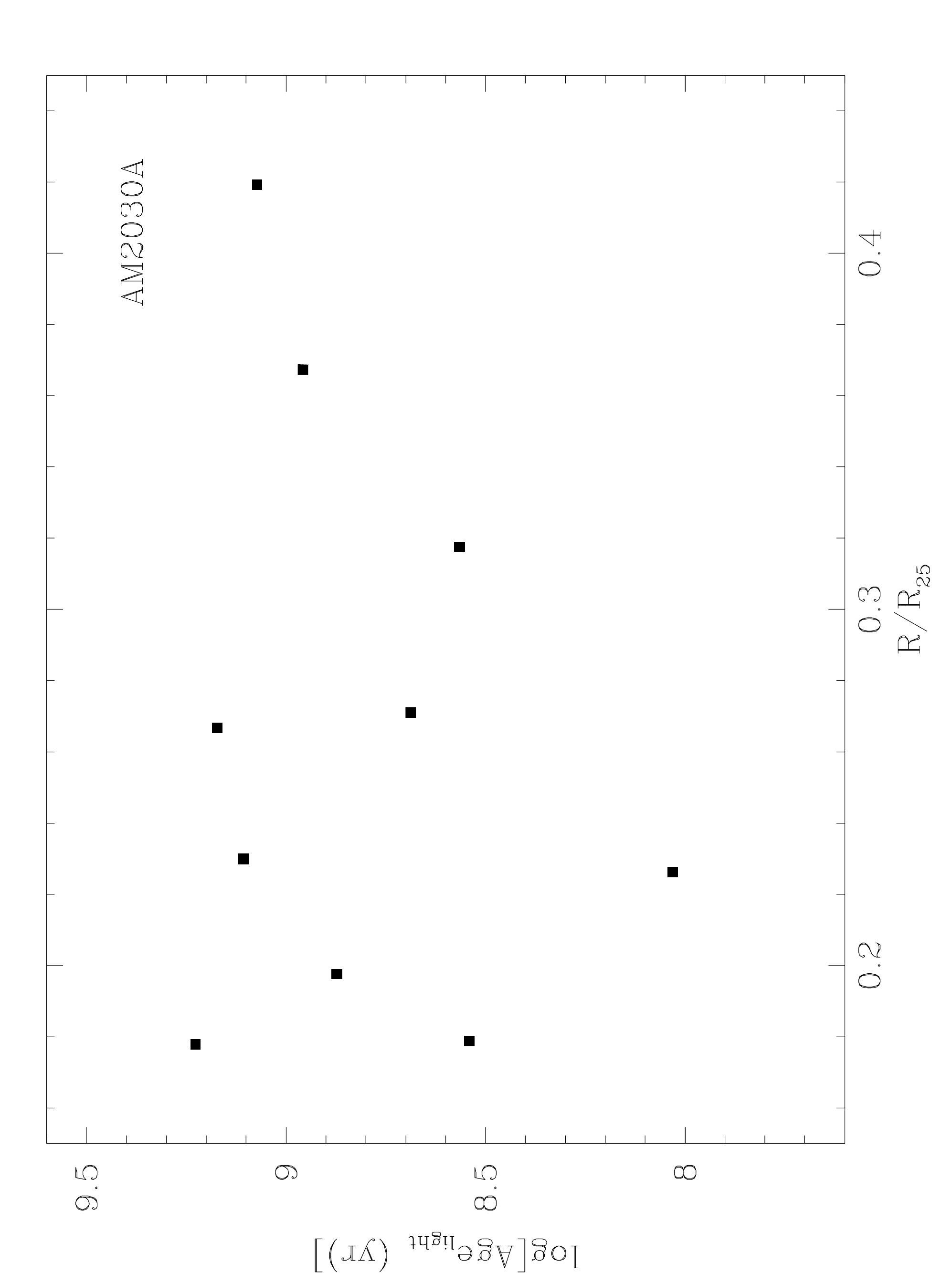}\hspace{0.4cm}
\includegraphics*[angle=-90,width=0.45\columnwidth]{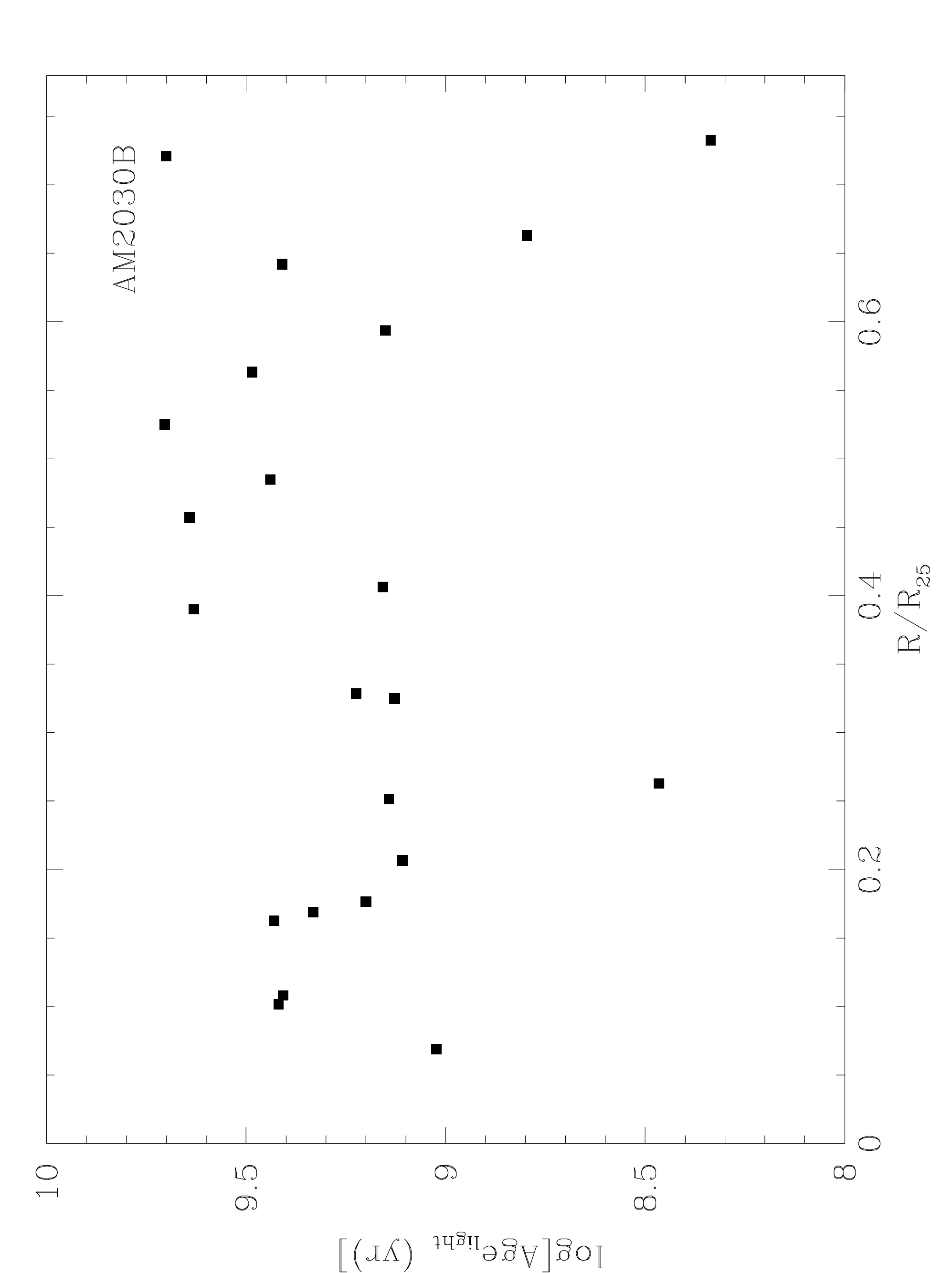}
\includegraphics*[angle=-90,width=0.45\columnwidth]{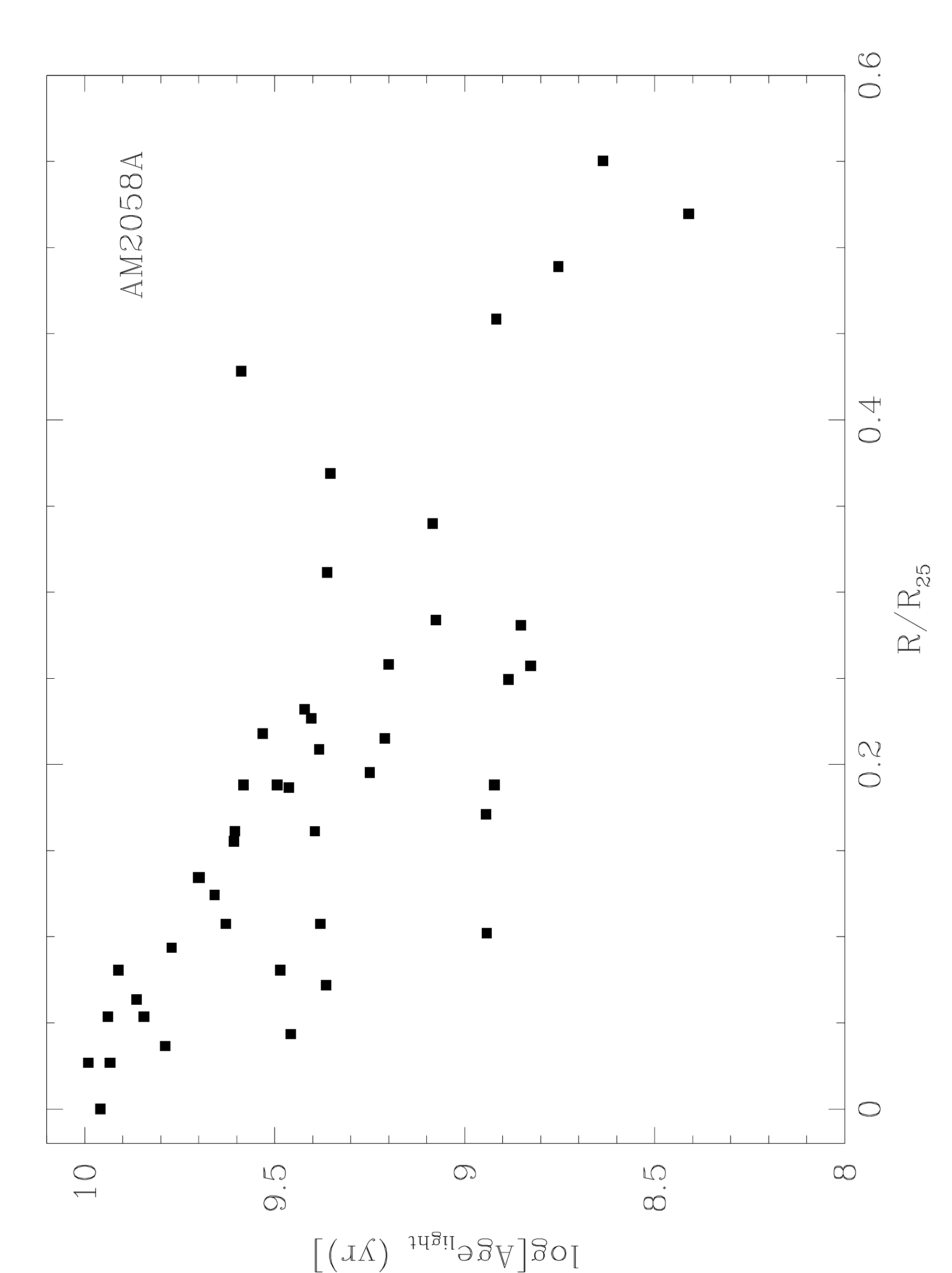}\hspace{0.4cm}
\includegraphics*[angle=-90,width=0.45\columnwidth]{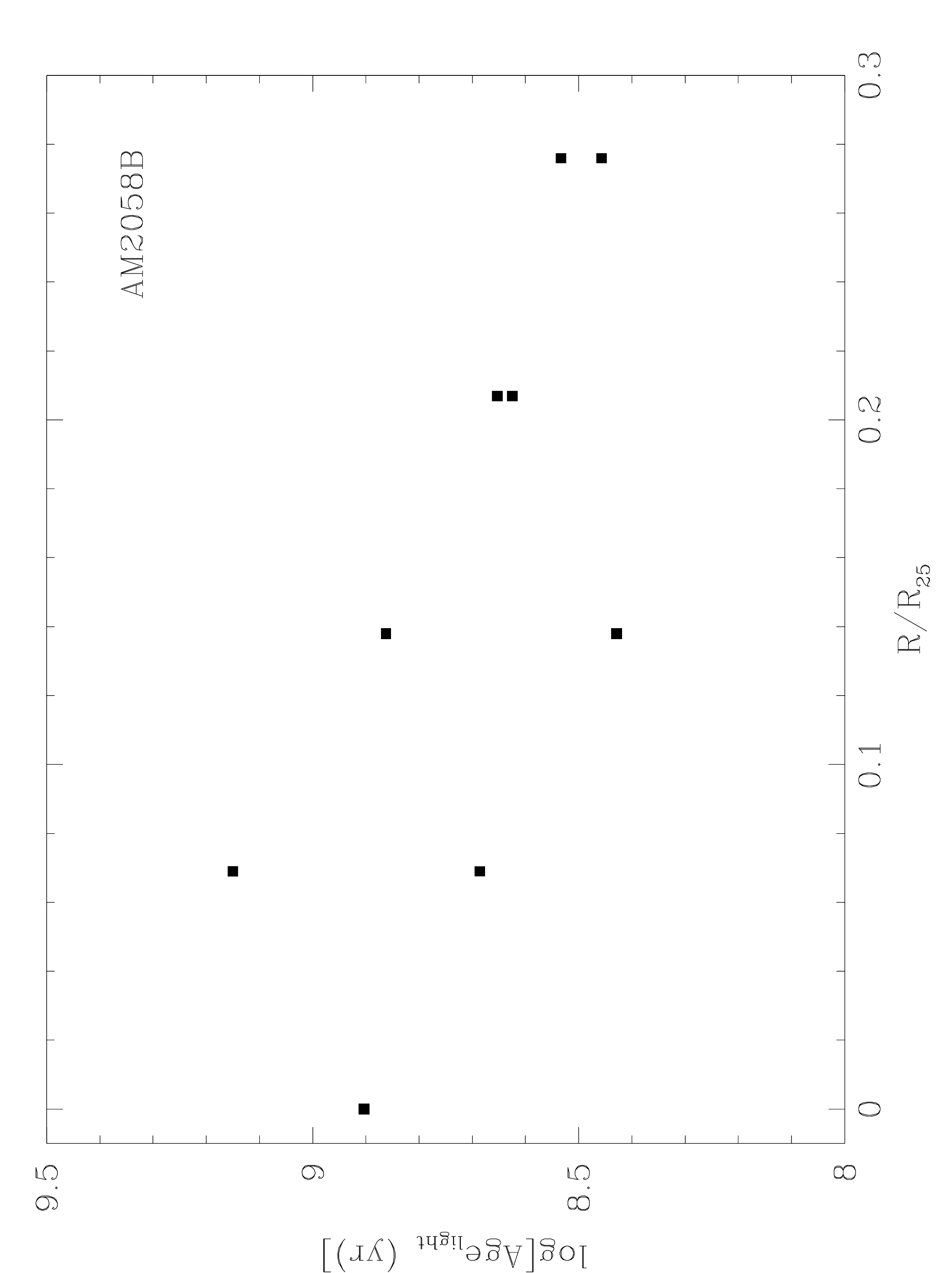}\\\vspace{0.4cm}
\includegraphics*[angle=-90,width=0.45\columnwidth]{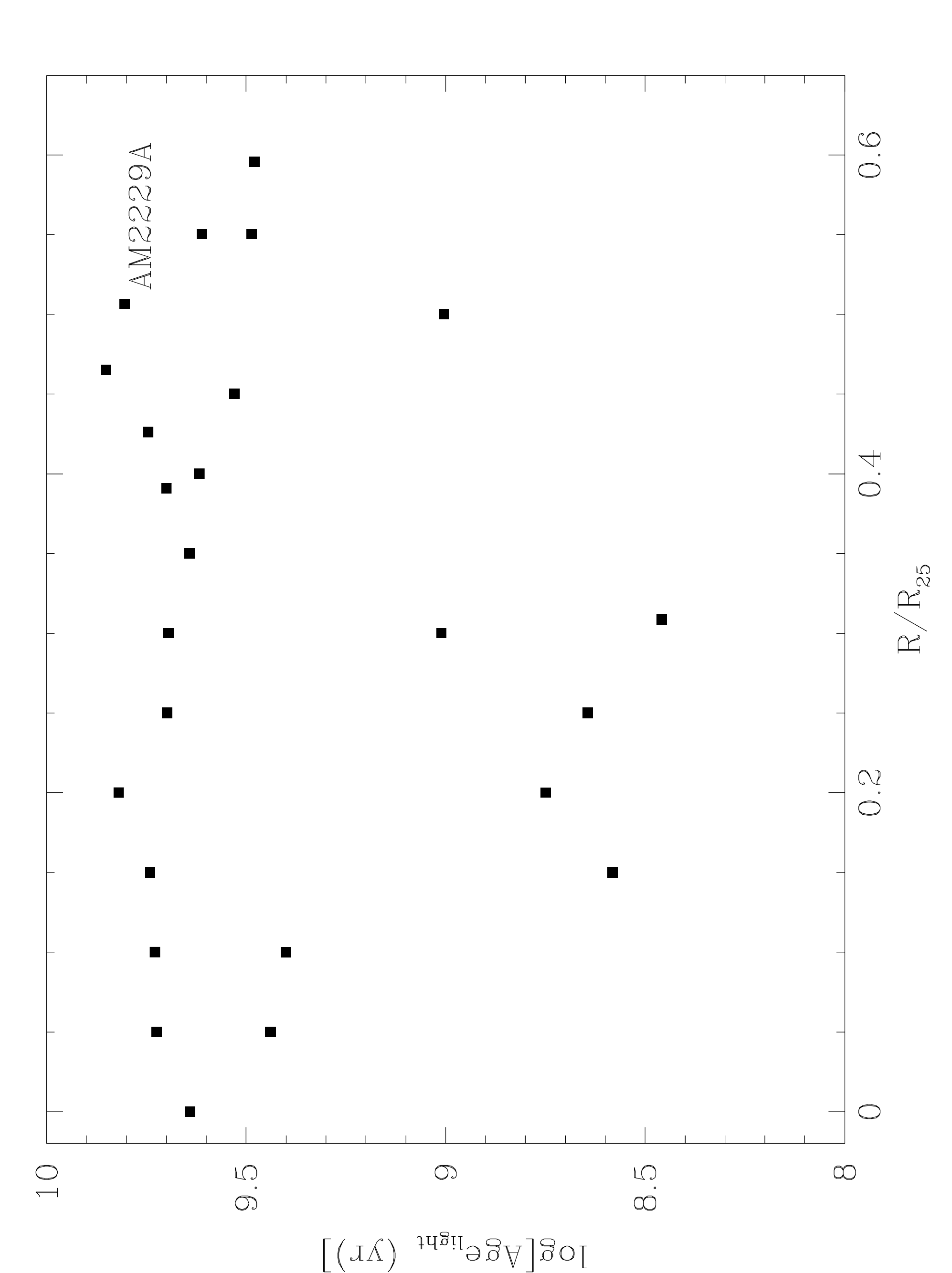}\hspace{0.4cm}
\includegraphics*[angle=-90,width=0.45\columnwidth]{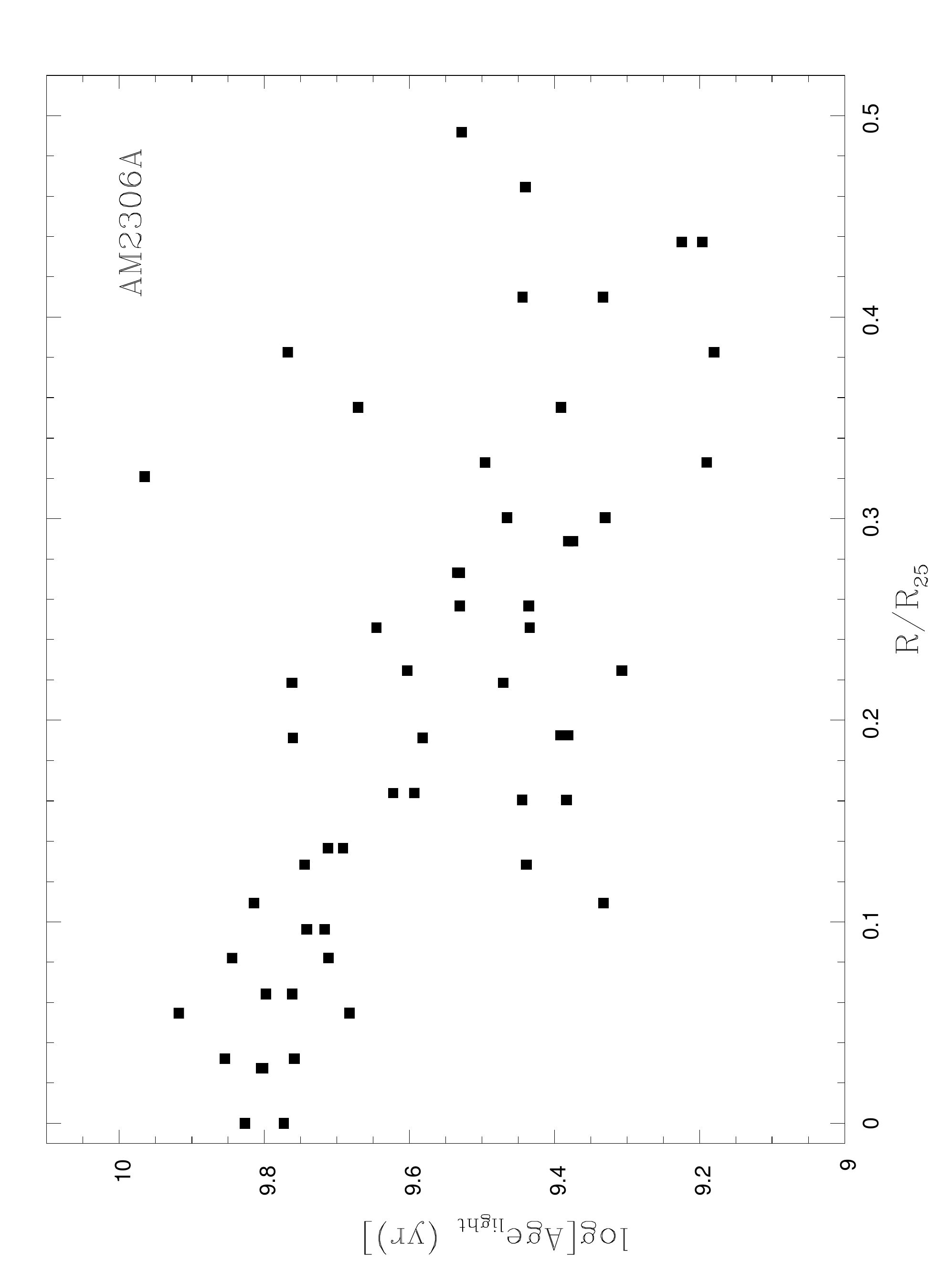}\hspace{0.4cm}
\caption{Such as Fig.~\ref{cen1} but for other objects as indicated.} 
\label{cen2}
\end{appendixfig}

\begin{appendixfig} 
\includegraphics*[angle=-90,width=0.45\columnwidth]{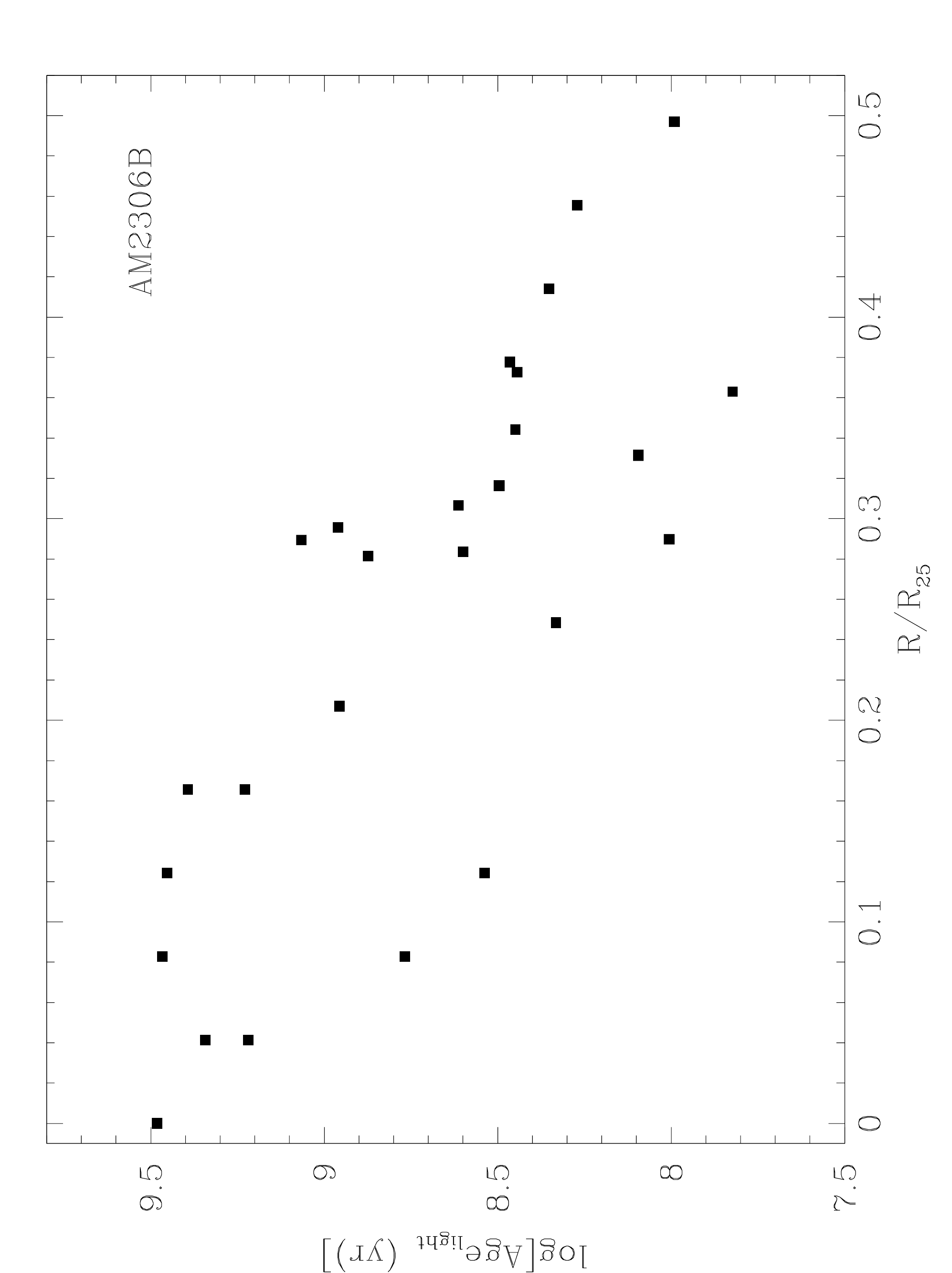}\hspace{0.4cm}
\includegraphics*[angle=-90,width=0.45\columnwidth]{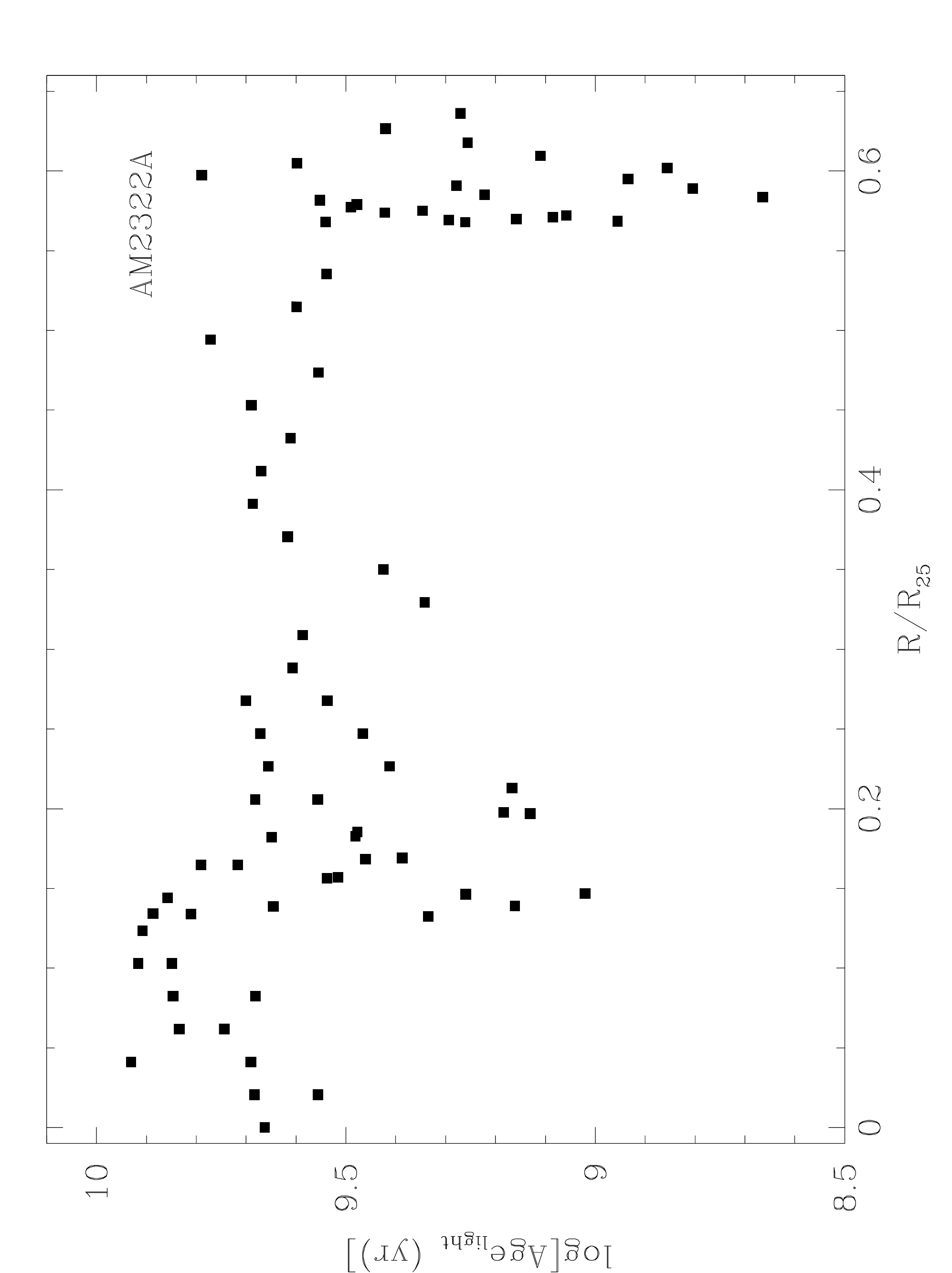}\\\vspace{0.4cm}
\includegraphics*[angle=-90,width=0.45\columnwidth]{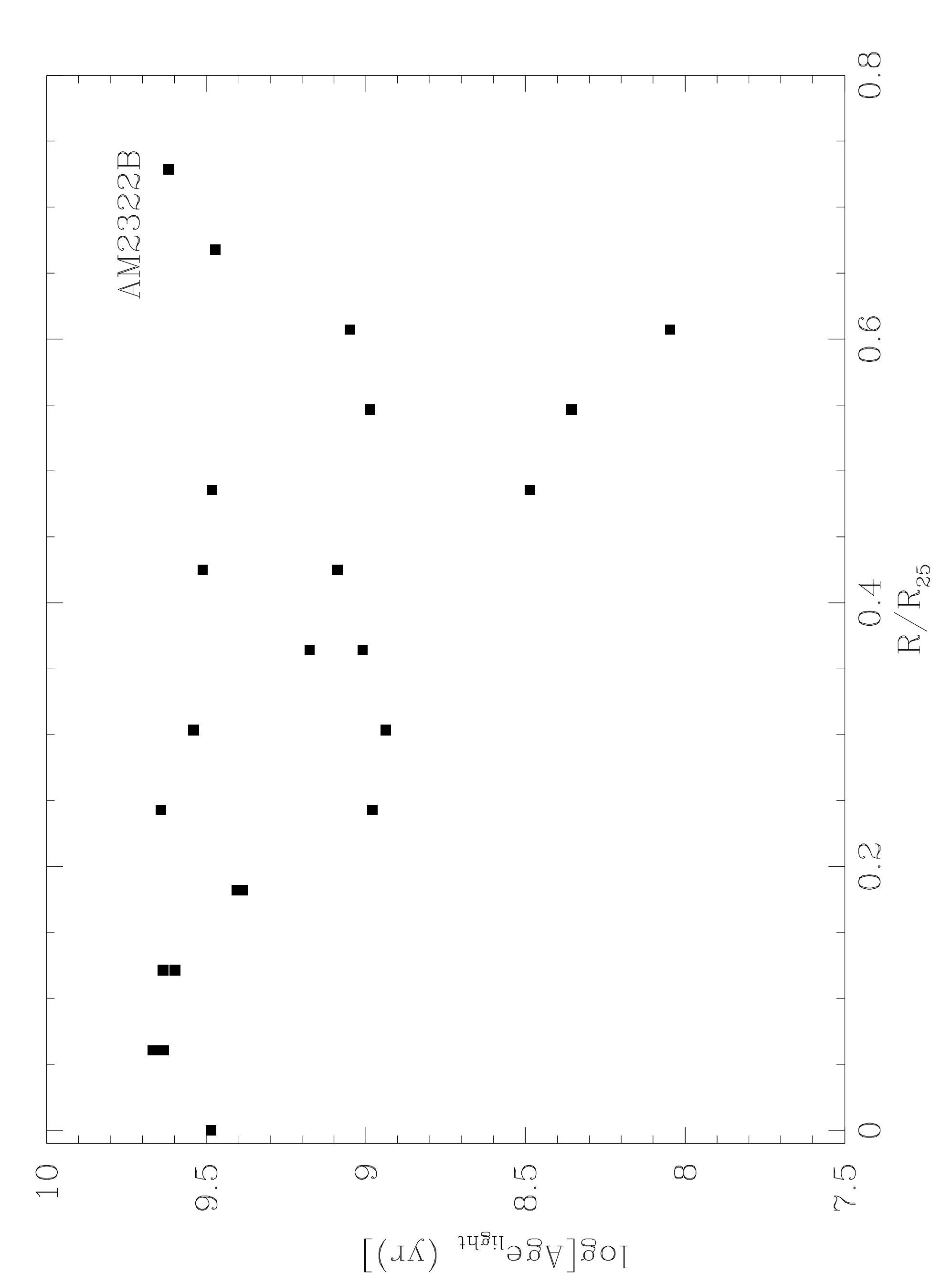}\hspace{0.4cm}
\caption{Such as Fig.~\ref{cen1} but for other objects as indicated.} 
\label{cen4}
\end{appendixfig}

\newpage

\section{Gas and stellar parameters}

\begin{appendixfig}
\includegraphics[angle=270,width=0.7\columnwidth]{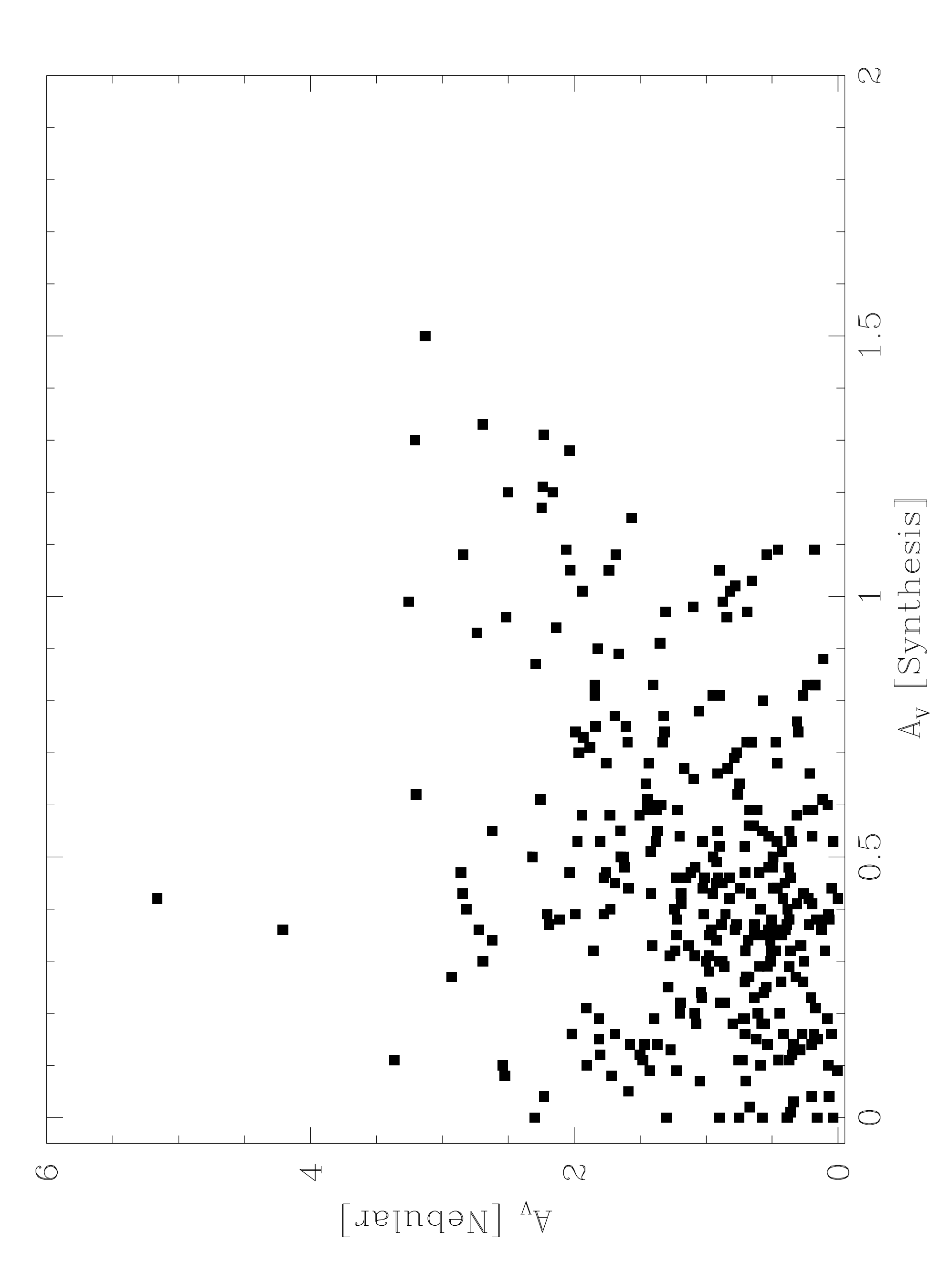}
\caption{Comparison between the extinction obtained from the stellar populations synthesis (A$_{V}$[Synthesis]) with the nebular 
extinction estimated from the H$\alpha$/H$\beta$ emission-line ratio (A$_{V}$\,[Nebular]) for the objects in our sample.}
\label{extinc}
\end{appendixfig}

\section{Gas and stellar parameters}

\begin{appendixfig}
\includegraphics[angle=270,width=0.7\columnwidth]{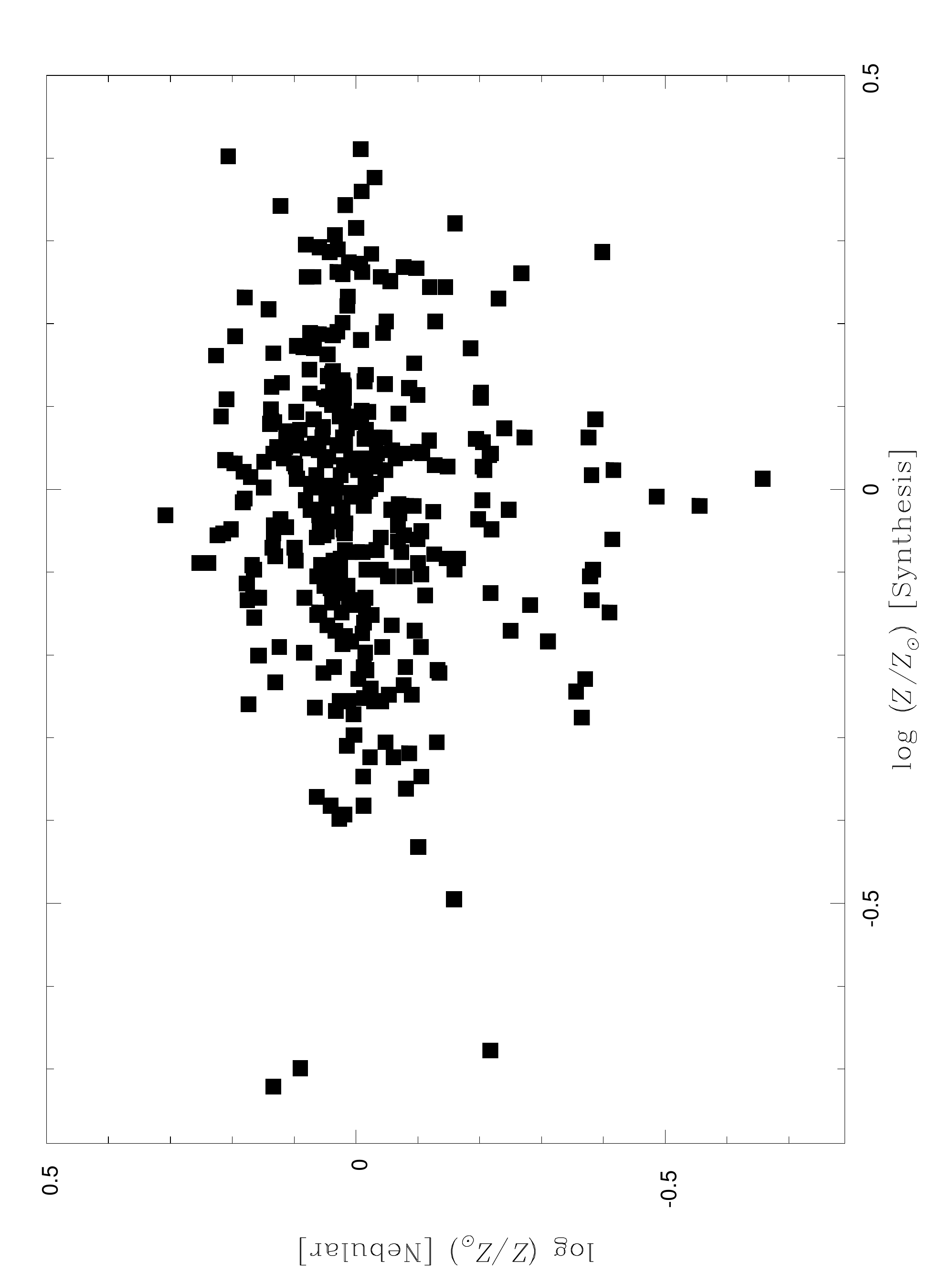}
\caption{Comparison between the stellar and nebular metallicities.}
\label{gradiente_gas_estelar}
\end{appendixfig}

\begin{appendixfig}
\includegraphics*[angle=270,width=0.45\columnwidth]{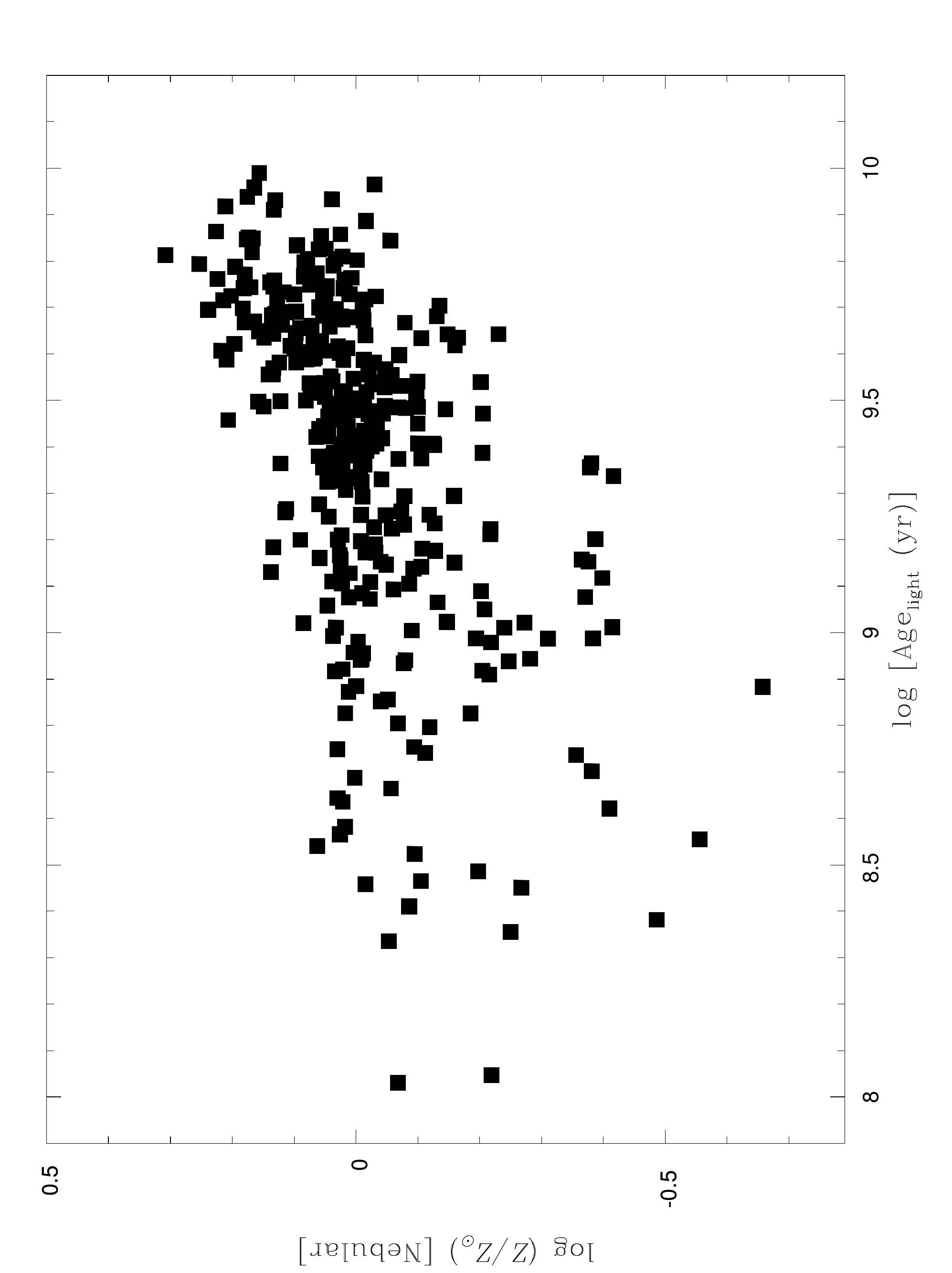}
\includegraphics*[angle=270,width=0.45\columnwidth]{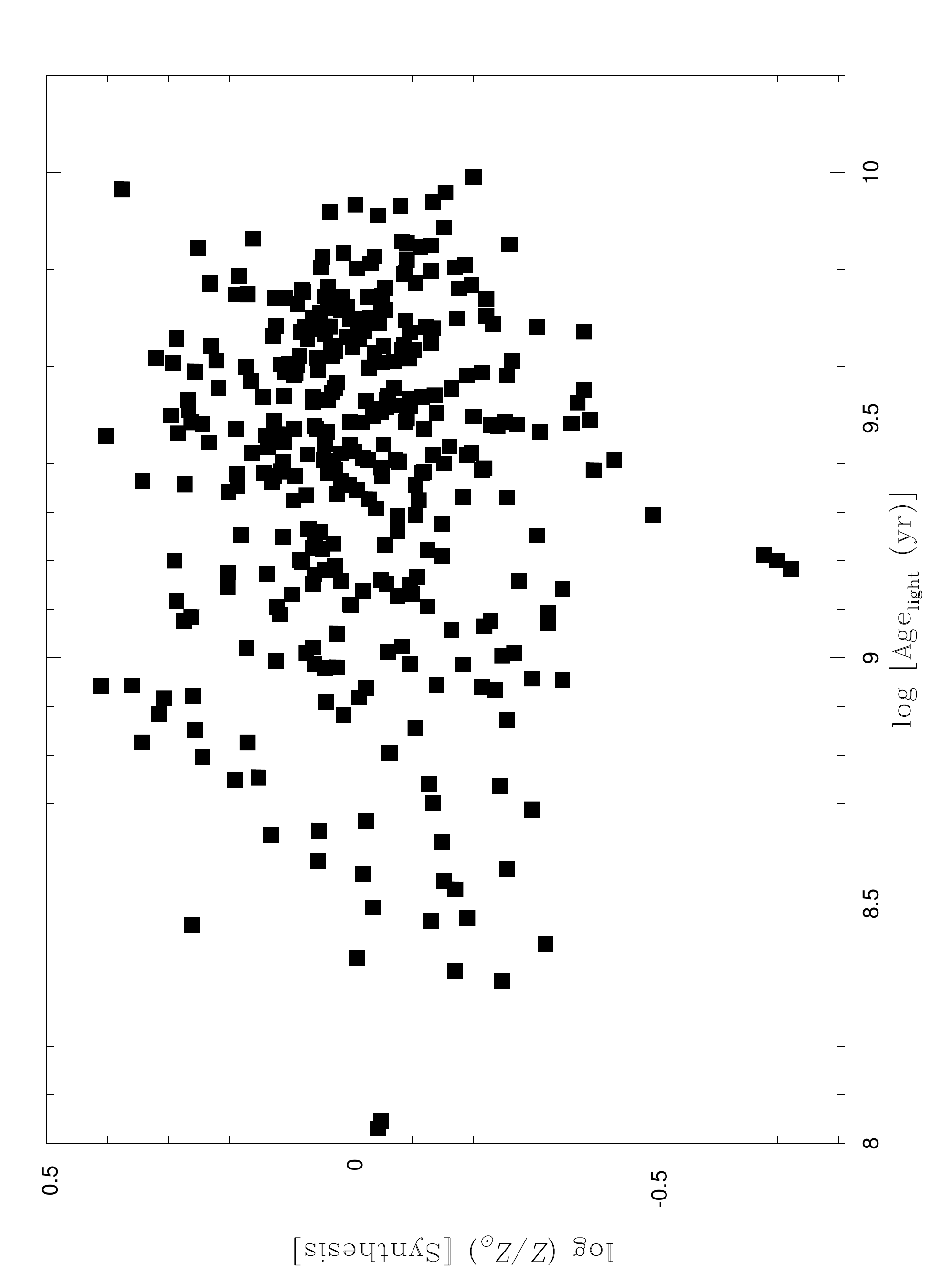}
\caption{Comparison between nebular (right) and stellar (left) metallicities  versus the stellar age.}
\label{gradiente_age_gas_stellar}
\end{appendixfig}

\section{Stellar population synthesis}

\label{ap_synth1}

\begin{figure*}
\centering
\includegraphics*[angle=-90,width=0.45\columnwidth]{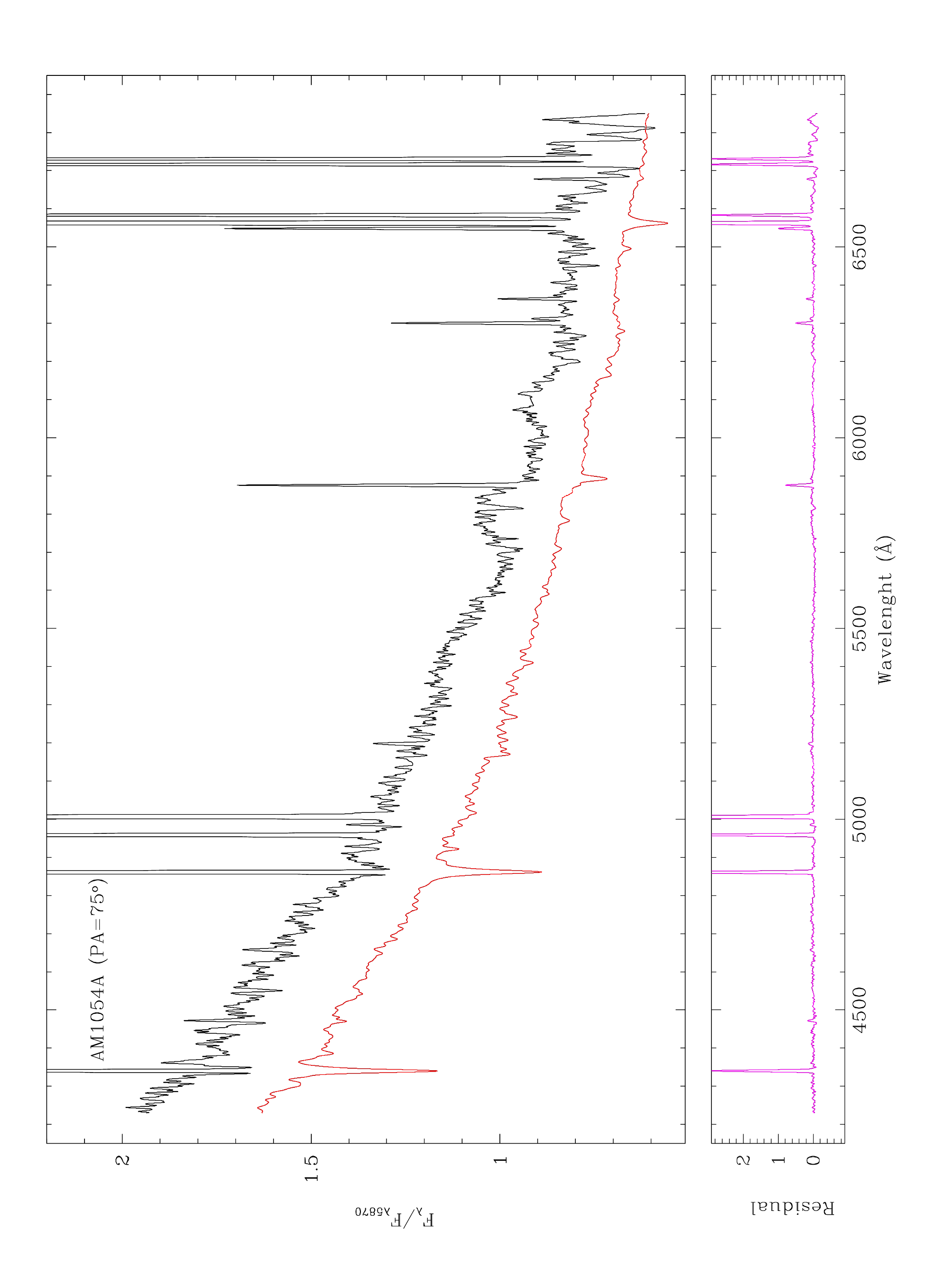}
\includegraphics*[angle=-90,width=0.45\columnwidth]{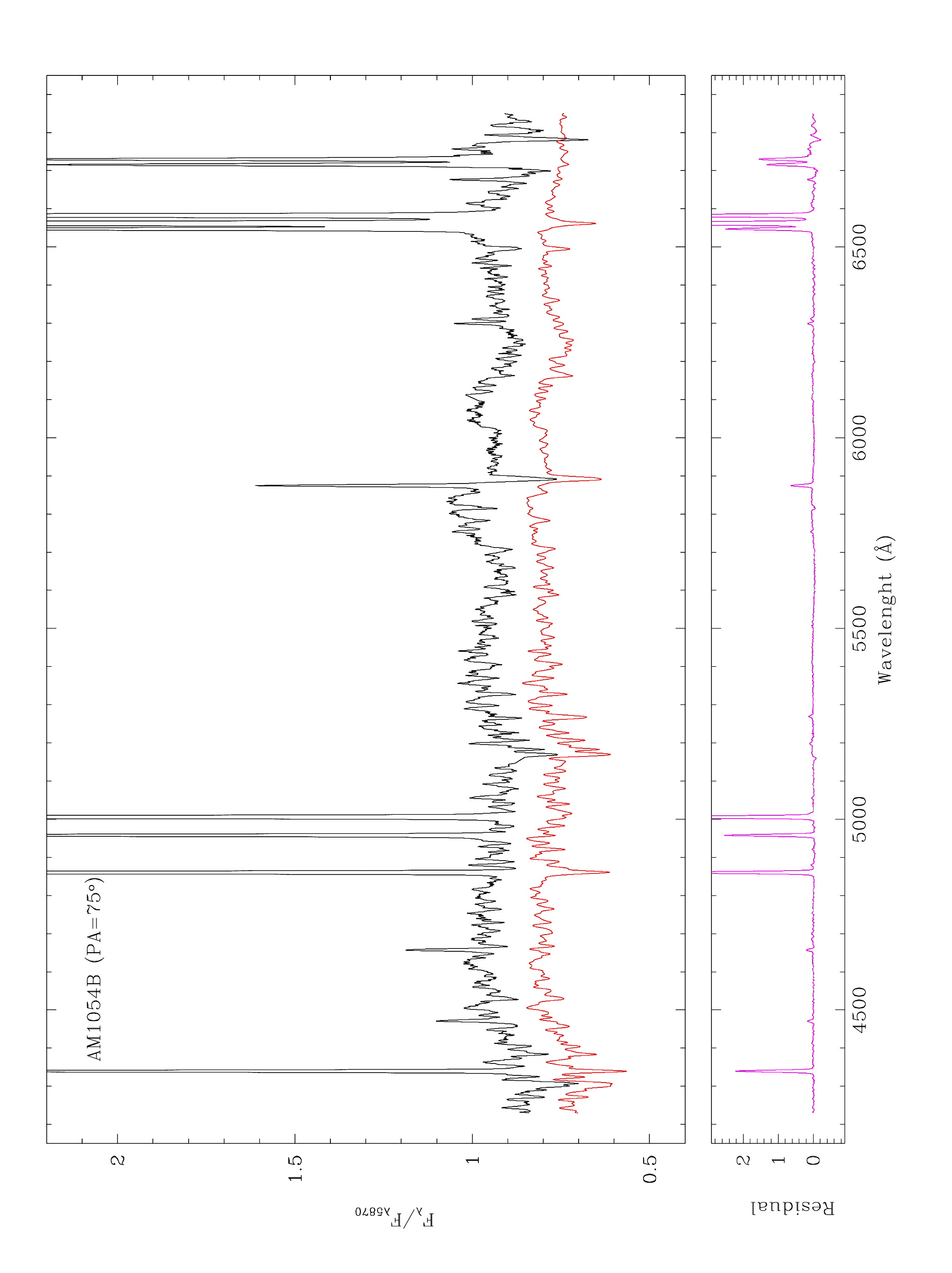}
\includegraphics*[angle=-90,width=0.45\columnwidth]{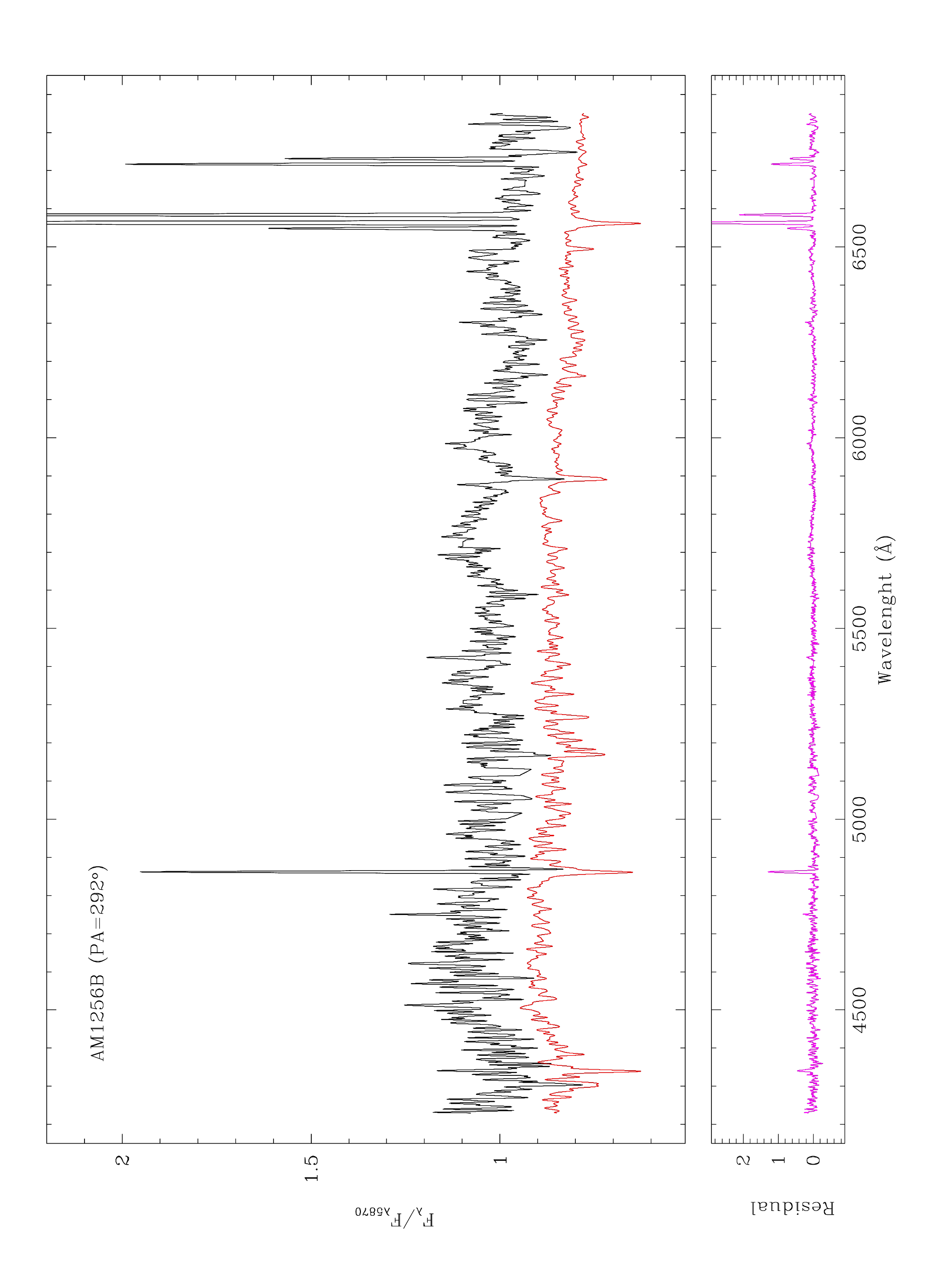}
\includegraphics*[angle=-90,width=0.45\columnwidth]{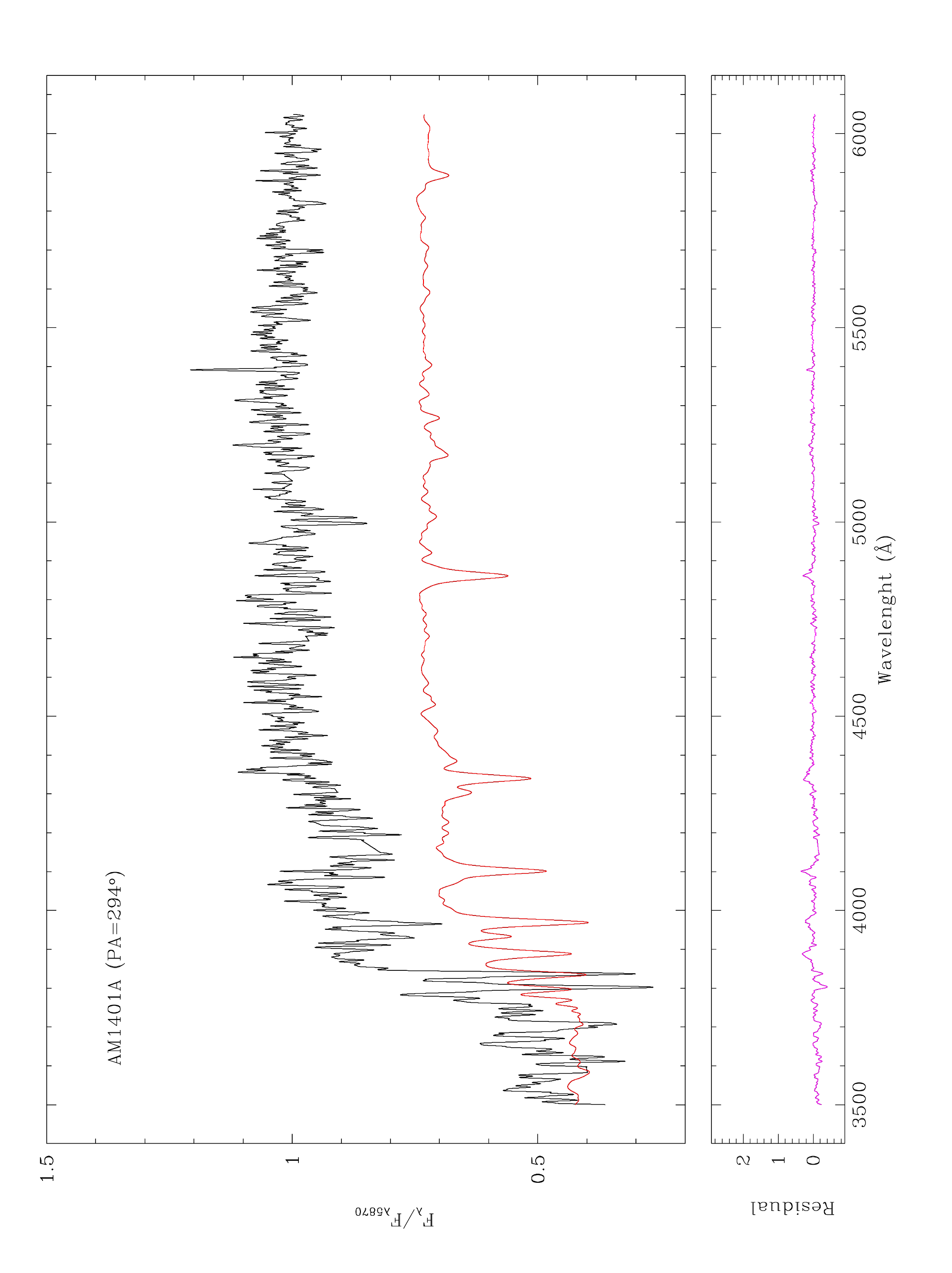}
\includegraphics*[angle=-90,width=0.45\columnwidth]{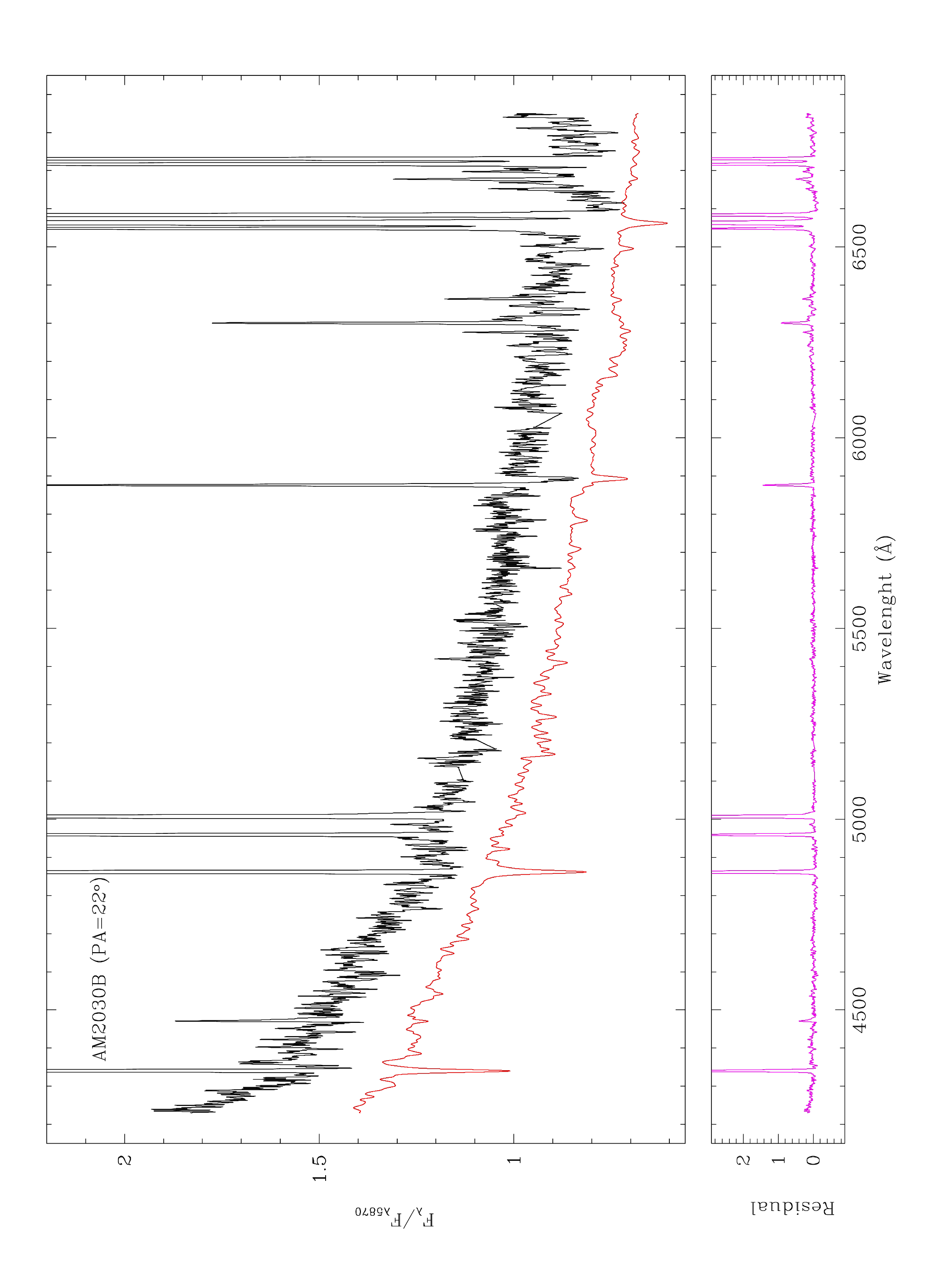}
\includegraphics*[angle=-90,width=0.45\columnwidth]{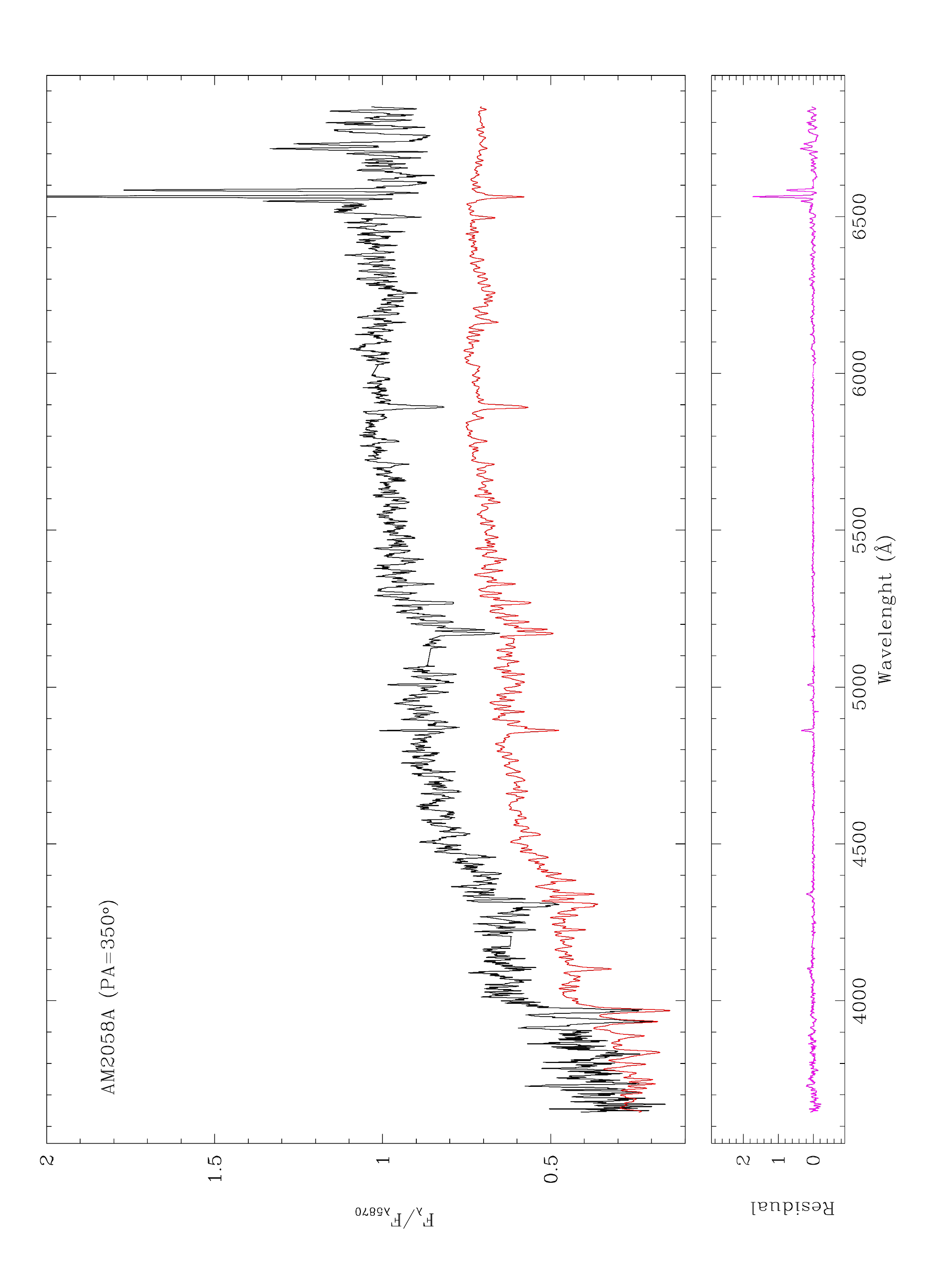}
\caption{Stellar population synthesis for the nuclear region of AM\,1054A, AM\,1054B, AM\,1256B, AM\,1401A, AM\,2030B and AM\,2058A.
Top panel: The observed spectrum is plotted as a black line and the synthesized
spectrum as a red line.The main absorption and emission features have been identified. Bottom
panel: pure emission spectrum corrected for reddening.}
\label{sintese_003}
\end{figure*}

\begin{figure*}
\centering
\includegraphics*[angle=-90,width=0.45\columnwidth]{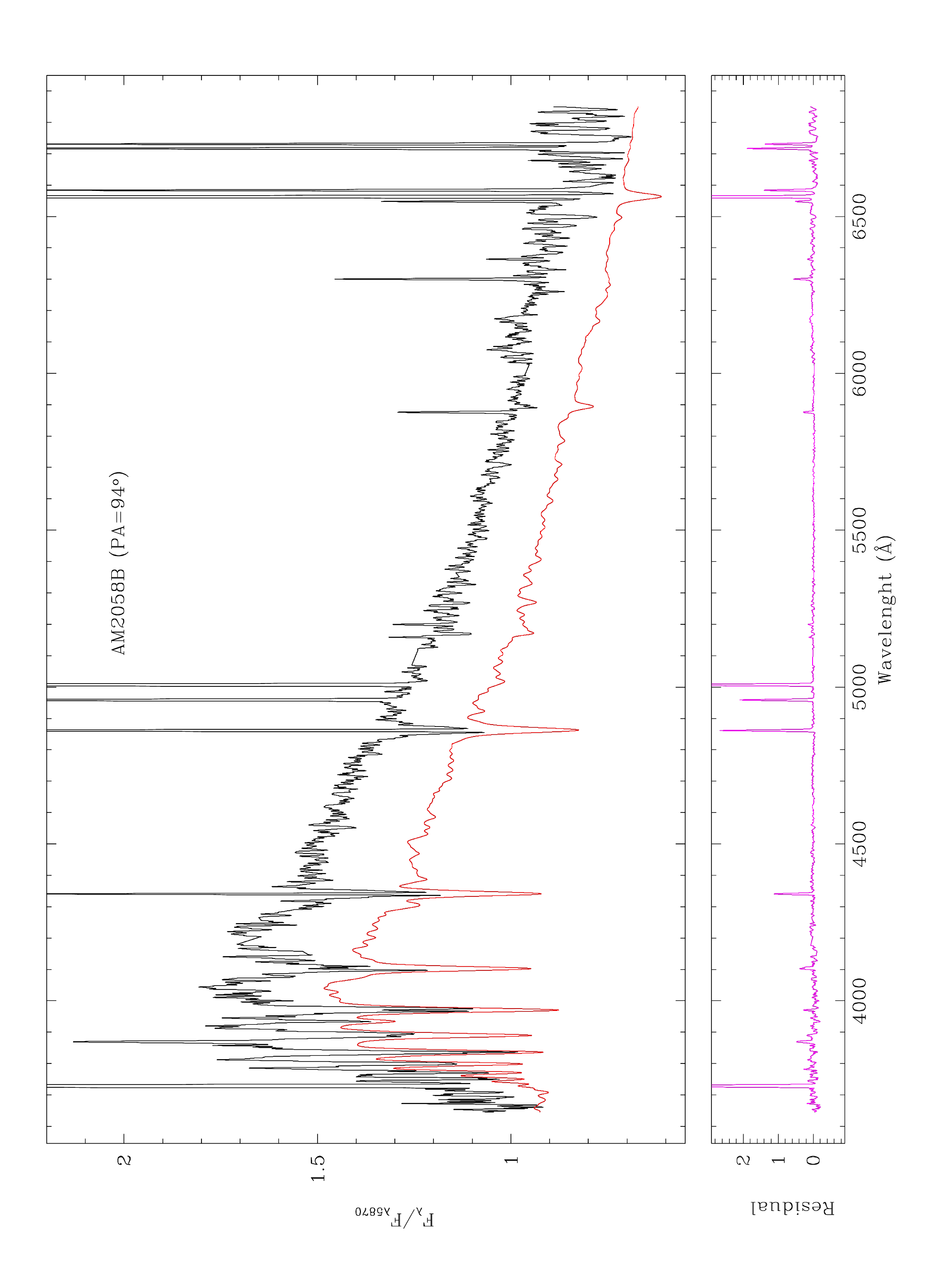}
\includegraphics*[angle=-90,width=0.45\columnwidth]{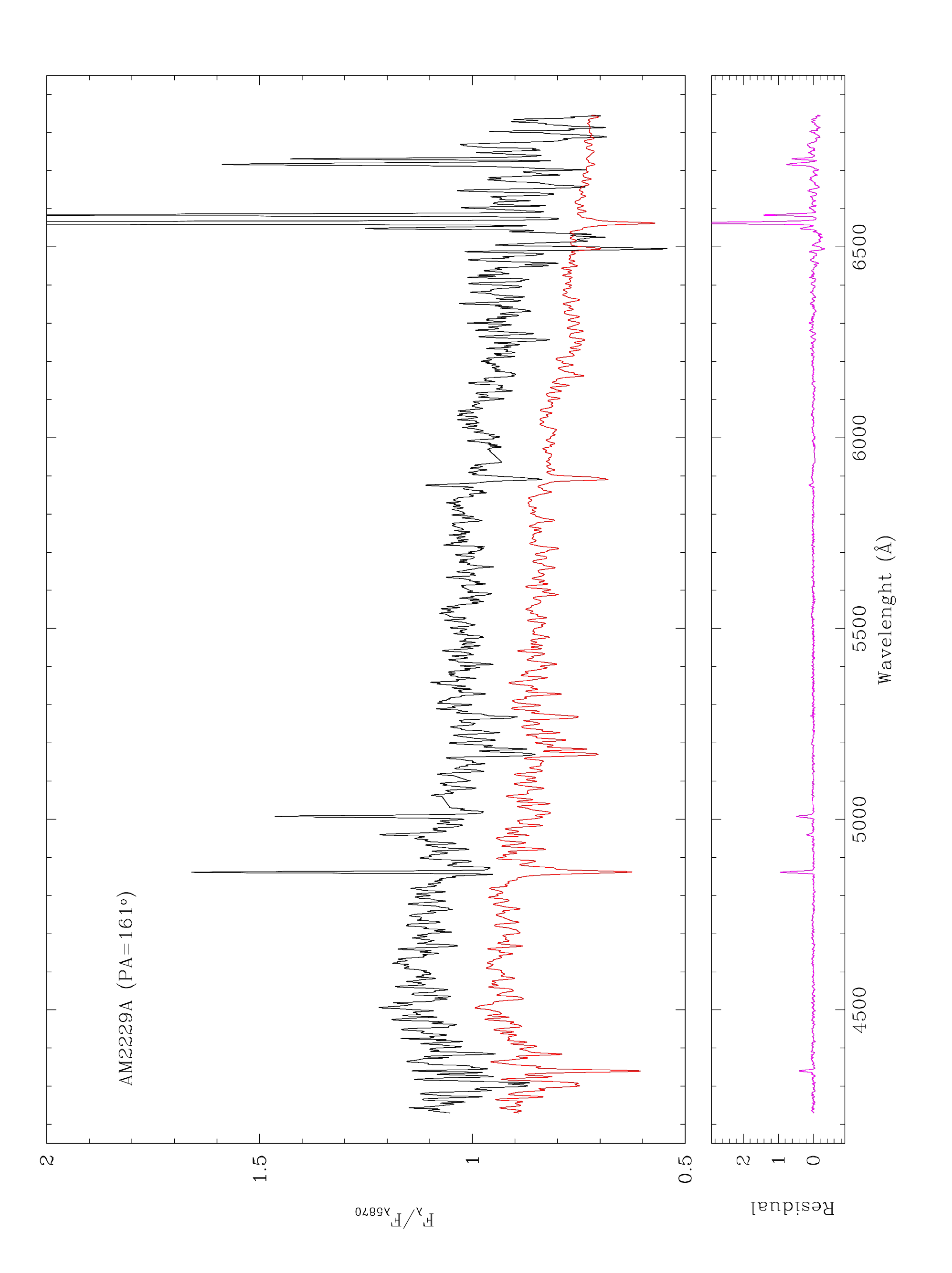}
\includegraphics*[angle=-90,width=0.45\columnwidth]{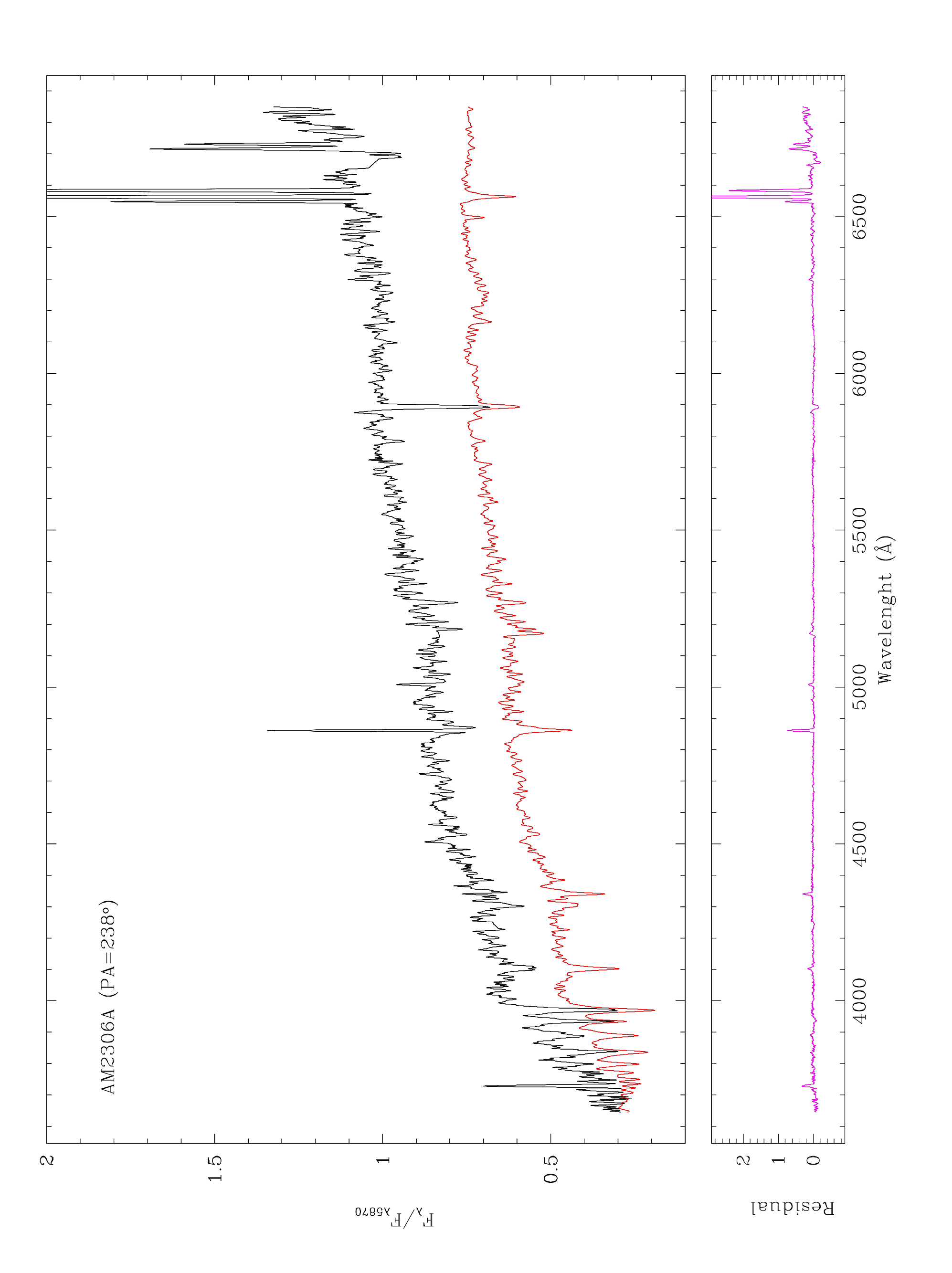}
\includegraphics*[angle=-90,width=0.45\columnwidth]{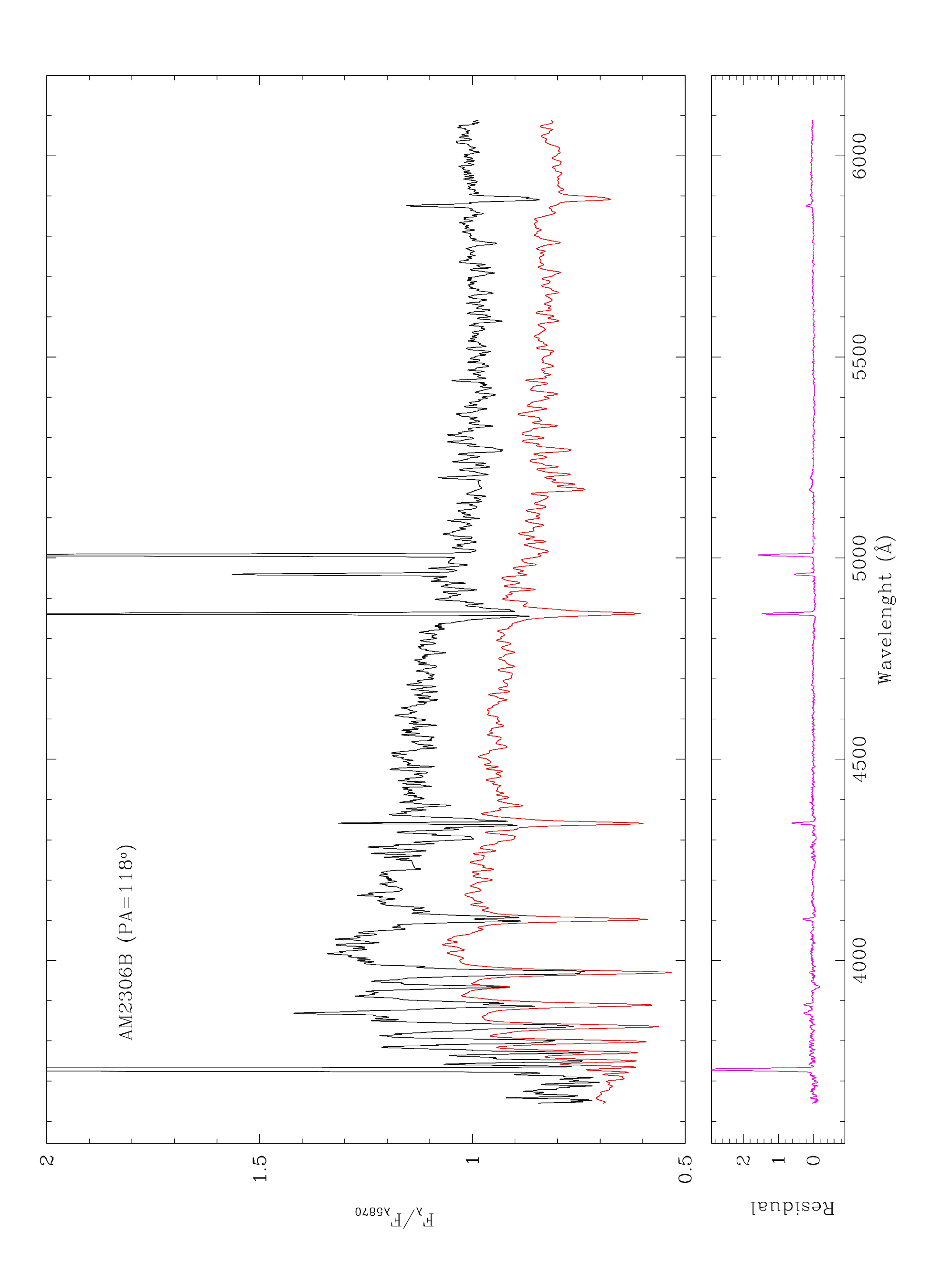}
\includegraphics*[angle=-90,width=0.45\columnwidth]{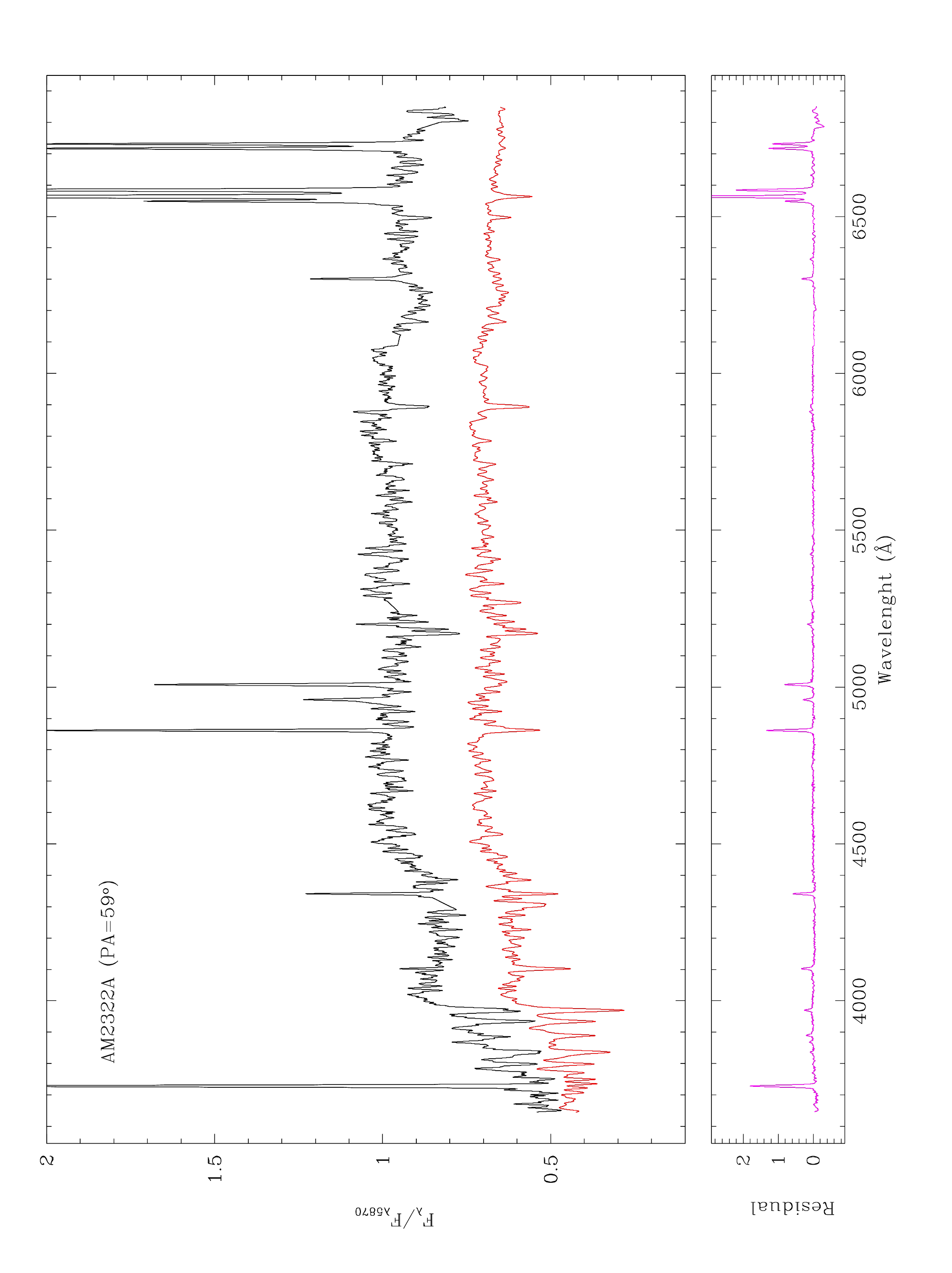}
\includegraphics*[angle=-90,width=0.45\columnwidth]{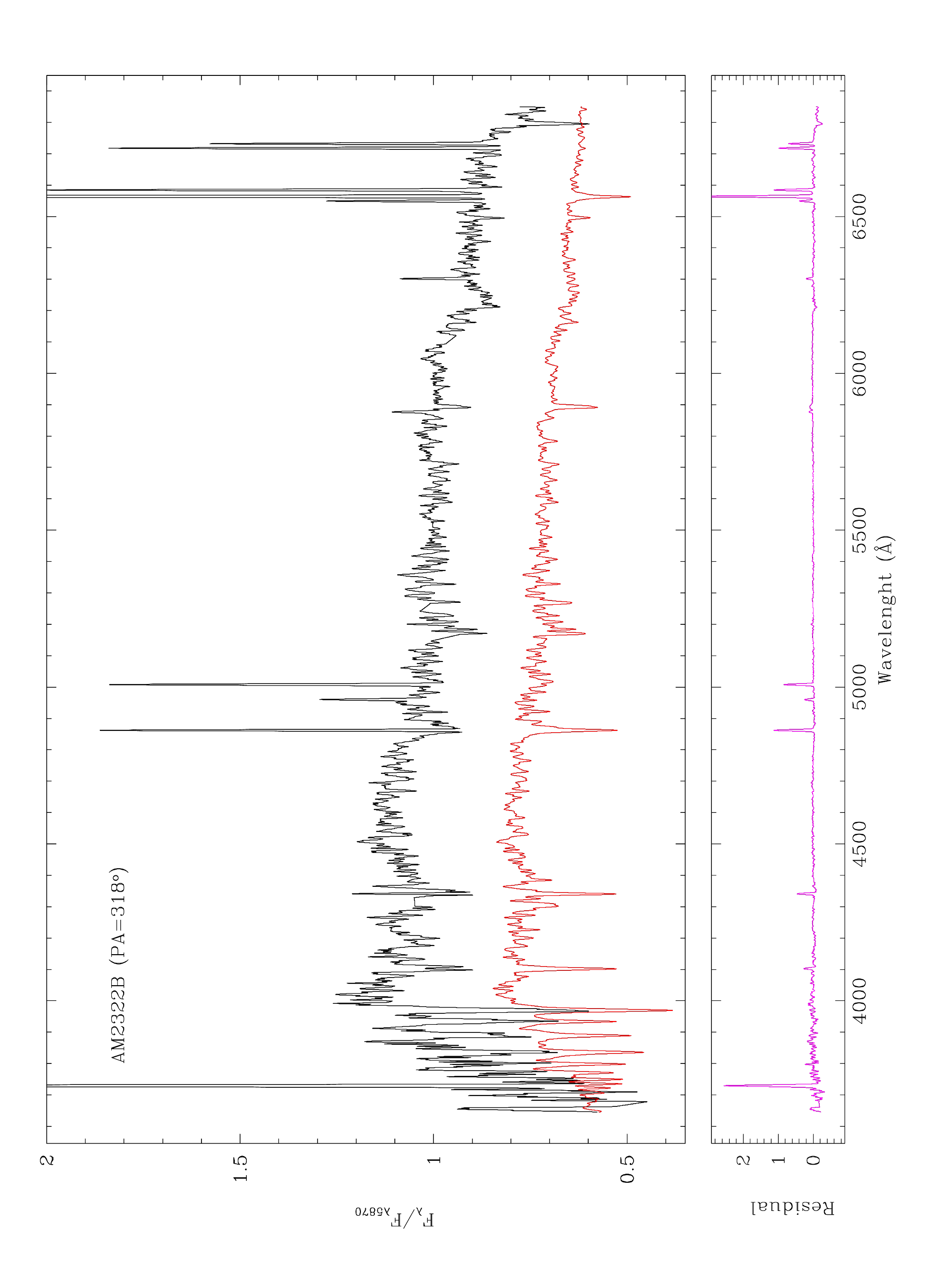}
\caption{The same as Fig.~\ref{sintese_002}, but for AM\,2058B, AM\,2229A, AM\,2306A, AM\,2306B, AM\,2322A and AM\,2322B.}
\label{sintese_004}
\end{figure*}

%\newpage
%\section{Spatial profile of the stellar population components}
%\subsection{Comparative contributions of the populations with different ages}
\label{ap_synth2}

\begin{figure*}
 \includegraphics*[angle=0,width=0.45\columnwidth]{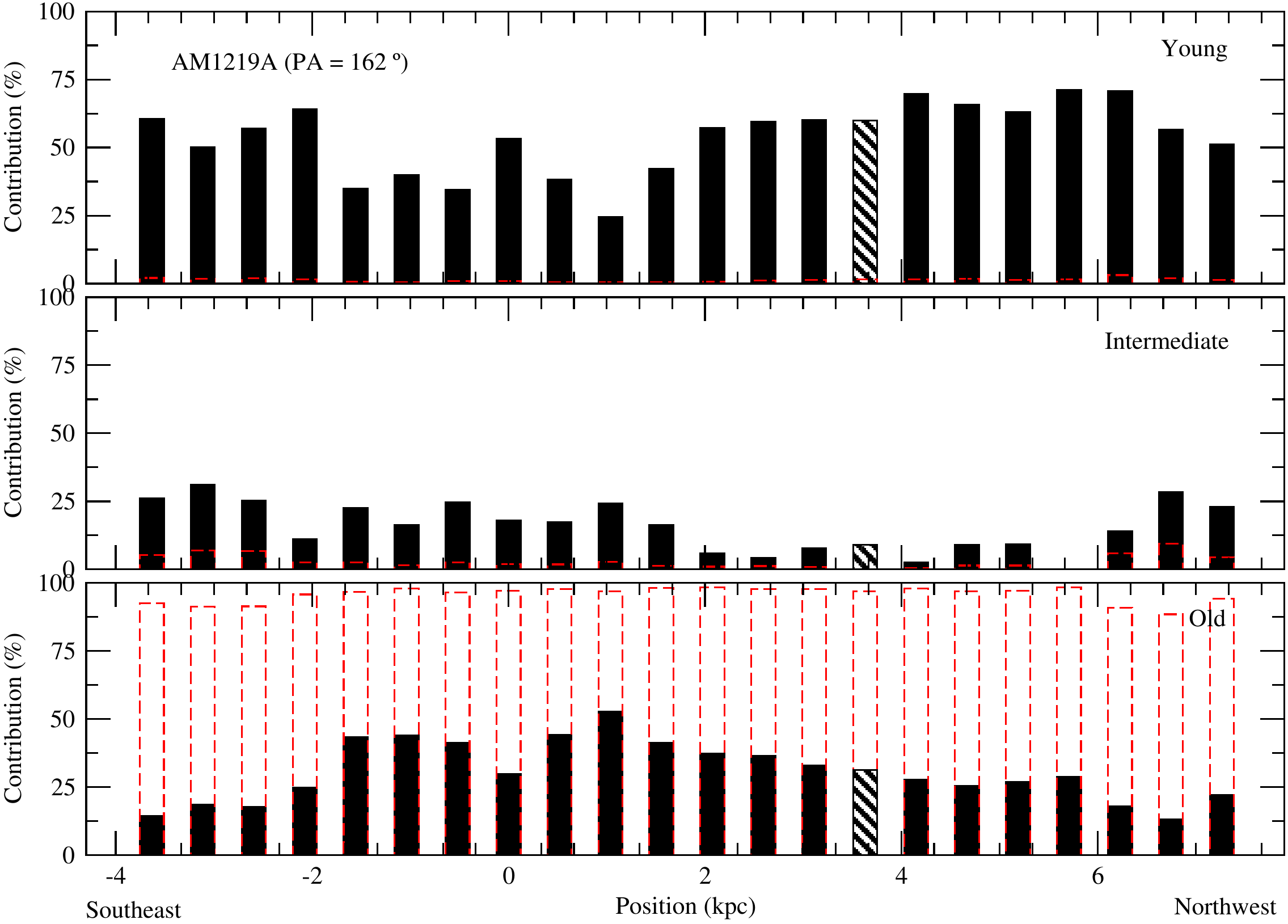} 
 \includegraphics*[angle=0,width=0.45\columnwidth]{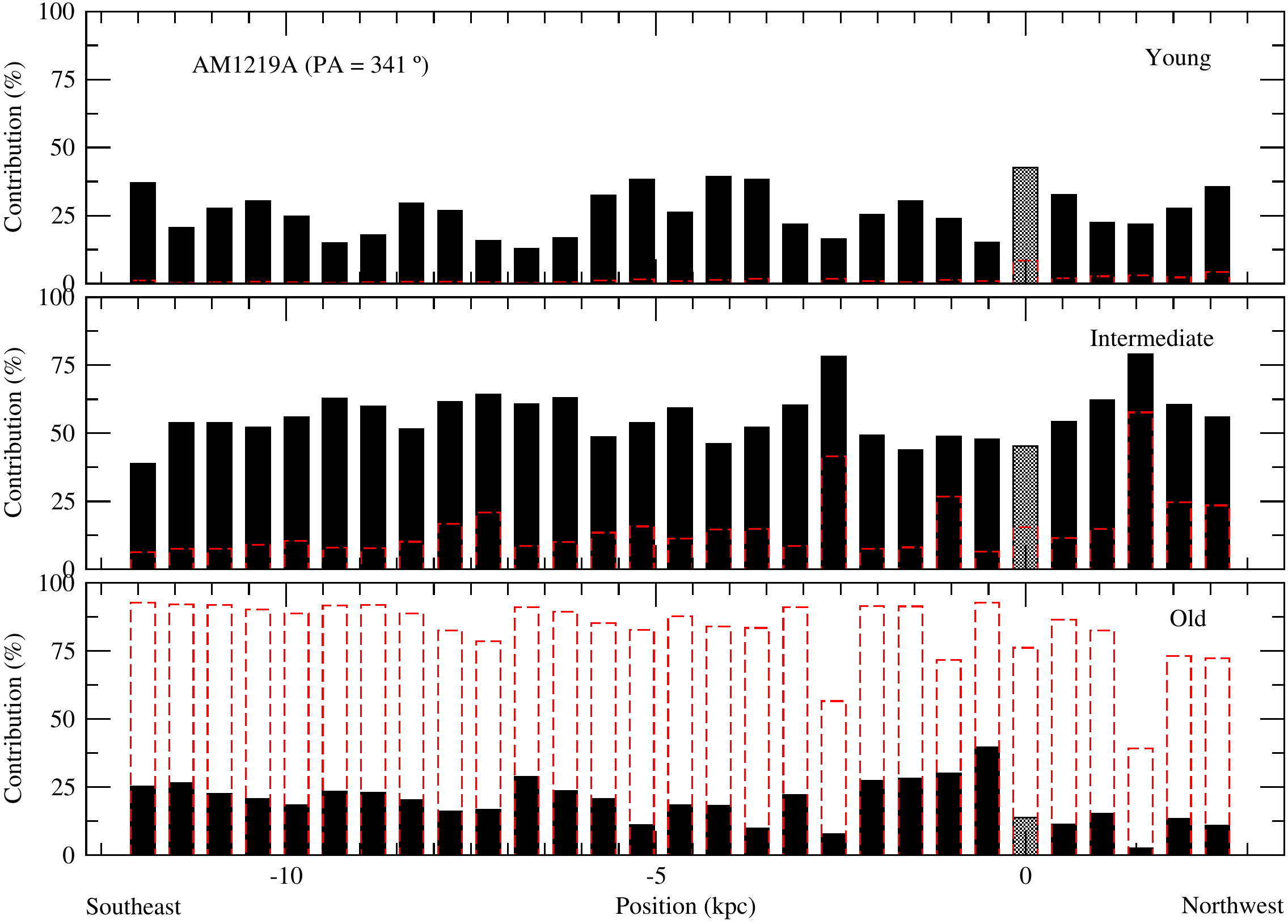}
 \includegraphics*[angle=0,width=0.45\columnwidth]{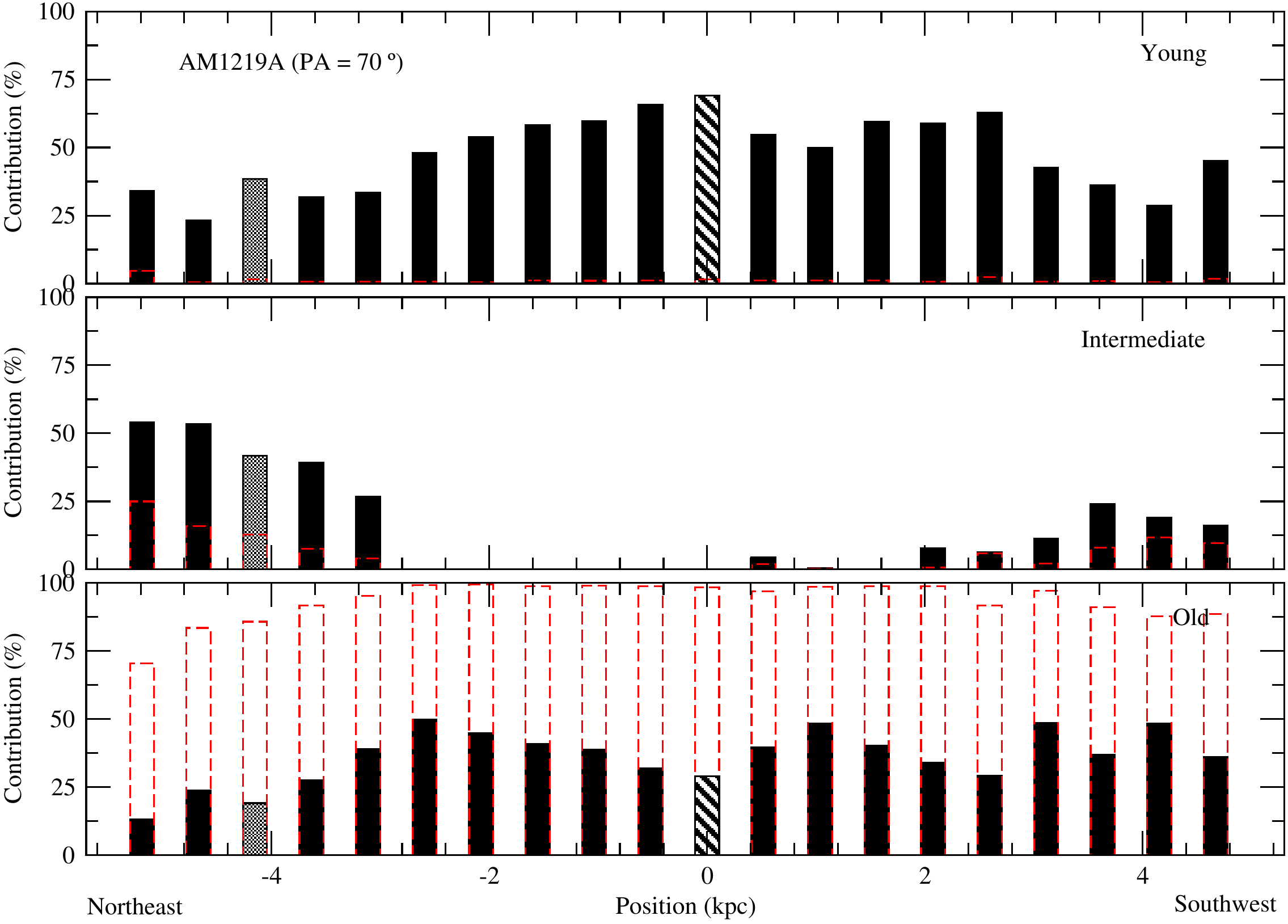}
 \includegraphics*[angle=0,width=0.45\columnwidth]{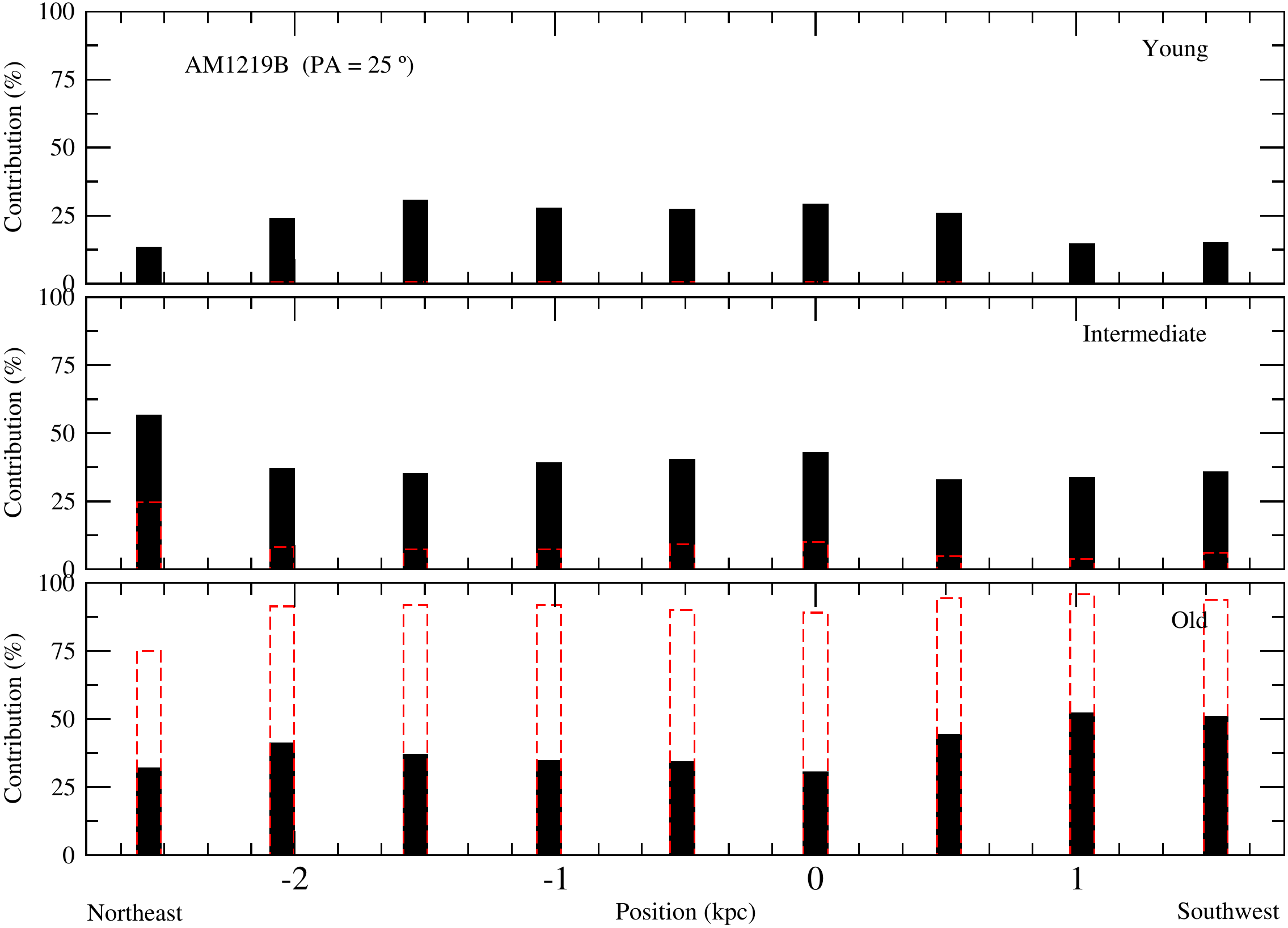}
\caption{Same as Fig.~\ref{sin_1054AB}, but for AM\,1219A along PA\,=\,162\degr, 341\degr, 70\degr and PA\,=\,25\degr for companion AM\,1219B.}
\label{sin_1219AB}
\end{figure*}

\begin{figure*} 
\includegraphics*[angle=0,width=0.45\columnwidth]{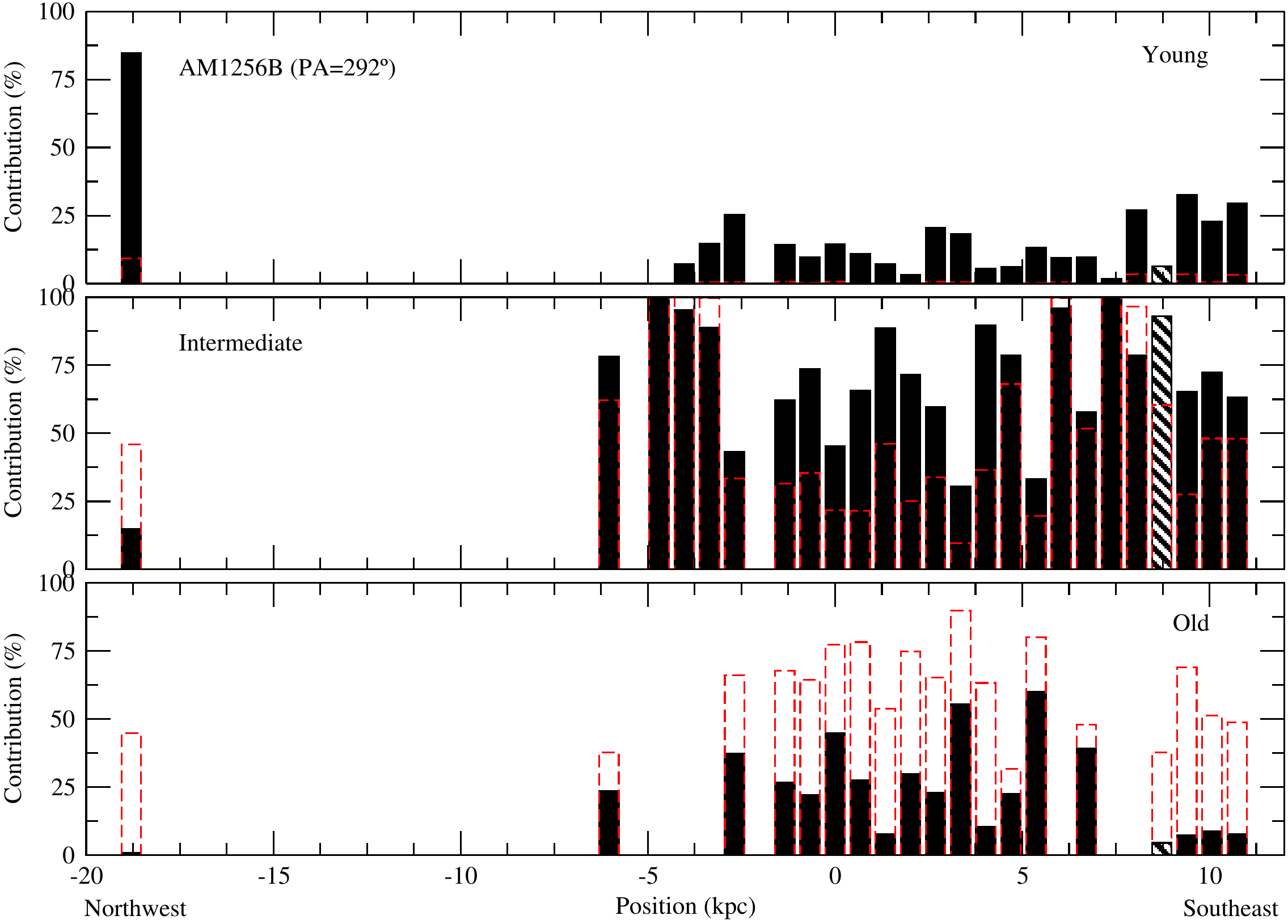}
 \includegraphics*[angle=0,width=0.45\columnwidth]{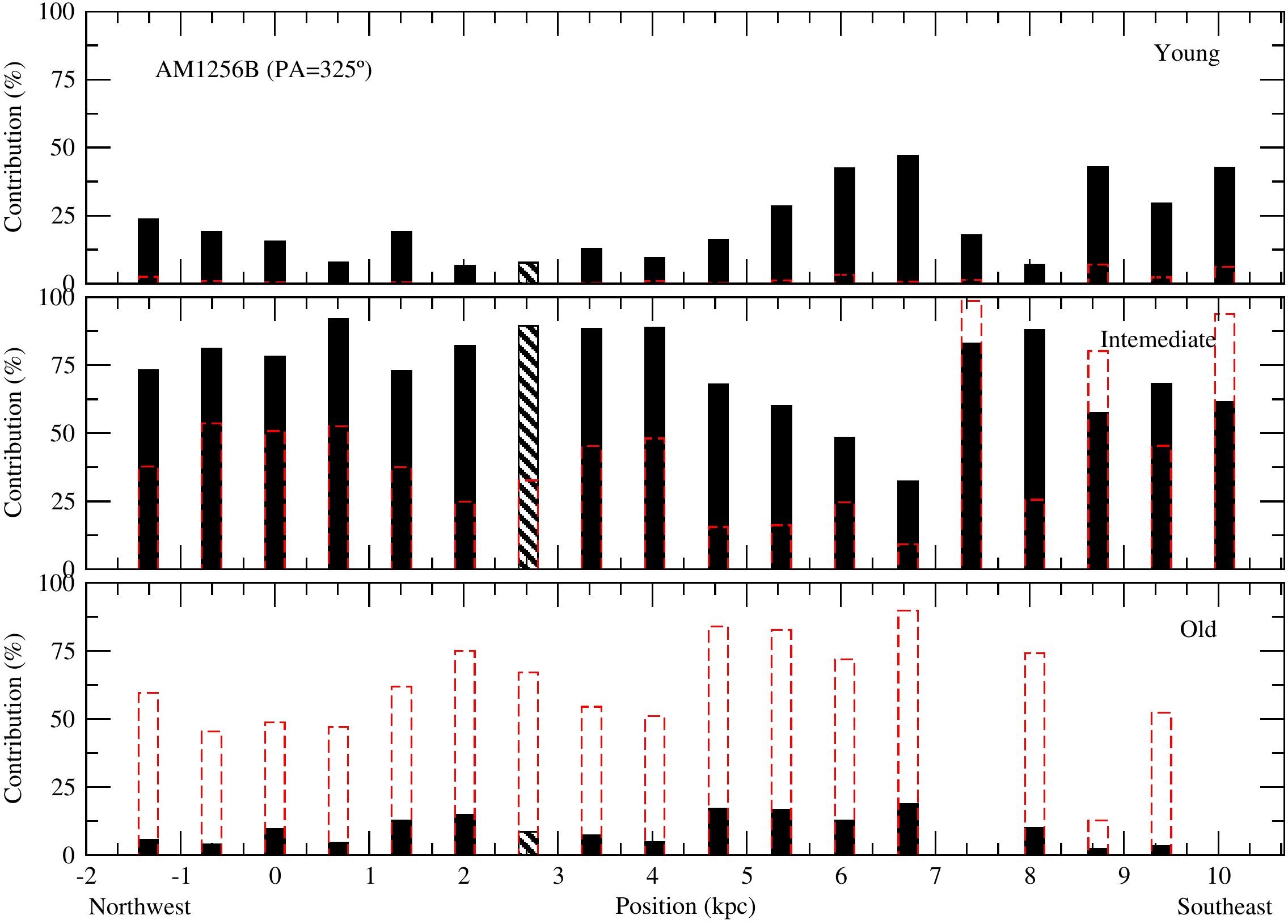}
\caption{Same as Fig.~\ref{sin_1054AB}, but for AM\,1256B along the two slit positions observed.}
\label{sin_1256B}
\end{figure*}

\begin{figure*} 
 \includegraphics*[angle=0,width=0.45\columnwidth]{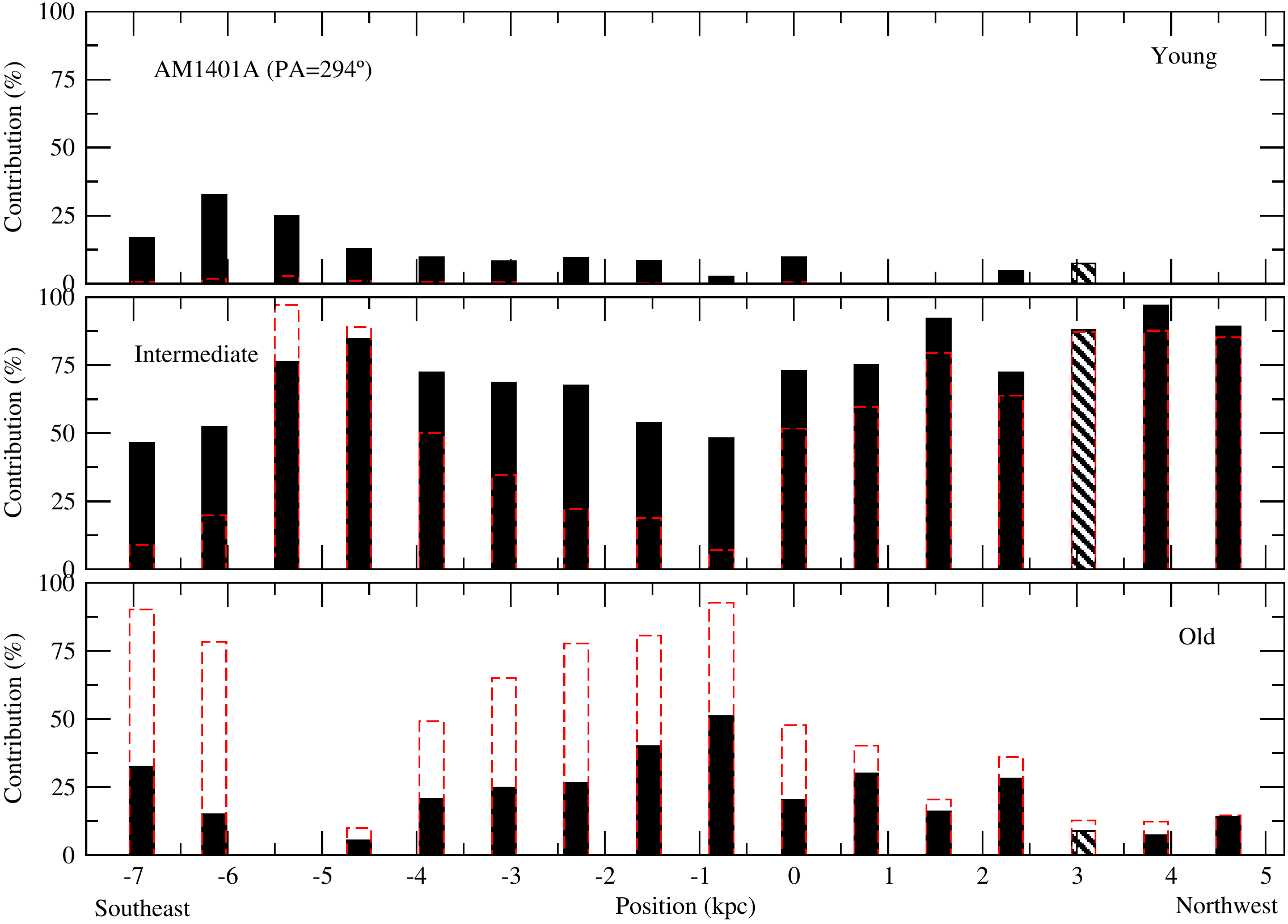} 
 \includegraphics*[angle=0,width=0.45\columnwidth]{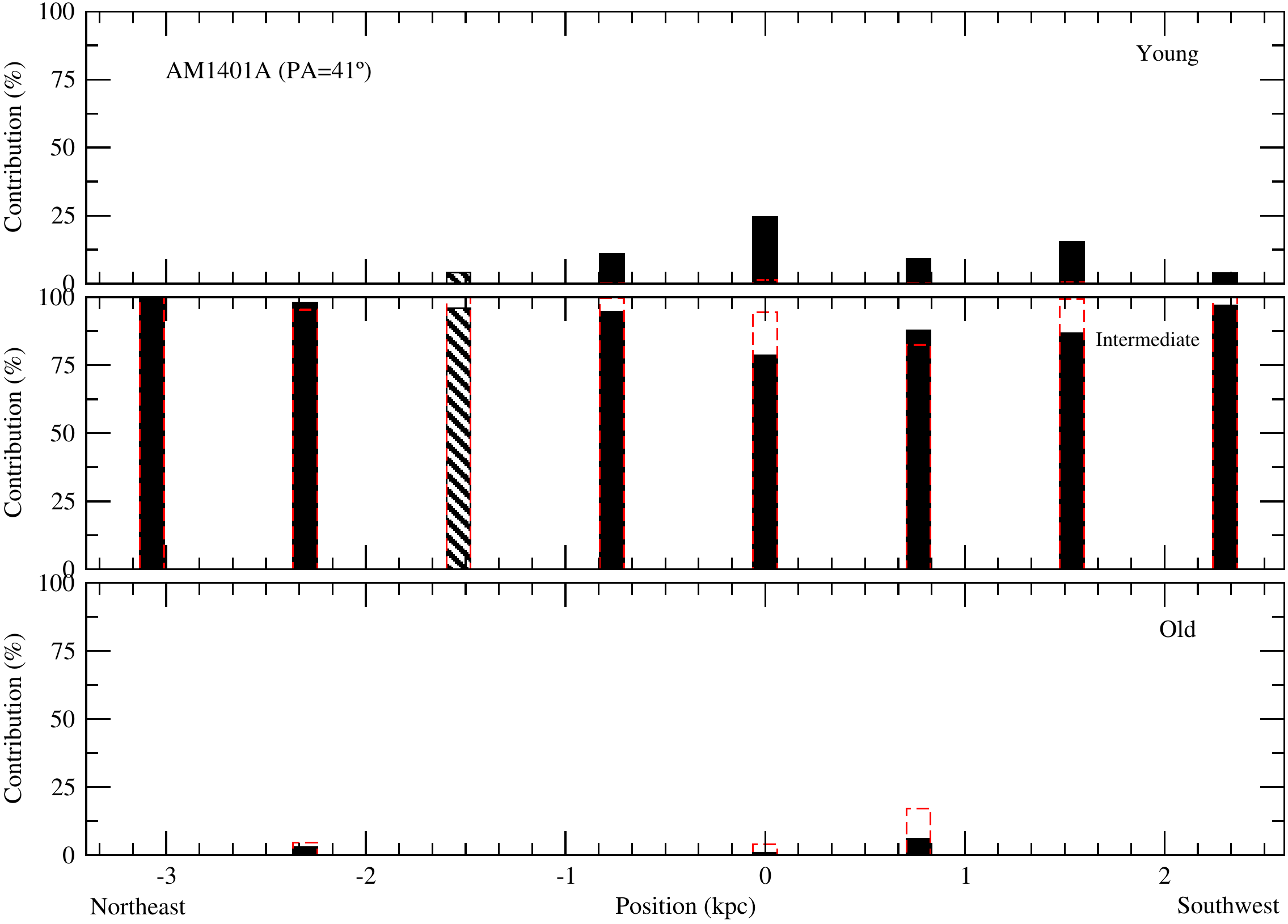}
\caption{Same as Fig.~\ref{sin_1054AB}, but for AM\,1401A along the two slit positions observed.}
\label{sin_1401A}
\end{figure*}

\begin{figure*}
\includegraphics*[angle=0,width=0.45\columnwidth]{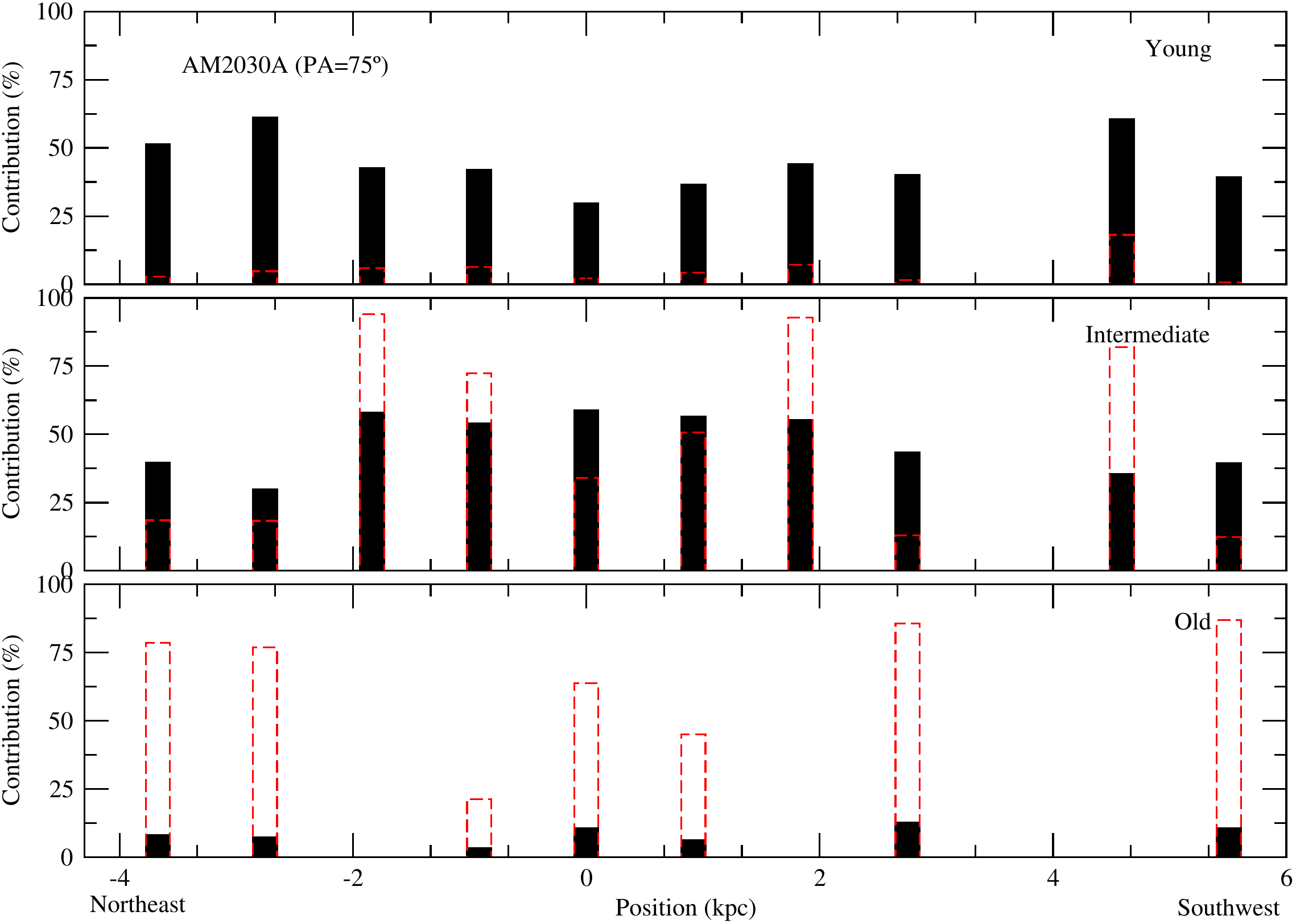}
\caption{Same as Fig.~\ref{sin_1054AB}, but for AM\,2030A.}
\label{sin_2030A}
\end{figure*}

\begin{figure*} 
\includegraphics*[angle=0,width=0.45\columnwidth]{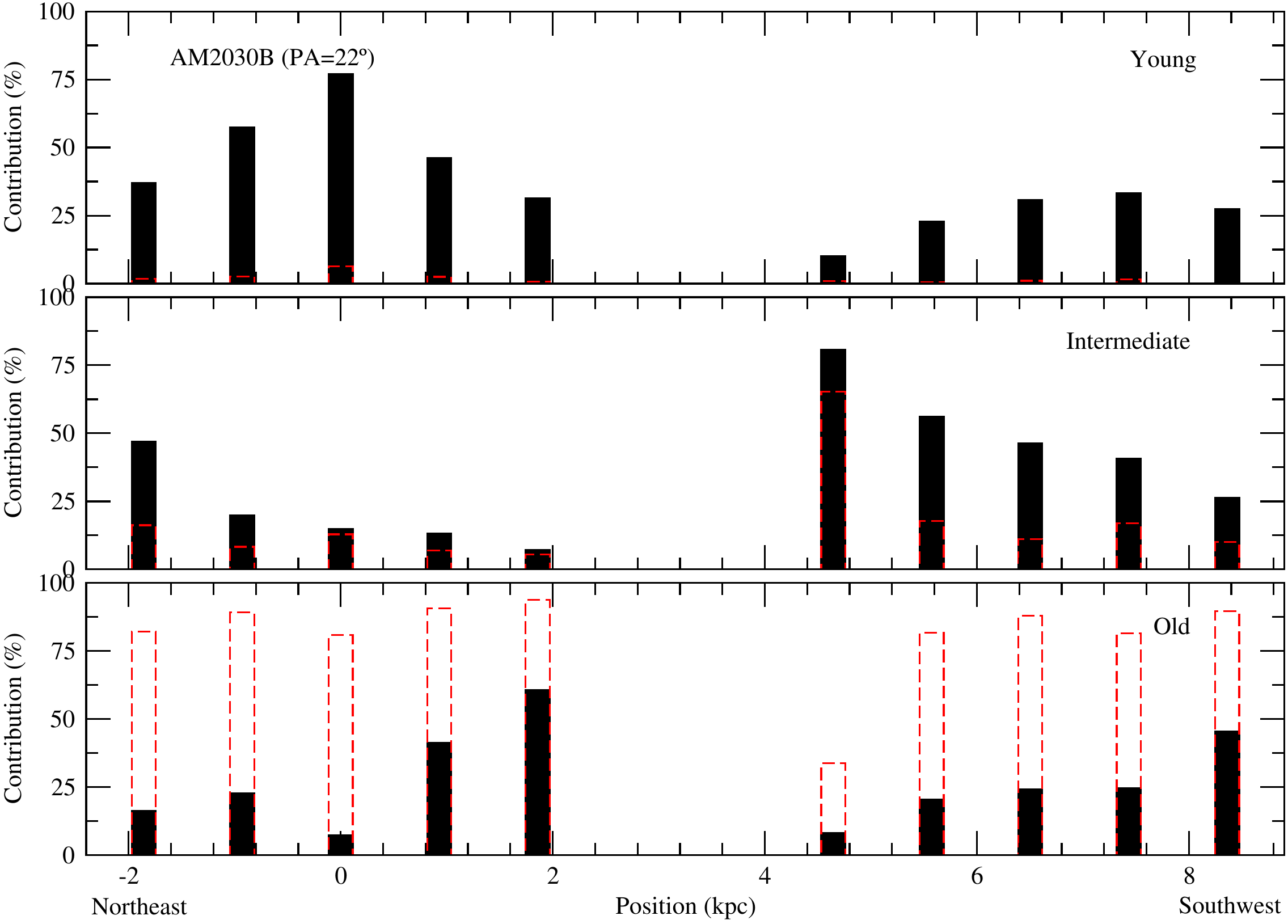} 
\includegraphics*[angle=0,width=0.45\columnwidth]{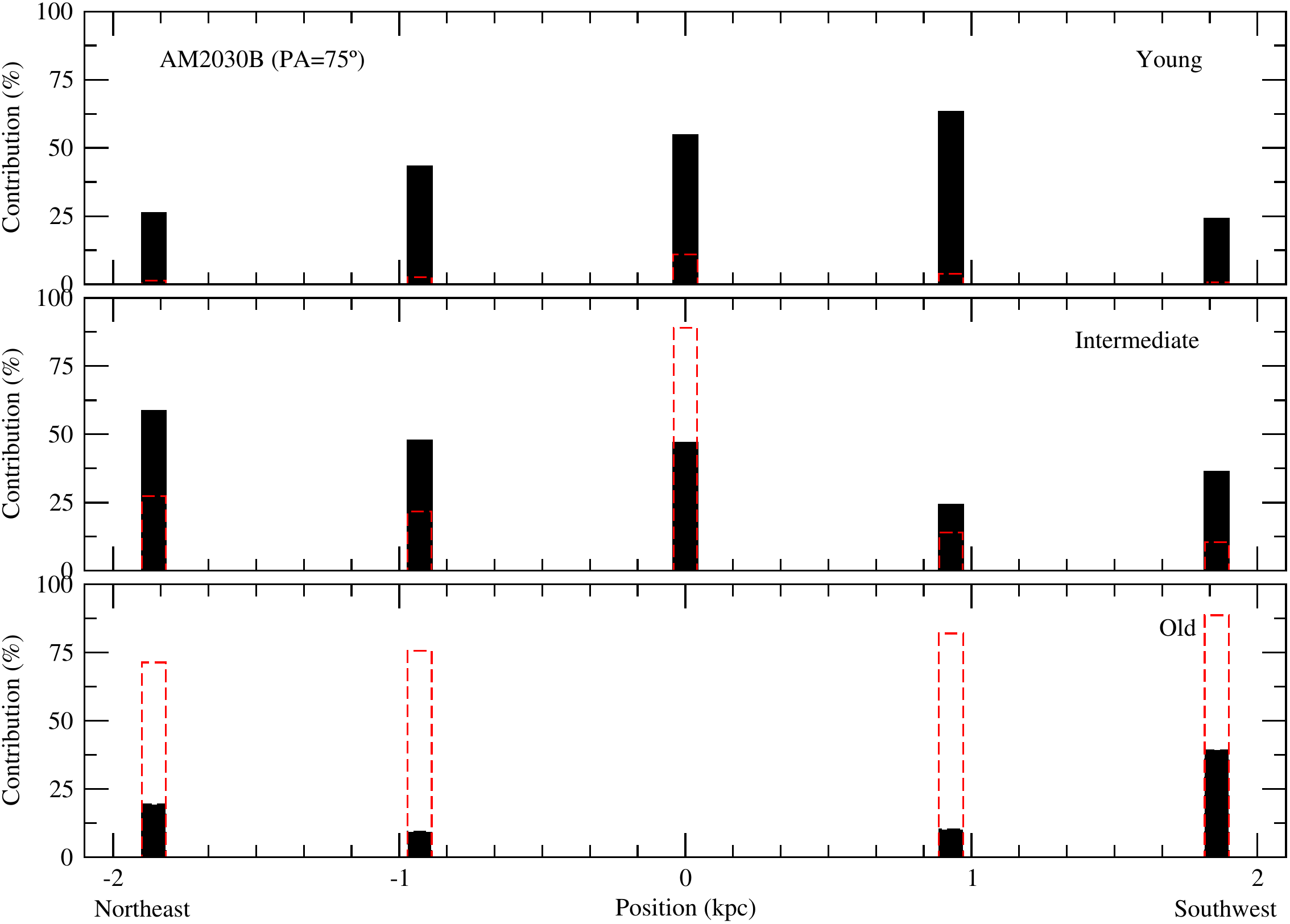}
\caption{Same as Fig.~\ref{sin_1054AB}, but for AM\,2030B along the two slit positions observed.}
\label{sin_2030B}
\end{figure*}

\begin{figure*} 
\includegraphics*[angle=0,width=0.45\columnwidth]{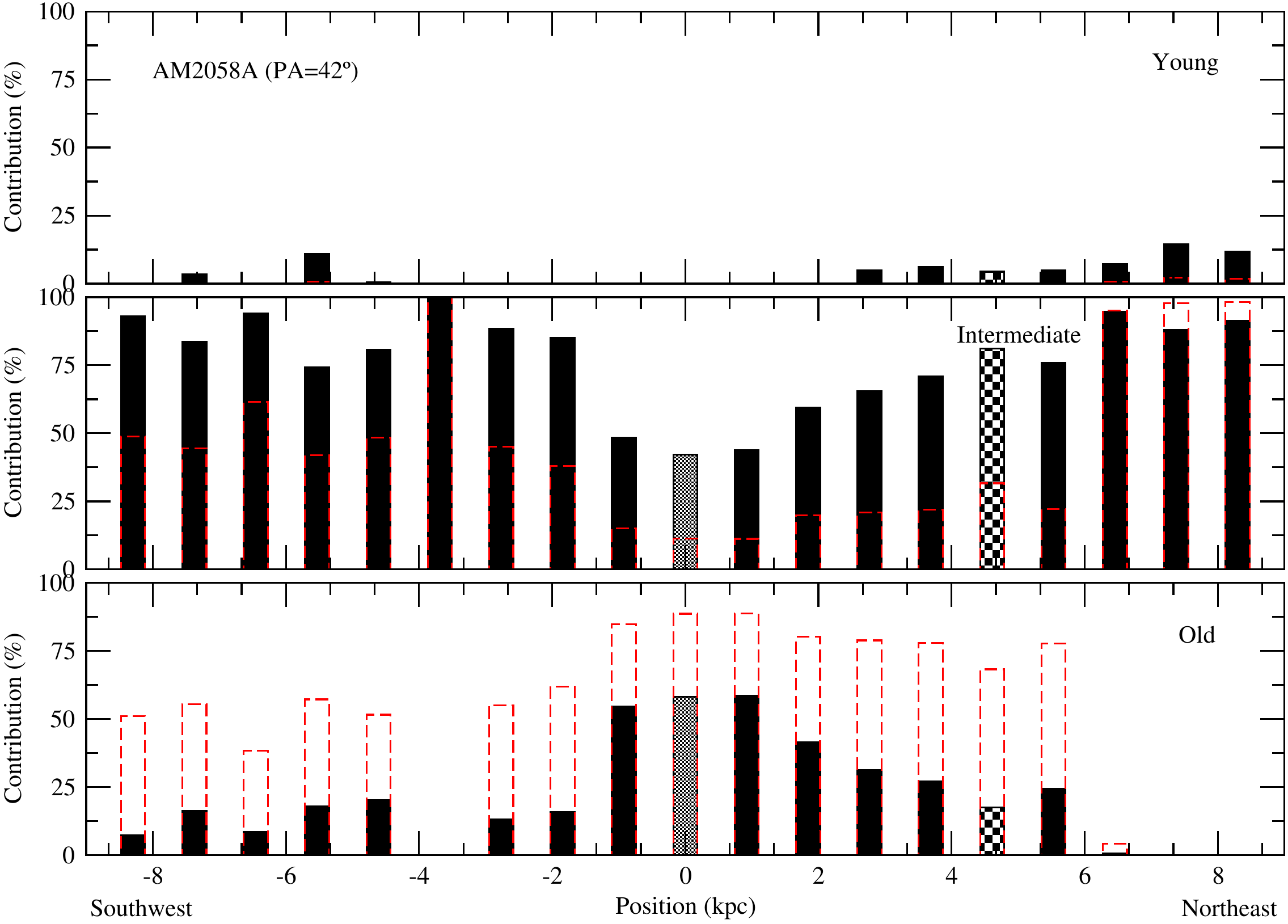}
\includegraphics*[angle=0,width=0.45\columnwidth]{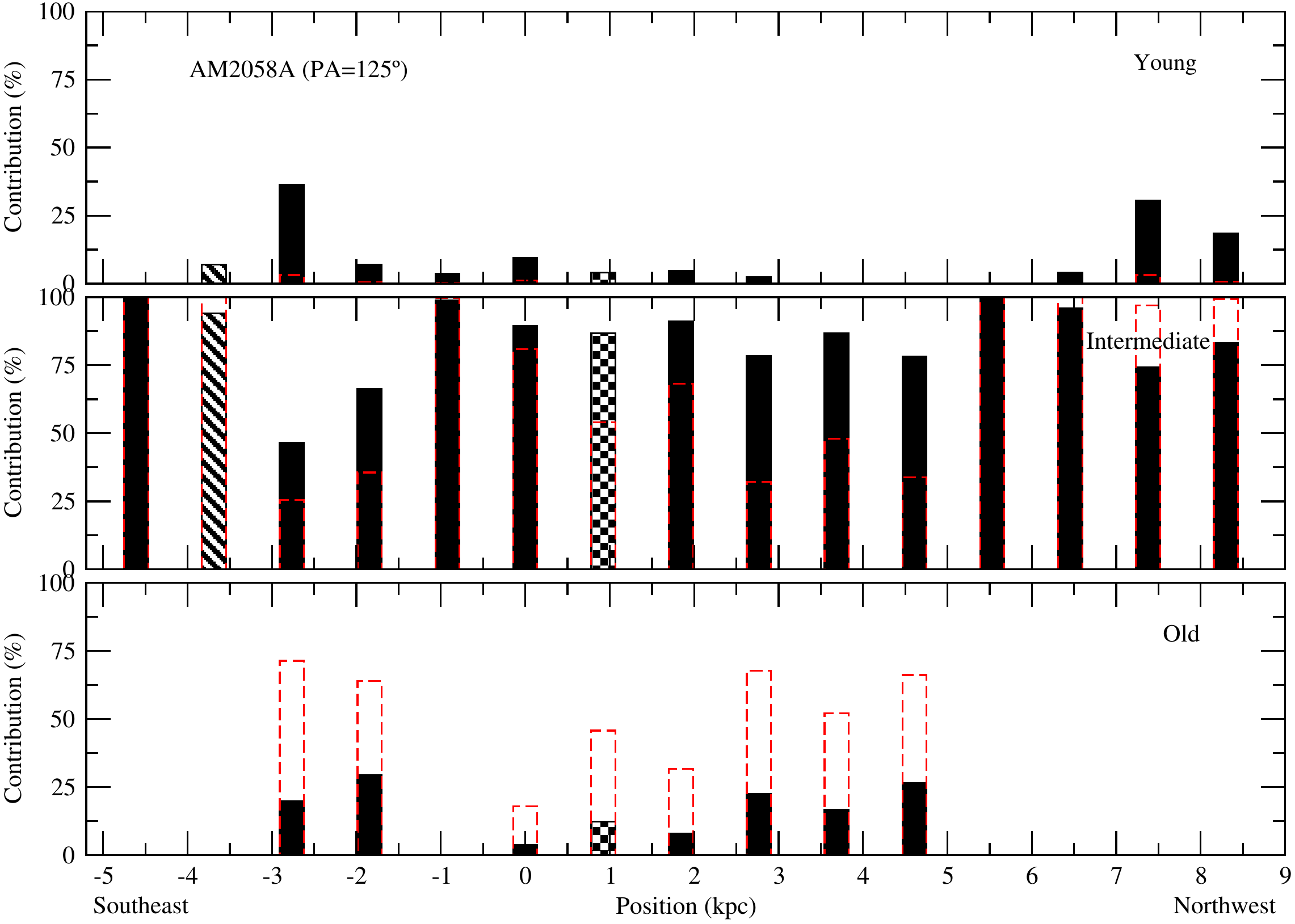}
\includegraphics*[angle=0,width=0.45\columnwidth]{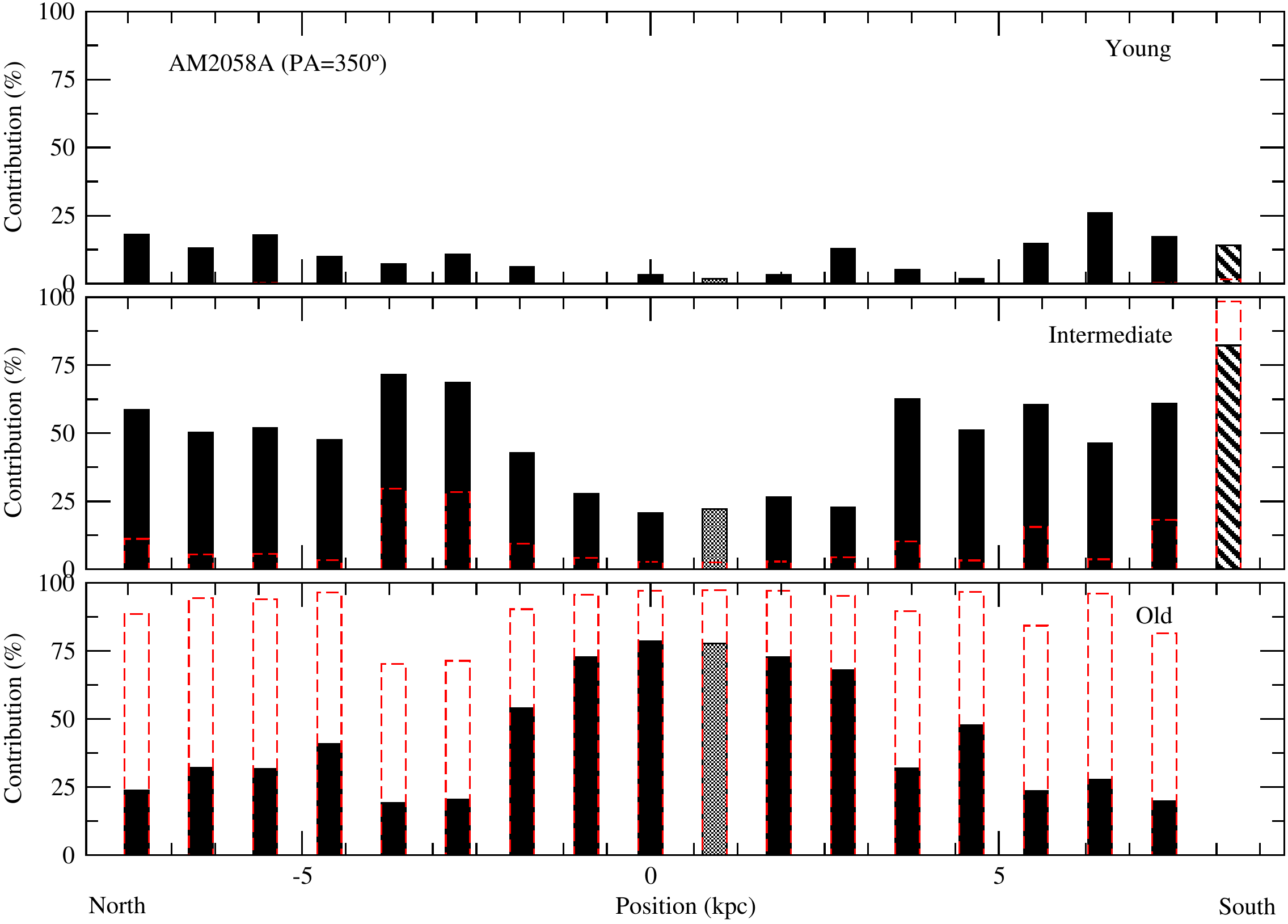}
\includegraphics*[angle=0,width=0.45\columnwidth]{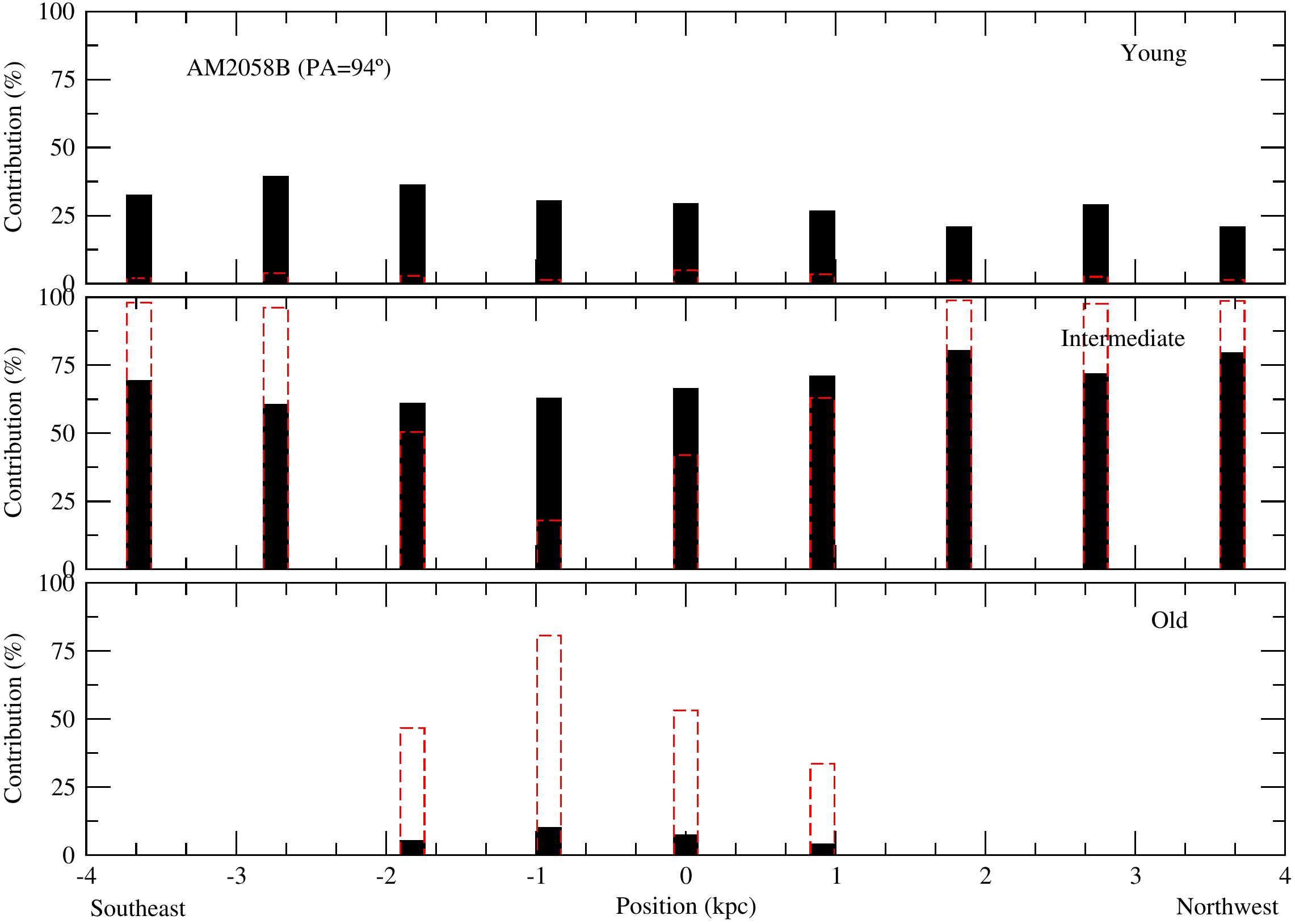}
\caption{Same as Fig.~\ref{sin_1054AB}, but for AM\,2058A and AM\,2058B (bottom right panel).}
\label{sin_2058A}
\end{figure*}

\begin{figure*} 
 \includegraphics*[angle=0,width=0.45\columnwidth]{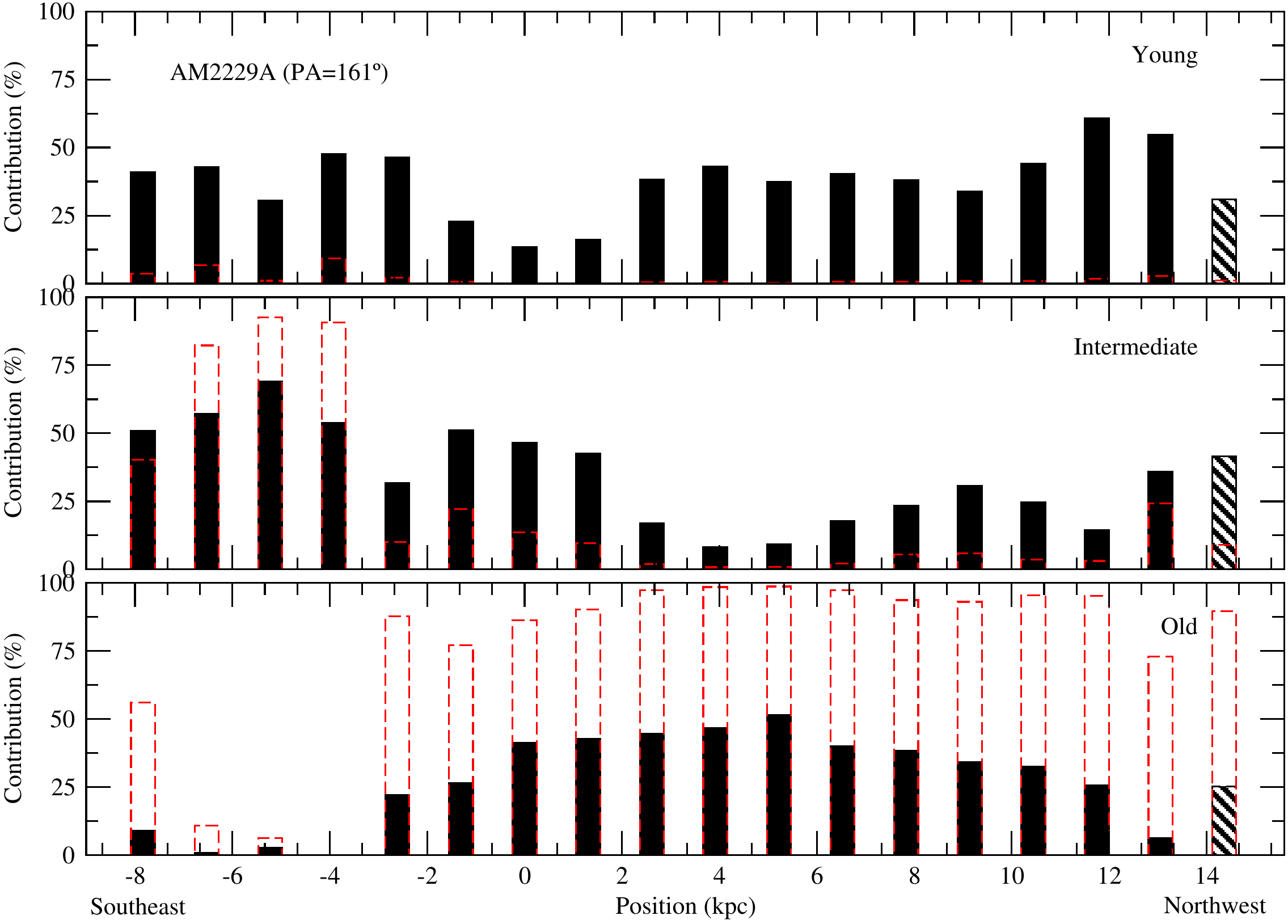}
 \includegraphics*[angle=0,width=0.45\columnwidth]{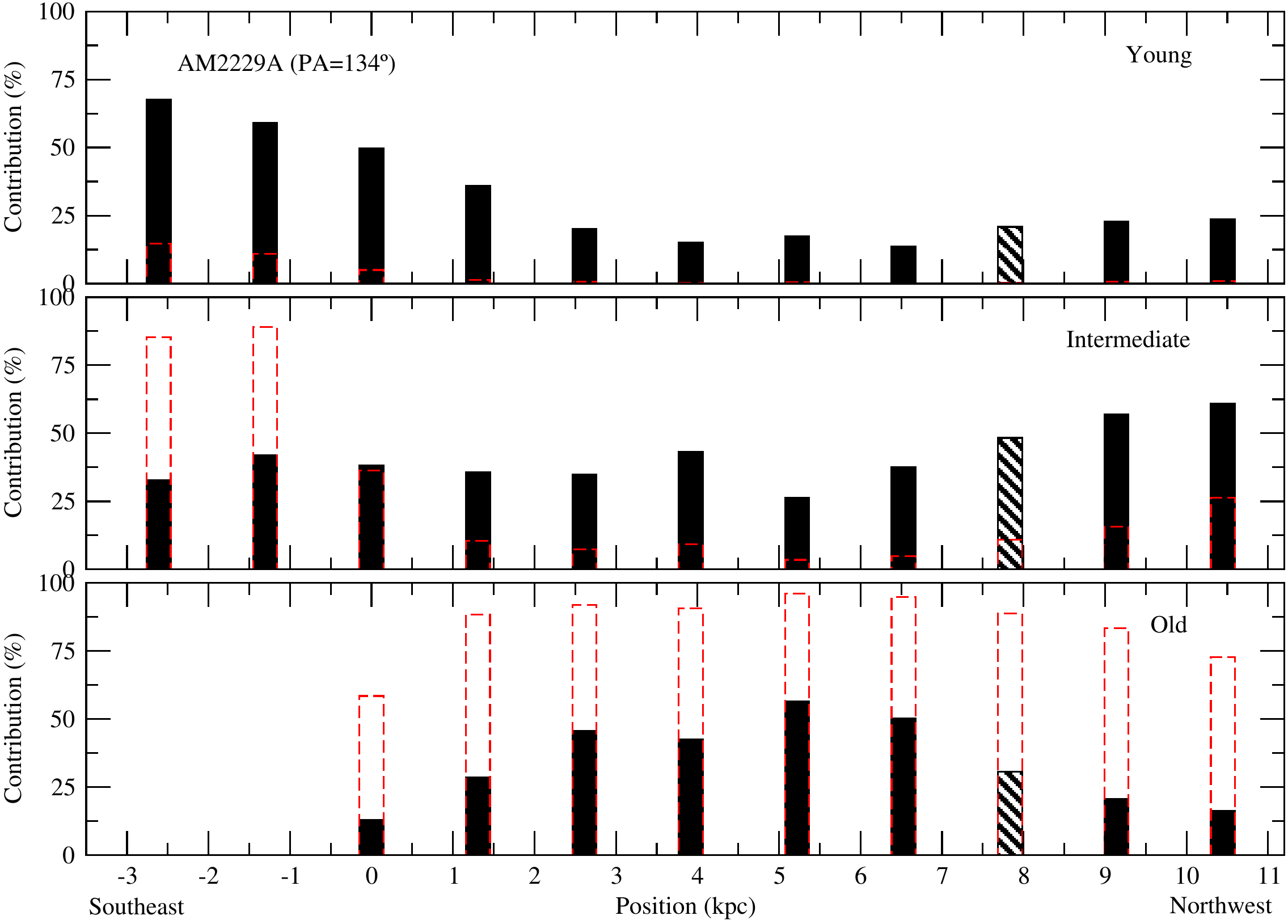}
\caption{Same as Fig.~\ref{sin_1054AB}, but for AM\,2229A along the two slit positions observed.}
\label{sin_2229A}
\end{figure*}

\begin{figure*}
\includegraphics*[angle=0,width=0.45\columnwidth]{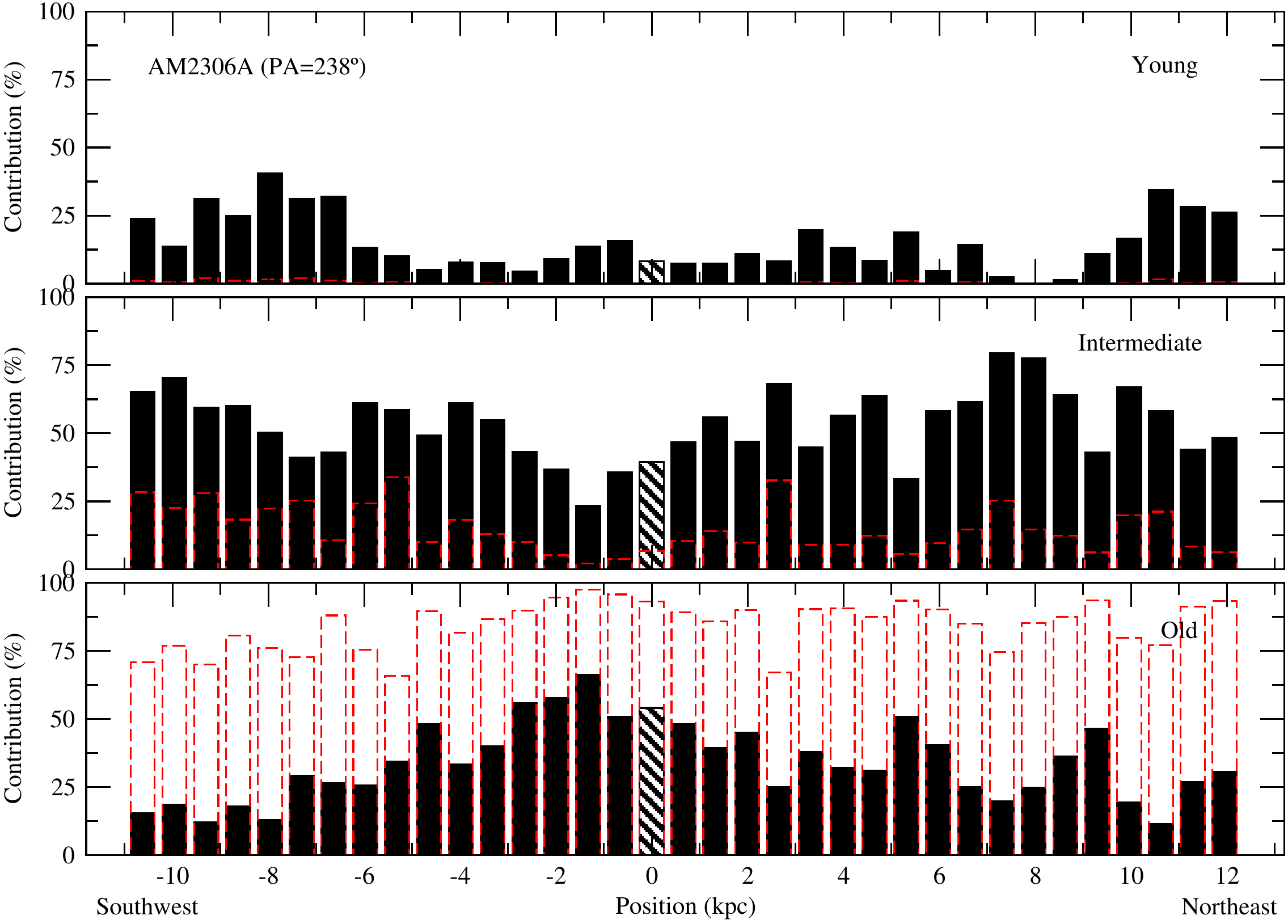}
\includegraphics*[angle=0,width=0.45\columnwidth]{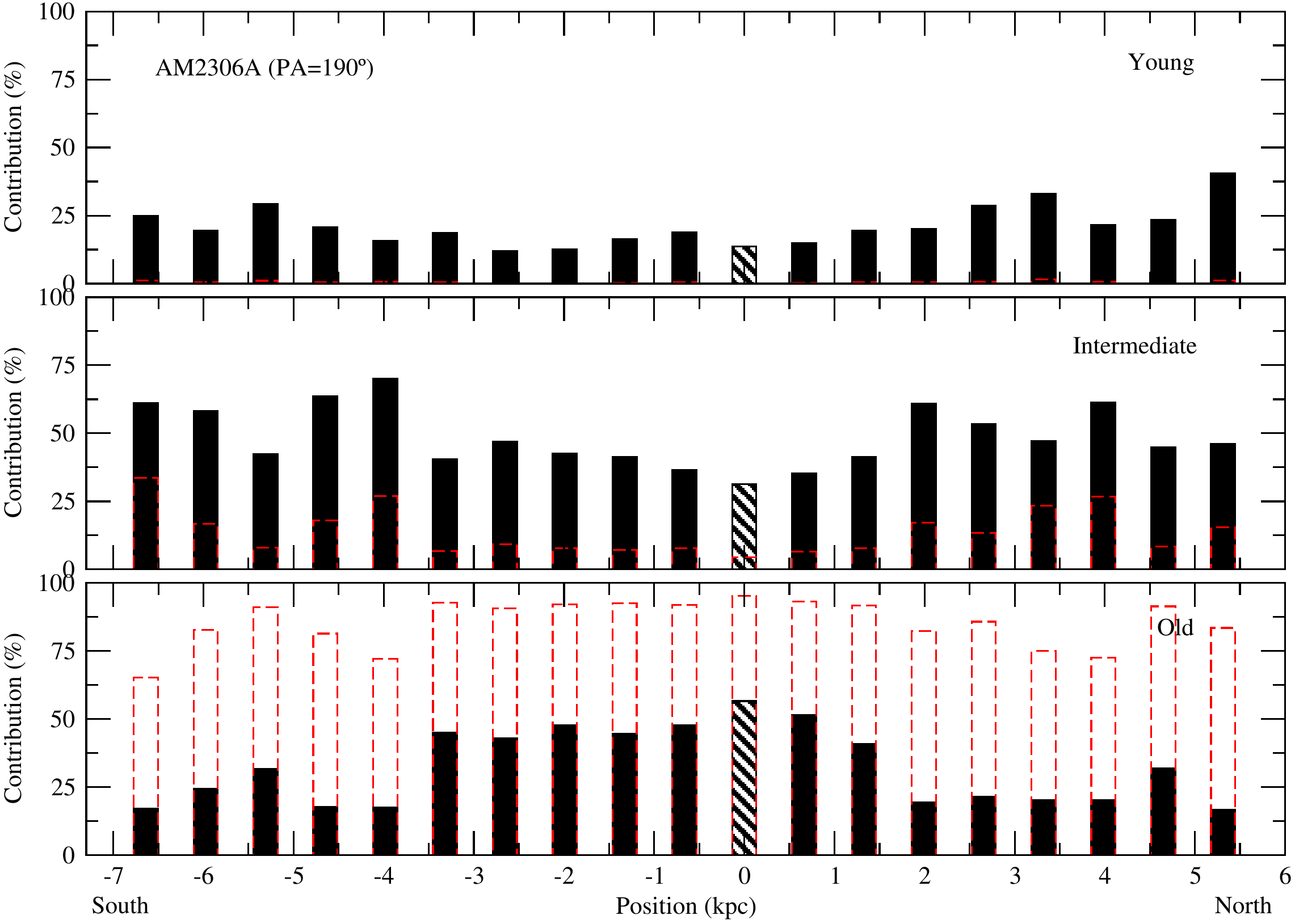}
\includegraphics*[angle=0,width=0.45\columnwidth]{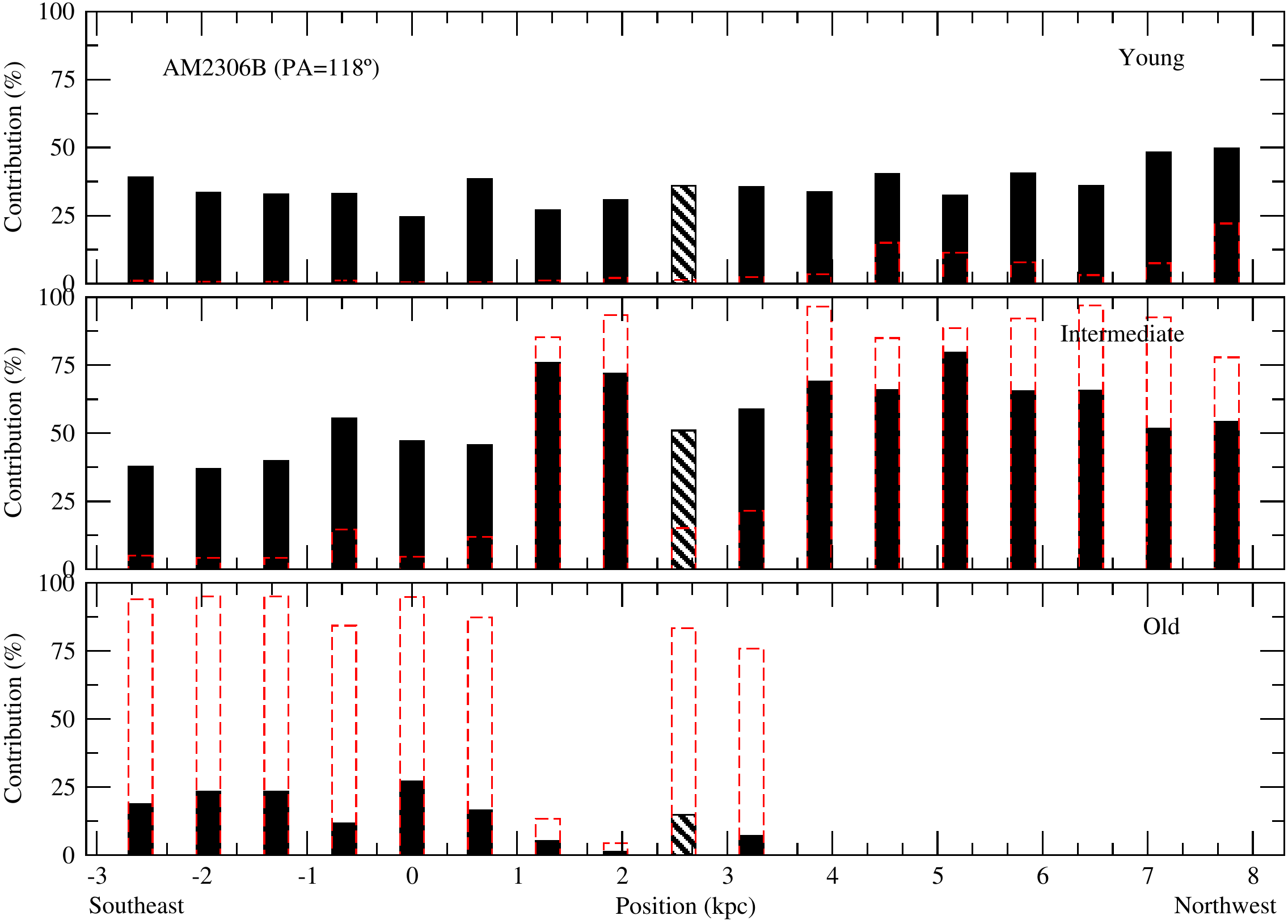}
\includegraphics*[angle=0,width=0.45\columnwidth]{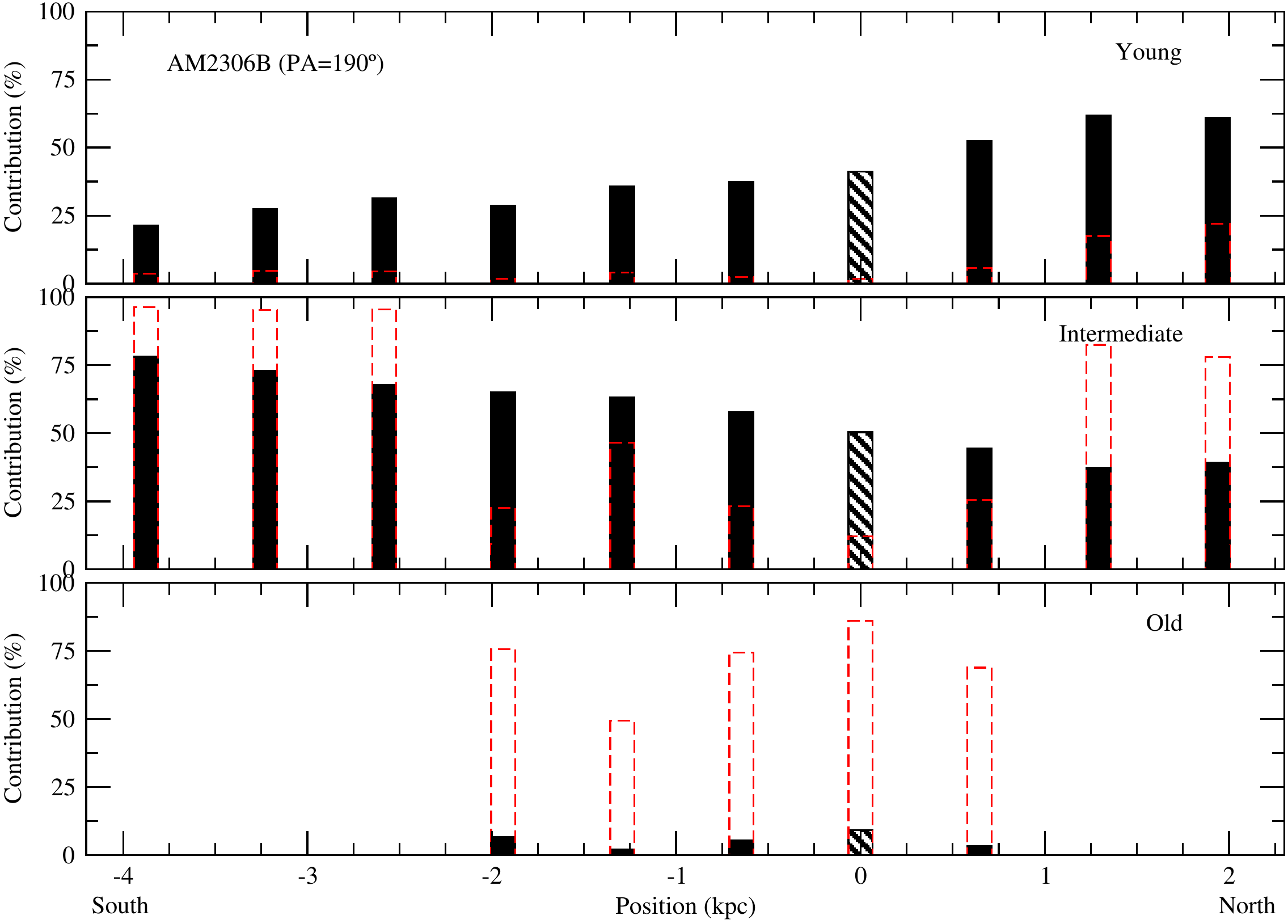}
\caption{Same as Fig.~\ref{sin_1054AB}, but for AM\,2306A (top panels) and AM\,2306B (bottom panels) along their corresponding slit positions.}
\label{sin_2306A}
\end{figure*}

\begin{figure*} 
\includegraphics*[angle=0,width=0.45\columnwidth]{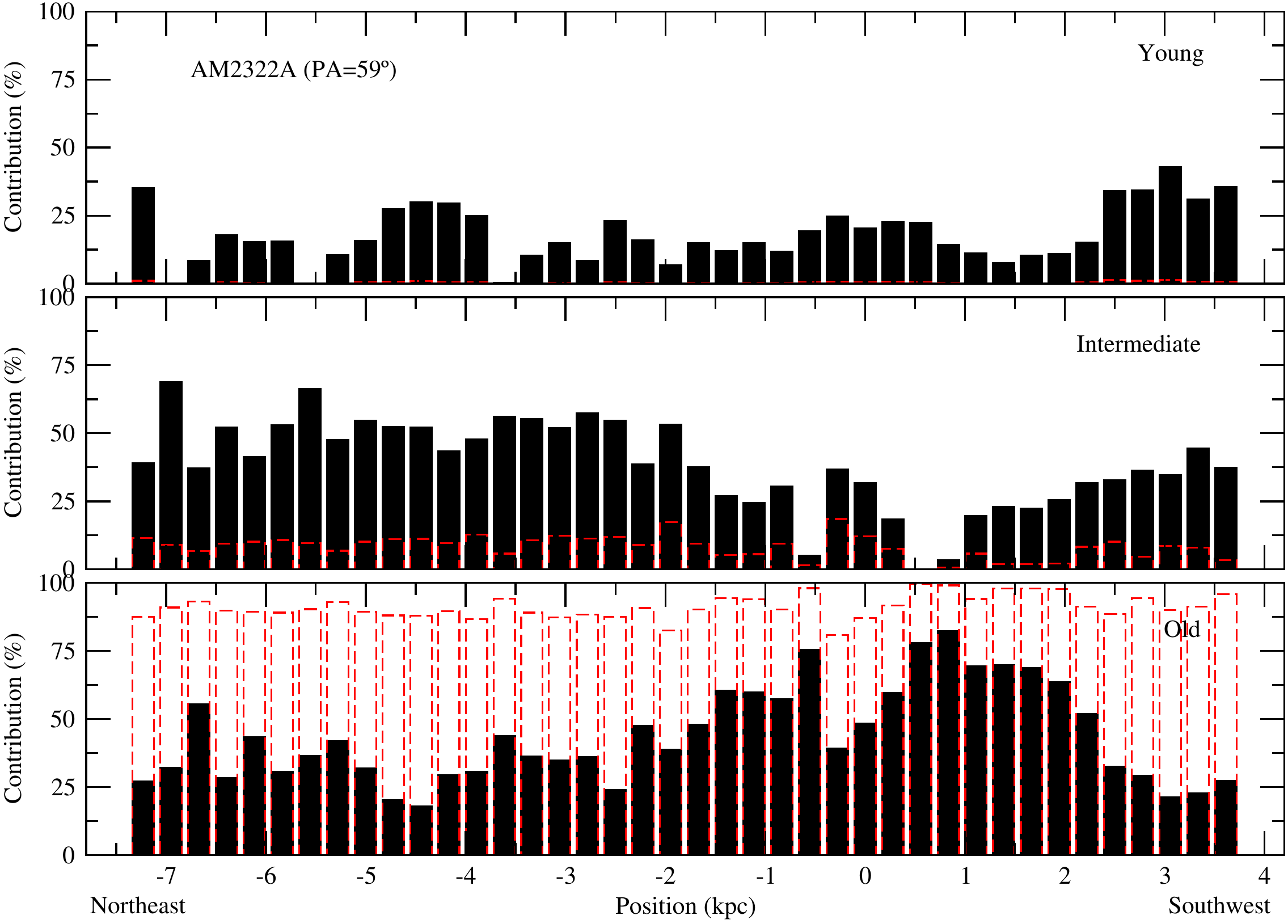}
\includegraphics*[angle=0,width=0.45\columnwidth]{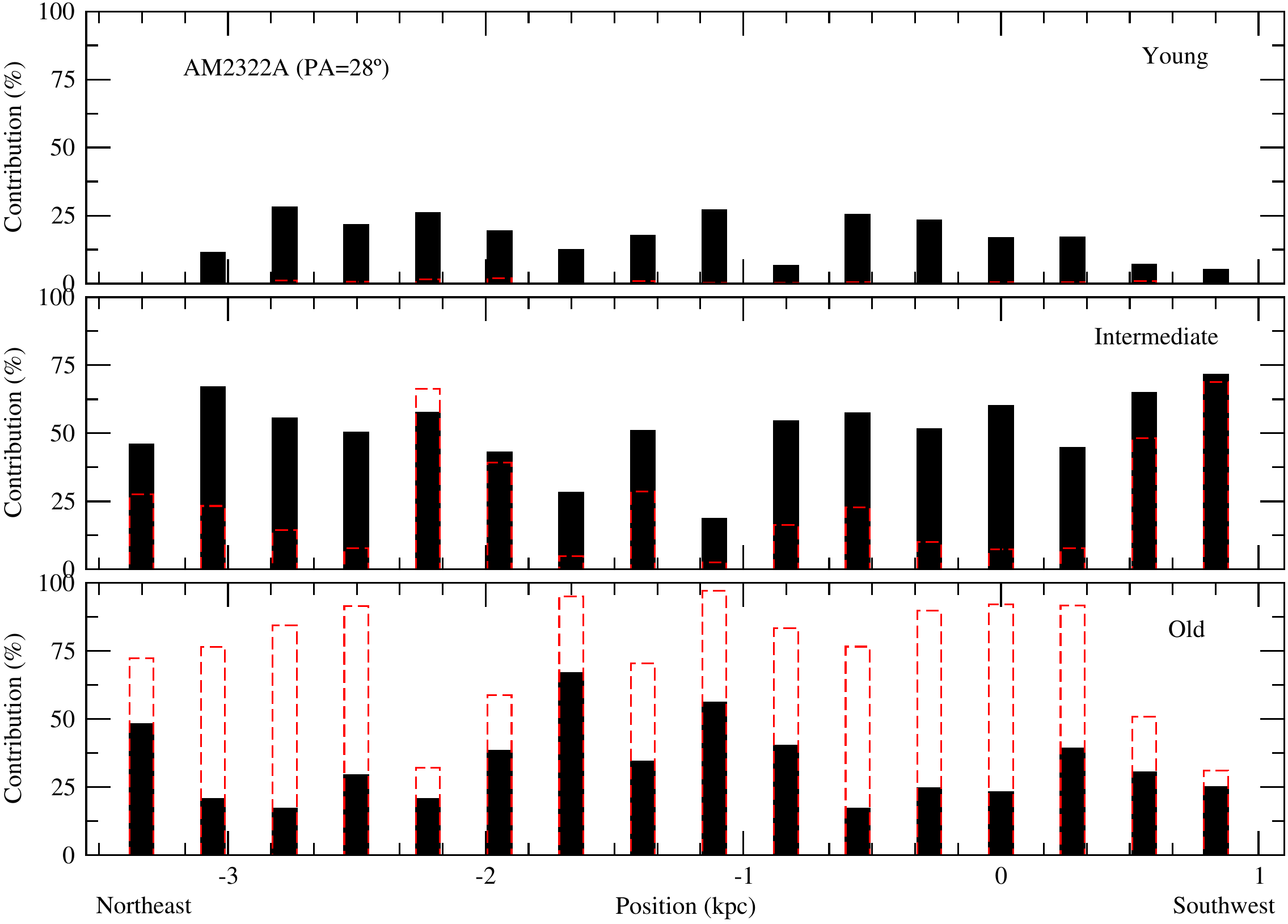}
\includegraphics*[angle=0,width=0.45\columnwidth]{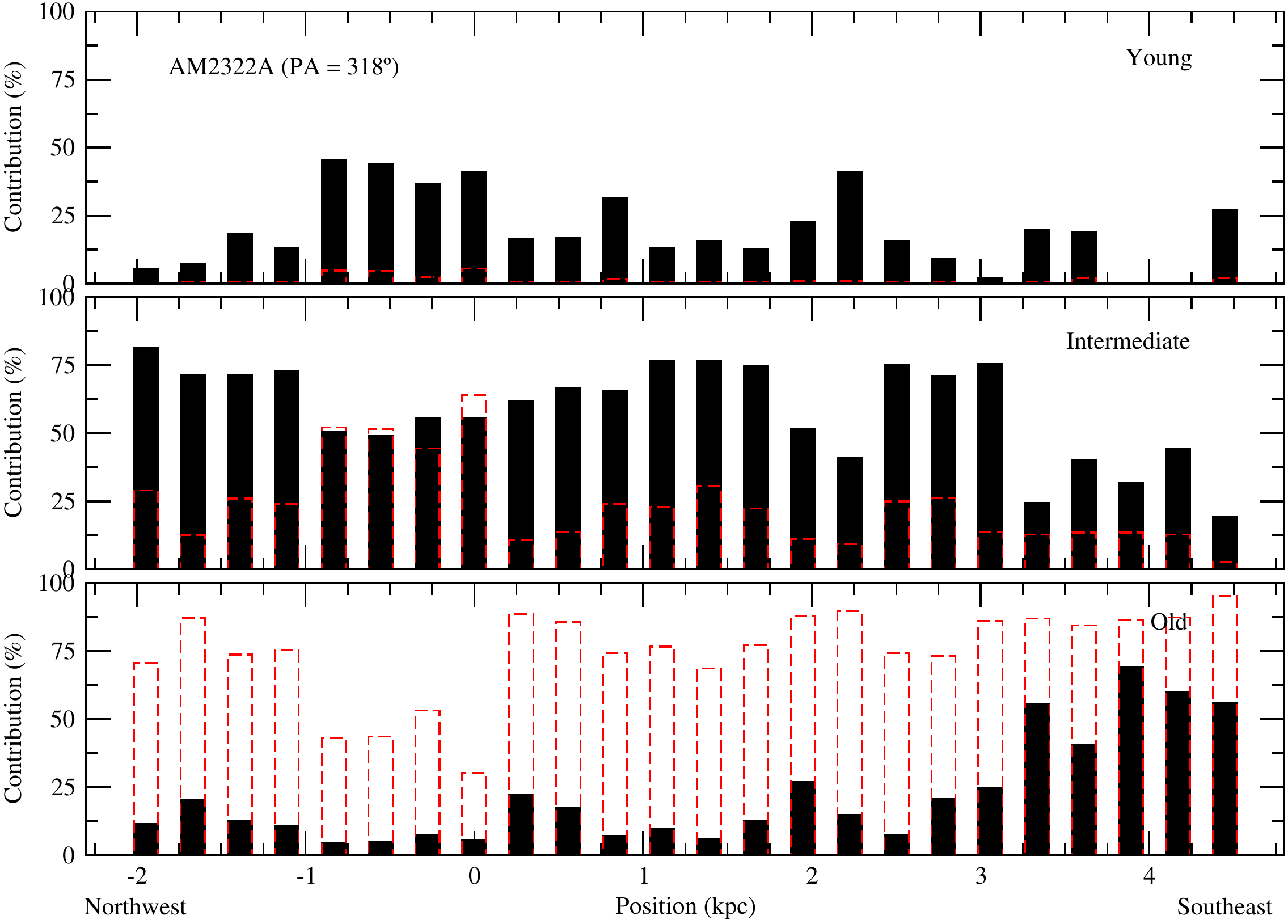}
\includegraphics*[angle=0,width=0.45\columnwidth]{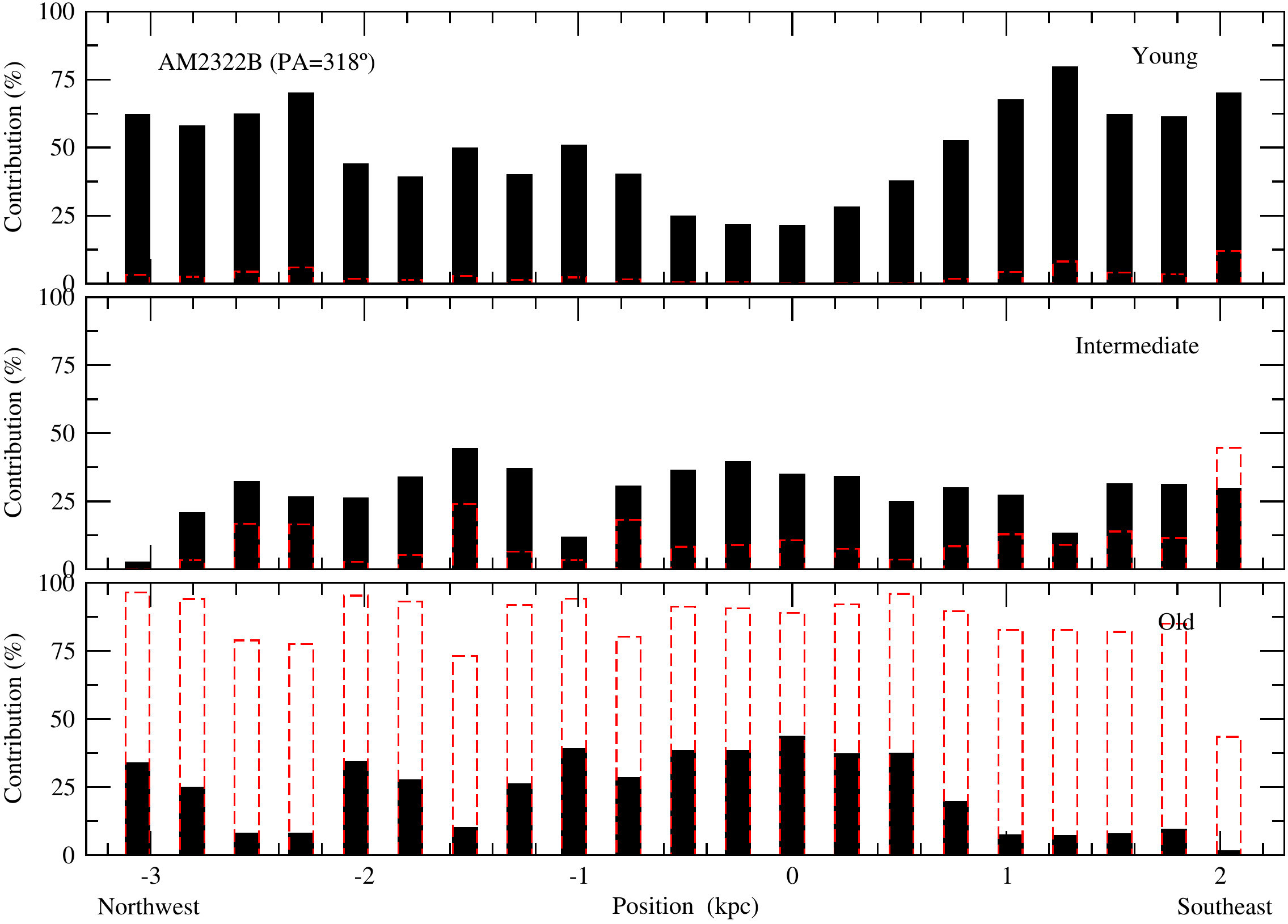}
\caption{Same as Fig.~\ref{sin_1054AB}, but for AM\,2322A and AM\,2322B.}
\label{sin_2322A}
\end{figure*}

\end{appendices}
%\end{onecolumn}
\label{lastpage}

\end{document}